\documentclass[12pt]{article}
\usepackage{amsfonts}
\usepackage{amssymb}
\usepackage{amsbsy}

\input epsf.sty

\newcommand{\BEQ}{\begin{equation}}     
\newcommand{\BEA}{\begin{eqnarray}}
\newcommand{\EEQ}{\end{equation}}       
\newcommand{\EEA}{\end{eqnarray}}
\newcommand{\eps}{\varepsilon}          
\newcommand{\del}{\delta}

\newcommand{\g}{{\mathfrak{g}}}
\newcommand{\so}{{\mathfrak{so}}}
\newcommand{\h}{{\mathfrak{h}}}

\newcommand{\s}{{\mathfrak{s}}}
\newcommand{\n}{{\mathfrak{n}}}

\newcommand{\p}{{\mathfrak{p}}}
\newcommand{\sv}{{\mathfrak{sv}}}

\newcommand{\kk}{{\mathfrak{k}}}

\newcommand{\md}{{\mathfrak{md}}}

\newcommand{\slin}{{\mathfrak{sl}}}

\newcommand{\sch}{{\mathfrak{sch}}}

\newcommand{\Sch}{{\mathrm{Sch}}}
\newcommand{\conf}{{\mathfrak{conf}}}

\newcommand{\Ad}{{\mathrm{Ad}}}
\newcommand{\Id}{{\mathrm{Id}}}
\newcommand{\Ker}{{\mathrm{Ker}}}
\newcommand{\gal}{{\mathfrak{gal}}}
\newcommand{\Gal}{{\mathrm{Gal}}}
\newcommand{\eucl}{{\mathfrak{eucl}}}

\newcommand{\eop}{$\Box$}
\newcommand{\Vect}{{\mathrm{Vect}}}
\newcommand{\diag}{{\mathrm{diag}}}
\newcommand{\Vir}{{\mathrm{Vir}}}
\newcommand{\vir}{{\mathfrak{vir}}}
\newcommand{\wid}{{\mathrm{wid}}}
\newcommand{\ad}{{\mathrm{ad}}}
\newcommand{\fsv}{{\mathfrak{fsv}}}
\newcommand{\tsv}{{\mathfrak{tsv}}}
\newcommand{\Diff}{{\mathrm{Diff}}}

\newcommand{\al}{\alpha}

\newcommand{\half}{{1\over 2}}
\newcommand{\Del}{\Delta}

\newcommand{\R}{\mathbb{R}}
\newcommand{\N}{\mathbb{N}}
\newcommand{\C}{\mathbb{C}}
\newcommand{\Z}{\mathbb{Z}}
\newcommand{\T}{\mathbb{T}}
\catcode`\@=11
\def\numberbysection{\@addtoreset{equation}{section}
        \def\theequation{\thesection.\arabic{equation}}}
\numberbysection

\begin{document}

\begin{titlepage}

\vskip 1.5 cm
\begin{center}
{\Large \bf The Schr\"odinger-Virasoro Lie group and algebra : from geometry to representation theory.}
\end{center}

\vskip 2.0 cm
   
\centerline{ {\bf Claude Roger}$^a$ and {\bf J\'er\'emie Unterberger}$^b$}
\vskip 0.5 cm
\centerline {$^a$ Institut Camille Jordan, \footnote{Unit\'e associ\'ee
au CNRS UMR 5208} 
Universit\'e Claude Bernard Lyon 1,} 
\centerline{21 avenue Claude Bernard, 69622 Villeurbanne Cedex, France}
\vskip 0.5 cm
\centerline {$^b$Institut Elie Cartan,\footnote{Laboratoire 
associ\'e au CNRS UMR 7502} Universit\'e Henri Poincar\'e Nancy I,} 
\centerline{ B.P. 239, 
54506 Vand{\oe}uvre l\`es Nancy Cedex, France}

\begin{abstract}

This article is concerned with  an extensive study of a
 infinite-dimensional Lie
algebra $\sv$, introduced in \cite{Henk94} in the context 
of non-equilibrium statistical
physics, containing as subalgebras both the Lie algebra of
 invariance of the free
Schr\"odinger equation and the central charge-free Virasoro 
algebra $\Vect(S^1).$
We call $\sv$ the {\it Schr\"odinger-Virasoro Lie algebra}. 
We choose to present $\sv$
from a Newtonian geometry point of view first, and then in
 connection with conformal
and Poisson geometry. We turn afterwards to its representation 
theory: realizations
as Lie symmetries of field equations, coadjoint representation,
 coinduced representations
in connection with Cartan's prolongation method (yielding analogues
 of the tensor
density modules for $\Vect(S^1)$), and finally Verma modules with a
 Kac determinant
formula. We also present a detailed cohomogical study, providing
 in particular a
classification of deformations and central extensions; there appears 
a non-local cocycle.

\vskip 3 cm
{\centerline{\it {\small in memory of Daniel Arnaudon}}}

\end{abstract}
\end{titlepage}

\setcounter{section}{-1}
\section{Introduction}

There is, in the physical literature of the past decades - without mentioning
the pioneering works of Wigner for instance -, a deeply rooted
belief that physical systems - macroscopic systems for statistical physicists,
quantum particles and fields for high energy physicists - could and should be classified
according to which group of symmetries act on them and how this group
acts on them.

Let us just point at two very well-known examples : elementary  particles on 
the $(3+1)$-dimensional Minkowski space-time, and two-dimensional conformal field theory.

From the point of view of 'covariant quantization', introduced at the time
of Wigner, elementary  particles of relativistic quantum mechanics
(of positive mass, say) may be described as
irreducible unitary representations of the Poincar\'e group $\p_4\simeq
\so(3,1)\ltimes\R^4$, which is the semi-direct product of the Lorentz group
of rotations and relativistic boosts by space-time translations : that is to say, the physical states of a particle of mass $m>0$ and spin $s\in\half\N$ are in bijection with the states of the Hilbert space corresponding to the associated
irreducible representation of $\p_4$; the indices $(m,s)$ characterizing
positive square mass representations come from the two Casimir of the
enveloping algebra ${\cal U}(\p_4)$.

This 'covariant quantization' has been revisited by the school of Souriau 
in the 60'es and 70'es as a particular case of geometric quantization; most
importantly for us, the physicists J.-J. L\'evy-Leblond and C. Duval introduced the
so-called {\it Newton-Cartan} manifolds (which provide the right geometric
frame for Newtonian mechanics, just as Lorentz manifolds do for relativistic
mechanics) and applied the tools of geometric quantization to construct wave
equations in a geometric context.

\vskip 10 pt

Two-dimensional conformal field theory is an attempt at understanding the 
universal behaviour of two-dimensional statistical systems at equilibrium and at the critical
temperature, where they undergo a second-order phase transition. Starting from the basic
assumption of translational and rotational invariance, together with the fundamental hypothesis
(confirmed by the observation of the fractal structure of the systems and
the existence of long-range correlations, and made into a cornerstone of
renormalization-group theory) that scale invariance holds at criticality, one
is \footnote{for systems with sufficiently short-ranged interactions} 
naturally led to the idea that invariance under the whole conformal group
Conf$(d)$ should also hold. This group is known to be finite-dimensional as
soon as the space dimension $d$ is larger than or equal to three, so physicists
became very interested in dimension $d=2$, where local conformal transformations 
are given by holomorphic or anti-holomorphic functions. A systematic investigation
of the theory of representations of the Virasoro algebra (considered as a central extension of the algebra of infinitesimal holomorphic transformations) in the 80'es led to introduce a class
of physical models (called {\it unitary minimal models}), corresponding to the unitary highest weight 
representations of the Virasoro algebra with central charge less than one. Miraculously, covariance alone is
enough to allow  the computation of the statistic correlators -- or so-called '$n$-point functions' --
for these highly constrained models.

\vskip 10 pt

We shall  give here a tentative mathematical foundation (though very sketchy
at present time, and not pretending to have physical applications!) to closely
related ideas, developed   since the mid-nineties (see short
survey \cite{Henk05a}), and applied to two related fields : 
strongly anisotropic critical systems  and out-of-equilibrium
statistical physics (notably ageing phenomena). Theoretical studies and numerical models coming from both
fields have been developed, in which invariance under space rotations and
{\it anisotropic dilations} $(t,r)\to (e^{\lambda z} t,e^{\lambda} r)$
 ($\lambda\in\R$) plays a central r\^ole. Here $r\in\R^d$ is considered as a space coordinate and $t\in\R$ is (depending on the context) either the time
coordinate or an extra (longitudinal, say) space coordinate; the parameter
$z\not=1$ is called the {\it anisotropy} or {\it dynamical exponent}.

We shall here restrict (at least most of the time) to the value $z=2$. Then the simplest wave equation 
invariant under translations, rotations {\it and} anisotropic dilations 
 is the free Schr\"odinger equation
$2{\cal M}\partial_t \psi=\Delta_d \psi$, where $\Delta_d:=\sum_{i=1}^d
\partial_{r_i}^2$ is the Laplacian in spatial coordinates. So it is natural
to believe that this equation should play the same r\^ole as the Klein-Gordon
equation in the study of relativistic quantum particles, or the Laplace equation
in conformal field theory, whose maximal group of Lie symmetries is the conformal group; in other words, one may also say that we are looking for symmetry
groups arising naturally in a non-relativistic setting, while hoping that
their representations might be applied to a classification of non-relativistic
systems, or, more or less equivalently, to $(z=2)$ anisotropic systems.

This program, as we mentioned earlier in this introduction, was partially
carried out by Duval, K\"unzle and others through the 70'es and 80'es
(see for instance [Duv1,Duv2,Duv3,Duv4]). We
shall discuss it briefly in the first part and see how the maximal group
of Lie symmetries of the Schr\"odinger group, $\Sch_d$,  called
the {\it Schr\"odinger group}, appears in some sense as the natural substitute for the conformal group in Newtonian mechanics. Unfortunately, it is finite-dimensional for every value of $d$, and its unitary irreducible representations
are well-understood and classified (see \cite{Per}), giving very interesting though
partial informations on two- and three-point functions in anisotropic and
out-of-equilibrium statistical physics at criticality
that have been systematically pursued in the past ten years of so (see \cite{Unt1,Unt2,Unt3,Henk02,Henk04,Bau05})
 but relying on
rather elementary mathematics, so this story could well
have stopped here short of further arguments. 

Contrary to the conformal group though, which corresponds to a rather 'rigid'
riemannian or Lorentzian geometry, the Schr\"odinger group is only one of
the groups of symmetries that come out of the much more 'flexible' Newtonian
geometry, with its loosely related time and space directions. In particular (restricting
here to one space dimension for simplicity, although there are straightforward generalizations
in higher dimension),
there arises a new group SV, which will be our main
object of study, and that we shall call the {\it Schr\"odinger-Virasoro} group for reasons
that will become clear shortly.
Its Lie algebra $\sv$ was originally introduced by M. Henkel in 1994 (see 
\cite{Henk94}) as a by-product of the computation of $n$-point functions
that are covariant under the action of the Schr\"odinger group.
It is given abstractly as
\BEQ
\sv=\langle L_n\rangle_{n\in\Z}\oplus\langle Y_m\rangle_{m\in\half+\Z}\oplus
\langle M_p\rangle_{p\in\Z}
\EEQ with relations
\BEQ
[L_n,L_p]=(n-p)L_{n+p}
\EEQ
\BEQ
[L_n,Y_m]=({n\over 2}-m)Y_{n+m},\ [L_n,M_p]=-pM_p
\EEQ
\BEQ
[Y_m,Y_{m'}]=(m-m')M_{m+m'}, \ [Y_m,M_p]=0,\ [M_n,M_p]=0 
\EEQ

\noindent ($n,p\in\Z, m,m'\in\half+\Z$). Denoting by $\Vect(S^1)=\langle L_n\rangle_{n\in\Z}$ the Lie algebra of
vector fields on the circle (with brackets $[L_n,L_p]=(n-p)L_{n+p}$), 
$\sv$ may be viewed as a semi-direct product $\sv\simeq\Vect(S^1)\ltimes\h$,
 where
$\h=\langle Y_m\rangle_{m\in\half+\Z}\oplus\langle M_p\rangle_{p\in\Z}$ is
a two-step nilpotent Lie algebra, isomorphic to ${\cal F}_{\half}\oplus{\cal F}_0$ as a $\Vect(S^1)$-module (see Definition 1.3 for notations). 

\vskip 10 pt

This articles aims at motivating the introduction of the $\sv$-algebra and
related objects, and at studying them from a mathematical point of view. 
It is not an easy task to choose the best order of exposition since, as usual
in mathematics, the best motivation for introducing a new object is often provided by what can be done with it, and also because the $\sv$ algebra actually
appears in many contexts (not only in the geometric approach chosen in this
introduction), and it is actually a matter of taste to decide which definition
is most fundamental. Generally speaking, though, we shall be more concerned with giving various definitions of $\sv$ in the two or three first parts, and with
the study of $\sv$ proper and its representations in the rest of the article.

Here is the plan of the article.

Chapter 1 will be devoted to a geometric introduction to $\sv$ and $SV$ in the
frame of Newton-Cartan manifolds, as promised earlier in this introduction.

In Chapter 2, we shall prove that $\sch^d=Lie(\Sch^d)$ appears as a real
subalgebra of $\conf(d+2)_{\C}$, extending results contained in a  previous
article of M. Henkel and one of the authors (see \cite{Unt1}). We shall also give a 'no-go'
theorem proving that this embedding cannot be extended to $\sv$.

In Chapter 3, we shall decompose $\sv$ as a sum of 
tensor density modules for $\Vect(S^1)$. Introducing its central extension
$\widehat{\sv}\simeq \Vir\ltimes\h$ which contains both the Virasoro algebra
and the Schr\"odinger algebra (hence its name!), we shall study its
coadjoint action on its regular dual $(\widehat{\sv})^*$. We shall also study
 the
action of $\sv$ on a certain 
space of Schr\"odinger-type  operators and on  some other spaces
of operators related to field equations. 

In Chapter 4, we shall see that 'half of $\sv$' can be interpreted as a Cartan
prolongation $\oplus_{k=-1}^{\infty} \g_k$ with $\g_{-1}\simeq\R^3$ and $\g_0$
three-dimensional solvable, and study the related co-induced representations
by analogy with the case of the algebra of formal vector fields on $\R$, where this
method leads to the tensor density modules of the Virasoro representation
theory.

Chapter 5 is  devoted to a systematic study of deformations and central
extensions of $\sv$.

Finally, we shall study in Chapter 6 the  Verma
 modules of $\sv$ and of some related algebras, and  the associated Kac determinant formulae.

Let us remark, for the sake of completeness, that we volontarily skipped a promising 
construction of $\sv$ and a family of supersymmetric extensions of $\sv$
 in terms of quotients
of the Poisson algebra on the torus or of the algebra of pseudo-differential
operators on the line that appeared elsewhere (see \cite{Unt3}). Also, representations
of $\sv$ into vertex algebras have been investigated (see \cite{Unt4}). 

\section{Geometric definitions of $\sv$}

\subsection{From Newtonian mechanics to the Schr\"odinger-Virasoro
algebra}

Everybody knows, since Einstein's and Poincar\'e's discoveries in the
early twentieth century,
that Lorentzian geometry  lies at the heart of relativistic mechanics; the
geometric formalism
 makes it possible
to define the equations of general relativity
in a coordinate-free, covariant way, in all
generality.

Owing to the success of the theory of general relativity, it was
natural that one should also think of geometrizing Newtonian mechanics.
This gap was filled in about half a century later, by several authors,
 including
J.-M. L\'evy-Leblond, C. Duval, H.P. K\"unzle and others (see for instance
\cite{Bur,LL,Duv1,Duv2,Duv3,Duv4}), leading to a geometric
reformulation of Newtonian mechanics on the so-called Newton-Cartan manifolds,
and also to the discovery of new fundamental field equations for Newtonian
particles.

Most Lie algebras and groups that will constitute our object of study
in this article appear to be closely related to the Newton-Cartan
geometry. That is why - although we shall not produce any new
result in that particular field - we chose to give a very short
introduction to
Newton-Cartan geometry, whose main objective is to lead as quickly as
possible to a definition of the Schr\"odinger group and its
infinite-dimensional generalization, the {\it Schr\"odinger-Virasoro group}.

{\bf Definition 1.1}

{\it  A Newton-Cartan manifold of dimension $(d+1)$ is a $C^{\infty}$
manifold $M$ of dimension $(d+1)$
provided with a closed one-form $\tau$, a degenerate symmetric {\it
contravariant} non-negative two-tensor $Q\in {\mathrm{Sym}}(T^2 M)$ with
 one-dimensional kernel generated by $\tau$ and a connection $\nabla$
preserving $\tau$ and $Q$.}

A right choice of local coordinates makes it
 clear why these data provide the right framework for Newtonian mechanics.
Locally, one may put $\tau=dt$
for a certain time coordinate $t$, and $Q=\sum_{i=1}^d \partial_{r_i}^2$
for a choice of $d$ space coordinate
vector fields $(r_i)$ on the hyperplane orthogonal to $dt$. The standard
Newton-Cartan manifold is the
flat manifold  $\R\times\R^d$
with coordinates $(t,r_1,\ldots,r_d)$ and $\tau,Q$ given globally by the
above formulas.

The analogue of the Lie algebra of infinitesimal isometries in Riemannian
or Lorentzian geometry is here the  Lie algebra of infinitesimal vector
fields $X$ preserving $\tau,Q$ and
$\nabla$. In the case of the flat manifold $\R\times\R^d$, it is equal
to  the well-known Lie  algebra of Galilei transformations $\gal_d$, namely,
 can be written
\BEQ
\gal_{d}=\langle X_{-1}\rangle\oplus\langle
Y_{-\half}^i,Y_{\half}^i\rangle_{1\le i\le d} \oplus \langle
 {\cal R}_{ij}\rangle_{1\le i<j\le d},
 \EEQ

including the time and space translations $X_{-1}=- \partial_t$,
$Y_{-\half}^i=-  \partial_{r_i}$, the generators of motion with
constant speed $Y_{\half}^i=-
t\partial_{r_i}$ and rotations ${\cal
R}_{ij}=r_{i}\partial_{r_{j}}-r_{j}\partial_{r_{i}}.$

It was recognized, mainly by Souriau, that (from the point of view of
symplectic geometry) many  original features of classical mechanics
stem from the existence of the two-cocycle $c\in H^{2}(\gal_d,\R)$
of the Galilei Lie algebra defined by
\BEQ
c(Y_{-\half}^i,Y_{\half}^j)=\del_{i,j},\quad
i,j=1,\ldots,d
\EEQ
leading to the definition of the mass as a central charge. We shall
denote by $\widetilde{\gal}_{d}$  the centrally extended Lie algebra.
Its has a  one-dimensional center,  generated by
$M_{0}$, and a modified Lie bracket
such that $[Y_{\half}^i,Y_{-\half}^j]=\del_{i,j}M_{0},$
while all the other relations remain unmodified. Schur's Lemma implies
that the generator $M_{0}$ is scalar on
irreducible representations of $\widetilde{\gal_{d}}$; the next
Proposition shows that it is natural to call {\it mass} the value of
$M_{0}$.

{\bf Definition 1.2}
{\it The Schr\"odinger Lie algebra in $d$ dimensions, denoted by
$\sch_{d}$, is the Lie algebra with generators $$X_{-1},X_{0},X_{1},
Y_{-\half}^i, Y_{\half}^i \ (i=1,\ldots, d), \ {\cal R}_{ij}\
(1\le i<j\le d),$$ isomorphic to the semi-direct product
$\slin(2,\R)\ltimes \widetilde{\gal}_{d}\simeq\langle
X_{-1},X_{0},X_{1}\rangle\ltimes \widetilde{\gal}_{d}$, with the following
choice of relations for $\slin(2,\R)$
\BEQ
[X_{0},X_{-1}]=X_{1},[X_{0},X_{1}]=-X_{1}, [X_{1},X_{-1}]=2X_{0}
\EEQ
and an action of $\slin(2,\R)$ on $\gal_{d}$ defined by
\BEQ
[X_{n},Y^i_{-\half}]=({n\over 2}+\half)Y^i_{n-\half},
[X_{n},Y^i_{\half}]=({n\over 2}-\half)Y^i_{n+\half} \quad (n=-1,0,1)
\label{gl:defsch1};
\EEQ
\BEQ
[X_{n},{\cal R}^{ij}]=0, \quad [X_{n},M_{0}]=0\label{gl:defsch2}.
\EEQ }

Note that the generator $M_{0}$ remains central in the semi-direct
product: hence, it still makes sense to speak about the mass of an
irreducible representation of $\sch_{d}$.

The motivation for this definition (and also the reason for this name)
lies in the following classical
Proposition.

We denote by $\Del_{d}=\sum_{i=1}^d \partial_{r_{i}}^{2}$ the
usual Laplace operator on $\R^d$.

{\bf Proposition 1.1}

{\em

\begin{enumerate}
    \item
     {\em (see \cite{Duv2})} The Lie algebra of projective (i.e. conserving
geodesics)
     vector fields $X$ on $\R\times\R^d$ such that there exists a function
$\lambda\in C^{\infty}(\R\times\R^d)$ with
     \BEQ
{\cal L}_X(dt)=\lambda\ dt,\quad {\cal L}_X(\sum_{i=1}^d
\partial_{r_i}^2)=-\lambda\sum_{i=1}^d \partial_{r_i}^2
\label{gl:proj}
\EEQ
is generated by the  vector fields
$$
L_{-1}=-\partial_{t}, L_{0}=-t\partial_{t}-\half \sum_i r_i\partial_{r_i},
L_{1}=-t^{2} \partial_{t}-t\sum_i r_i\partial_{r_i}
$$
$$
Y_{-\half}^i=-\partial_{r_{i}}, Y_{\half}^i=-t\partial_{r_{i}}
$$
\BEQ
{\cal
R}^{ij}=r_{i}\partial_{r_{j}}-r_{j}\partial_{r_{i}},\quad 1\le
i<j\le d
\EEQ
that define a massless representation of the Schr\"odinger Lie algebra
$\sch_{d}$.

\item
{\em (see \cite{Fus93})} Let ${\cal M}\in\C$.
The Lie algebra of differential
operators $\cal X$ on $\R\times\R^d$, of order at most one, preserving
the space of solutions of the free Schr\"odinger equation
\BEQ
(2{\cal M}\partial_{t}-\Del_{d})\psi=0,
\EEQ that is, verifying
$$
(2{\cal M}\partial_{t}-\Del_{d}){\cal X}\psi=0
$$ for every function $\psi$ such that $(2{\cal
M}\partial_{t}-\Del_{d})\psi=0$, gives a representation $d\pi_{d/4}^d$ of mass
$-{\cal M}$ of the Schr\"odinger
algebra $\sch_{d}$, with the following realization:
$$
d\pi_{d/4}^d(L_{-1})=-\partial_{t},\ d\pi_{d/4}^d(
L_{0})=-t\partial_{t}-\half \sum_i r_i\partial_{r_i}-{d\over 4},\
d\pi_{d/4}^d(L_{1})=-t^{2} \partial_{t}-t\sum_i r_i\partial_{r_i}-\half {\cal M}r^{2}-\frac{d}{2}t
$$
$$
d\pi_{d/4}^d(Y_{-\half}^i)=-\partial_{r_{i}},\
d\pi_{d/4}^d(Y_{\half}^i)=-t\partial_{r_{i}}-{\cal M}r_{i},\
d\pi_{d/4}^d(M_{0})=-{\cal M}
$$
\BEQ
d\pi_{d/4}^d({\cal
R}^{ij})=r_{i}\partial_{r_{j}}-r_{j}\partial_{r_{i}},\quad 1\le
i<j\le d. \label{gl:sch}
\EEQ

\end{enumerate} }

Let us mention, anticipating on Section 1.2, that  realizing instead $L_0$ by the operator
$-t\partial_t-\half \sum_i r_i\partial_{r_i}-\lambda$ and $L_1$ by $-t^2
\partial_t-t\sum_i r_i\partial_{r_i}-\half {\cal M}r^2
-2\lambda t$ $(\lambda\in\R)$, leads to a family of representations
$d\pi_{\lambda}^d$ of $\sch_d$. This
accounts for the parameter ${d\over 4}$ in our definition of
$d\pi_{d/4}^d$. The parameter $\lambda$
may be interpreted physically as the {\it scaling dimension} of the field
on which $\sch_d$ acts,
$\lambda={d\over 4}$ for solutions of the free Schr\"odinger equation in
$d$ space dimensions.

We shall also frequently use the realization $d\tilde{\pi}_{\lambda}^d$
of the Schr\"odinger
algebra given by the Laplace transform of the above generators with
respect to the mass, which is formally equivalent to replacing the
parameter ${\cal M}$ in
the
above formulas by $\partial_{\zeta}$. This simple transformation leads
to a representation $d\tilde{\pi}_{\lambda}^d$ of $\sch^d$; let us consider
the particular case $\lambda=0$ for simplicity. Then ,
 $d\tilde{\pi}^d:=d\tilde{\pi}^d_0$ gives a realization of
$\sch^d$ by vector fields on $\R^{d+2}$ with coordinates $t,r_{i}$
$(i=1,\ldots,d)$ and $\zeta$. Let us write the action of the generators
for further reference (see \cite{Unt1}):

$$
d\tilde{\pi}^d(L_{-1})=-\partial_{t},d\tilde{\pi}^d(
L_{0})=-t\partial_{t}-\half \sum_i r_i\partial_{r_i},
d\tilde{\pi}^d(L_{1})=-t^{2} \partial_{t}-t\sum_i r_i\partial_{r_i}-\half
r^{2}\partial_{\zeta}
$$
$$
d\tilde{\pi}^d(Y_{-\half}^i)=-\partial_{r_{i}},
d\tilde{\pi}^d(Y_{\half}^i)=-t\partial_{r_{i}}-r_{i}\partial_{\zeta},
 d\tilde{\pi}^d(M_{0})=-\partial_{\zeta}
$$
\BEQ
d\tilde{\pi}^d({\cal
R}^{ij})=r_{i}\partial_{r_{j}}-r_{j}\partial_{r_{i}},\quad 1\le
i<j\le d. \label{gl:schzeta}
\EEQ

So, according to the context, one may use either the representation by
differential operators on $\R^{d+1}$
of order one, or  the representation by vector fields on $\R^{d+2}$. Both
points of view prove to
 be convenient. We shall see later  (see Section 1.2)  that these
representations extend to
representations of  the Schr\"odinger-Virasoro algebra, and give a more
intrinsic interpretation of
this formal Laplace transform.

{\bf Proposition 1.2.}

{\it The Lie algebra of vector fields $X$ on $\R\times\R^d$ - not
necessarily preserving the connection -  such that
equations (\ref{gl:proj}) are verified
 is generated by the
following set of transformations :
\begin{enumerate}
\item[(i)] $$L_{f}=-f(t)\partial_t-\half f'(t)\sum_{i=1}^d
r_i\partial_{r_i} \quad (
{\mathrm{Virasoro-like\ transformations}})$$

\item[(ii)] $$Y_{g_{i}}^i=-g_i(t)\partial_{r_i} \quad
({\mathrm{time-dependent\
space\ translations}}) $$

\item[(iii)] $${\cal R}_{h_{ij}}^{ij}=-h_{ij}(t)(r_i \partial_{r_j}-r_j
\partial_{r_i}) ,
\ 1\le i<j\le d
\quad
({\mathrm{time-dependent\ space\ rotations}})$$
\end{enumerate}
where $f,g_{i},h_{ij}$ are arbitrary functions of $t$. }

{\bf Proof.} Put $r=(r_1,\ldots,r_d)$. Let
$X=f(t,r)\partial_t+\sum_{i=1}^d g_i(t,r)\partial_{r_i}$ verifying
the conditions of Proposition 1.1.
Then ${\cal L}_X dt=df$ is compatible with the first condition if $f$
depends on time only, so $\lambda=f'$
is also a function of time only. Hence
\BEA
{\cal L}_X \left( \sum_{j=1}^d \partial_{r_j}^2\right)&=& -2 \sum_{i,j}
\partial_{r_j}\otimes
\partial_{r_j}g_i \ \partial_{r_i} \nonumber \\
&=& -2\sum_{j} \partial_{r_j}\otimes \partial_{r_j}g_j \ \partial_{r_j} -
2 \sum_{i\not=j} \partial_{r_j}\otimes
\partial_{r_j}g_i \ \partial_{r_i} \nonumber
\EEA
so $\partial_{r_j}g_i=-\partial_{r_i}g_j$ if $i\not=j$, which gives the
time-dependent rotations, and
$2\partial_{r_j}g_j=f'$ for $j=1,\ldots,d$, which gives the Virasoro-like
transformations and the time-dependent
translations. \eop

Note that the Lie algebra of Proposition 1.1 (1), corresponds to
$f(t)=1,t,t^2$,
$g_i(t)=1,t$ and $h_{ij}(t)=1$.

One easily sees that the ${\cal R}_{h_{ij}}^{ij}$ generate the algebra of
currents on $\so(d)$, while the ${\cal R}_{h_{ij}}^{ij}$ and the
$Y_{g_{i}}^i$ generate together the algebra of currents on the
Euclidean Lie algebra $\eucl(d)=\so(d)\ltimes\R^d.$ The transformations
$L_{f}$ generate a copy of the Lie algebra of tangent vector fields on
$\R$, denoted by $\Vect(\R)$, so one has
\BEQ
[L_{f},L_{g}]=L_{\{f,g\}}
\EEQ where $\{f,g\}=f'g-fg'.$ So this Lie algebra can be described
algebraically as a semi-direct product
$\Vect(\R)\ltimes\eucl(d)_{\R}$,
where $\eucl(d)_{\R}$ stands for the Lie algebra of currents with
values in $\eucl(d)$. In our realization, it is embedded as a
subalgebra
of $\Vect(\R\times\R^d).$

For both topological and algebraic reasons, we shall from now
on compactify the
 the $t$ coordinate. So we work now on
$S^{1}\times\R^d$, and $\Vect(\R)\ltimes\eucl(d)_{\R}$ is replaced
by
$\Vect(S^{1})\ltimes\eucl(d)_{S^{1}},$ where  $\Vect(S^{1})$
stands for the famous (centerless) Virasoro algebra.

It may be the right place to recall some well-known facts about the Virasoro
algebra, that we shall use throughout the article.

We represent an element of $\Vect(S^1)$ by the vector field $f(z)\partial_z$,
where $f\in\C[z,z^{-1}]$ is a Laurent polynomial. Vector field brackets
$[f(z)\partial_{z},g(z)\partial_z]=(fg'-f'g)\partial_z$, may equivalently
be rewritten in the basis $(l_n)_{n\in\Z}$, $l_n=-z^{n+1}\partial_z$
(also called Fourier components), which yields
$[\ell_n,\ell_m]=(n-m)\ell_{n+m}$. Notice the unusual choice of signs,
justified (among other arguments) by the precedence of \cite{Henk94}
 on our subject.

The Lie algebra $\Vect(S^1)$ has only one non-trivial
 central extension (see \cite{GuiRog05} or \cite{Kac} for instance), given by the
so-called Virasoro cocycle $c\in Z^2(\Vect(S^1),\R)$  defined by
\BEQ
c(f\partial_z,g\partial_z)=\int_{S^1} f'''(z)g(z)\ dz,
\EEQ
or, in Fourier components,
\BEQ
c(\ell_{n},\ell_{m})=\del_{n+m,0} {(n+1)n(n-1)}.
\EEQ
The resulting centrally extended  Lie algebra, called {\it Virasoro algebra},
 will be denoted by $\vir$.

The Lie  algebra $\Vect(S^{1})$ has a one-parameter
family of representations ${\cal F}_{\lambda},\lambda\in\R$.

{\bf Definition 1.3.}

{\it We denote by ${\cal F}_{\lambda}$ the representation of $\Vect(S^1)$
on $\C[z,z^{-1}]$ given by
\BEQ
\ell_n.z^m=(\lambda n-m)z^{n+m},\quad n,m\in\Z.
\EEQ }

 An
element of ${\cal F}_{\lambda}$ is naturally understood as a
$(-\lambda)$-density $\phi(z)dz^{-\lambda}$, acted by $\Vect(S^{1})$ as
\BEQ
f(z)\partial_z. \phi(z)dz^{-\lambda}=(f\phi'-\lambda f'\phi)(z) dz^{-\lambda}.
\EEQ
In the bases $\ell_n=-z^n\partial_z$ and $a_m=z^m dz^{-\lambda}$, one
gets $\ell_n . a_m=(\lambda n-m)a_{n+m}.$

Replacing formally $t$ by the compactified variable $z$ in the formulas
of Proposition 1.2, and putting $f(z)=-z^{n+1}, g_i(z)=-z^{n+\half},
h_{ij}(z)=-z^n,$ one gets a realization of $\Vect(S^1)\ltimes\eucl(d)_{S^1}$
as a Lie subalgebra of $\Vect(S^1\times\R^d)$ generated by $L_{n},
Y_{m_i}^i, {\cal R}_{p_{ij}}^{ij}$ (with integer indices $n$ and $p_{ij}$
and half-integer indices $m_i$), with
 the following set of relations:
$$
[L_n,L_p]=(n-p)L_{n+p}
$$
$$
[L_n,Y^i_m]=({n\over 2}-m)Y^i_{n+m},\ [L_n,{\cal R}^{ij}_p]=-p{\cal
R}^{ij}_{n+p}
$$
$$
[Y^i_m,Y^j_{m'}]=0, [{\cal R}^{ij}_p,Y^k_m]=\del_{j,k}
Y^i_{m+p}-\del_{i,k} Y^j_{m+p}
$$
\BEQ
 [{\cal R}^{ij}_n,{\cal R}^{kl}_p]=\del_{j,k}{\cal
R}^{il}_{n+p}+\del_{i,l} {\cal R}^{jk}_{n+p}-\del_{j,l}{\cal
R}^{ik}_{n+p}-\del_{i,k}{\cal R}^{jl}_{n+p}
\EEQ

With the above definitions, one sees immediately that, under the
action of $\langle L_{f}\rangle_{f\in C^{\infty}(S^{1})}\simeq
\Vect(S^{1})$, the $(Y^i_{m})_{m\in\half+\Z}$ behave as elements of
the module
${\cal F}_{\half}$, while the $({\cal R}^{ij}_{m})_{m\in\Z}$ and
the $(M_{m})_{m\in\Z}$ define several copies of ${\cal F}_{0}$.

The commutative Lie algebra generated by the $Y^i_n$ has an infinite
family of
central extensions. If we want to leave unchanged the action of
$\Vect(S^1)$ on the
$Y^i_n$ and to extend the action of $\Vect(S^1)$ to the central charges,
though, the
most natural possibility (originally discovered by M. Henkel, see \cite{Henk94}, by
extrapolating the relations (\ref{gl:defsch1}) and (\ref{gl:defsch2}) to
integer or
half-integer indices),
containing $\sch_d$
as a Lie subalgebra,
 is the Lie algebra $\sv^d$ defined as follows.

{\bf Definition 1.4.}
{\em We denote by $\sv^d$ the Lie algebra with generators
$
X_n,Y^i_m,M_n,{\cal R}^{ij}_n (n\in\Z,m\in\half+\Z)$ and
  following relations (where $n,p\in\Z,m,m'\in\half+\Z$) :
$$
[L_n,L_p]=(n-p)L_{n+p}
$$
$$
[L_n,Y^i_m]=({n\over 2}-m)Y^i_{n+m},\ [L_n,{\cal R}^{ij}_p]=-p{\cal
R}^{ij}_{n+p}
$$
$$
[Y^i_m,Y^j_{m'}]=(m-m')M_{m+m'}, [{\cal R}^{ij}_p,Y^k_m]=\del_{j,k}
Y^i_{m+p}-\del_{i,k} Y^j_{m+p}
$$
$$  [Y^i_m,M_p]=0, \ [{\cal R}^{ij}_n,M_p]=0,\ [M_n,M_p]=0
$$
\BEQ
 [{\cal R}^{ij}_n,{\cal R}^{kl}_p]=\del_{j,k}{\cal
R}^{il}_{n+p}+\del_{i,l} {\cal R}^{jk}_{n+p}-\del_{j,l}{\cal
R}^{ik}_{n+p}-\del_{i,k}{\cal R}^{jl}_{n+p}
\EEQ

One sees immediately that $\sv^d$  has a semi-direct product structure
$\sv^d\simeq \Vect(S^1)\ltimes \h^d$, with $\Vect(S^1)\simeq \langle
L_n\rangle_{n\in\Z}$ and $\h^d=\langle Y^i_m\rangle_{m\in\Z,i\le d}\oplus
\langle M_p\rangle_{p\in\Z}\oplus\langle {\cal R}^{ij}_m\rangle_{m\in\Z,
1\le i<j\le d}$. }

Note that the Lie  subalgebra $\langle
X_1,X_0,X_1,Y^i_{-\half},Y^i_{\half},
{\cal R}^{ij}_0,M_0
\rangle\subset \sv^d$ is isomorphic to $\sch_d$. So the following
Proposition gives
a positive answer to a most natural question.

{\bf Proposition 1.3} (see \cite{Henk94})
{\it
The
realization $d\tilde{\pi}^d$ of
$\sch_d$    (see  (\ref{gl:schzeta}))
extends to the following
 realization $d\tilde{\pi}^d$ of $\sv^d$ as vector fields on
$S^1\times\R^{d+1}$:

$$
d\tilde{\pi}^d(L_f)=-f(z)\partial_z-\half f'(z) (\sum_{i=1}^d
r_i\partial_{r_i})-
{1\over 4} f''(z) r^2 \partial_{\zeta}
$$
$$
d\tilde{\pi}^d(Y_{g_i}^i)=-g_i(z)\partial_{r_i}-
g'_i(z)  r_i \partial_{\zeta}
$$
$$
d\tilde{\pi}^d({\cal R}^{ij}_{k_{ij}})
=-k_{ij}(z) (r_i \partial_{r_j}-r_j \partial_{r_i})
$$
\BEQ
d\tilde{\pi}^d(M_h)=- h(z)\partial_{\zeta} \label{gl:repsvzeta}
\EEQ }

Let us rewrite this action in Fourier components for completeness. In
the following formulas, $n\in\Z$ while $m\in\half+\Z$:

$$
d\tilde{\pi}^d(L_n)=-z^{n+1}\partial_z-\half (n+1)z^n (\sum_{i=1}^d
r_i\partial_{r_i})-
{1\over 4} (n+1)nz^{n-1} r^2 \partial_{\zeta}
$$
$$
d\tilde{\pi}^d(Y_m)=-z^{m+\half}\partial_{r_i}-
(m+\half)z^{m-\half} r_i \partial_{\zeta}
$$
$$
d\tilde{\pi}^d({\cal R}^{ij}_n)=-z^n (r_i \partial_{r_j}-r_j \partial_{r_i})
$$
\BEQ
d\tilde{\pi}^d(M_n)=- z^n\partial_{\zeta}
\EEQ

We shall restrict to the case $d=1$ in the rest of the article and
write $\sv$ for $\sv^1$, $d\pi$ for $d\pi^1$, $\h$ for $\h^1$, $Y_f$
for $Y_f^1$, $Y_m$ for $Y_m^1$ to simplify notations.

Then (as one sees immediately) $\sv\simeq\langle L_n\rangle_{n\in\Z}\ltimes
\langle Y_m,M_p\rangle_{m\in\half+\Z,p\in\Z}$
 is generated by three fields, $L,Y$
and $M$, with commutators
$$
[L_n,L_p]=(n-p)L_{n+p},\ [L_n,Y_m]=({n\over 2}-m)Y_{n+m},\ [L_n,M_p]=-pM_{n+m}
$$
\BEQ
[Y_m,Y_{m'}]=(m-m')M_{m+m'},\ [Y_m,M_p]=0,\ [M_n,M_p]=0  \label{gl:sv}
\EEQ
where $n,p\in\Z,m,m'\in\half+\Z$,
and $\h=\langle Y_m,M_p\rangle_{m\in\half+Z,p\in\Z}$ is a
two-step nilpotent infinite dimensional Lie algebra.

\subsection{Integration of the Schr\"odinger-Virasoro algebra to a group}

We let $\Diff(S^1)$ be the group of orientation-preserving
$C^{\infty}$-diffeomorphisms of
the circle. Orientation is important since we shall need to consider the
square-root of the jacobian of
the diffeomorphism (see Proposition 1.5).

{\bf Theorem 1.4.}
{\it
\begin{enumerate}
\item
Let $H=C^{\infty}(S^1)\times C^{\infty}(S^1)$ be the product of two copies
of the space of
infinitely differentiable functions on the circle, with its group
structure modified as follows:
\BEQ
(\alpha_2,\beta_2).(\alpha_1,\beta_1)=(\alpha_1+\alpha_2,\beta_1+\beta_2+\half(
{
\alpha}'_1
\alpha_2-\alpha_1\alpha'_2)).
\EEQ
Then $H$ is a Fr\'echet-Lie group which integrates $\h$.

\item
Let $SV=\Diff(S^1)\ltimes H$ be the group with semi-direct product given by
\BEQ
(1;(\alpha,\beta)).(\phi;0)=(\phi;(\alpha,\beta))
\EEQ
and
\BEQ
(\phi;0).(1;(\alpha,\beta))=(\phi;
((\phi')^{\half}(\alpha\circ\phi),\beta
\circ\phi)).
\EEQ
Then $SV$ is a Fr\'echet-Lie group which integrates $\sv$.
\end{enumerate}
}

{\bf Proof.}

\begin{enumerate}
    \item
    From Hamilton (see \cite{Ham}), one easily sees that $H$ is a
    Fr\'echet-Lie group, its underlying manifold being the Fr\'echet
    space $C^{\infty}(S^{1})\times C^{\infty}(S^{1})$ itself.

    One sees moreover that its group structure is unipotent.

    By computing commutators
    \BEQ
     (\alpha_{2},\beta_{2})(\alpha_{1},\beta_{1})
     (\alpha_{2},\beta_{2})^{-1}(\alpha_{1},\beta_{1})^{-1}=
     (0,\alpha'_{1}\alpha_{2}-\alpha_{1}\alpha'_{2})
    \EEQ
    one recovers the formulas for the nilpotent Lie algebra $\h$.

    \item
    It is a well-known folk result that the Fr\'echet-Lie group
    Diff($S^{1}$) integrates the Lie algebra $\Vect(S^{1})$ (see
    Hamilton \cite{Ham}, or \cite{GuiRog05}, chapter 4, for details). Here the
    group $H$ is realized
(as Diff($S^{1}$)-module) as a product of
     modules of densities ${\cal F}_{\half}\times {\cal F}_{0}$,
     hence the semi-direct product $\Diff(S^{1})\ltimes H$ integrates
     the semi-direct product $\Vect(S^{1})\ltimes\h.$
\end{enumerate} \eop

The representation $d\tilde{\pi}$, defined in Proposition 1.3, can be
exponentiated into a representation of
$SV$, given in the
following Proposition :

{\bf Proposition 1.5.} (see \cite{Unt1})

{\it \begin{enumerate}
\item Define $\tilde{\pi}:SV\to \Diff(S^1\times \R^2)$ by
$$\tilde{\pi}(\phi;(\alpha,\beta))=\tilde{\pi}(1;(\alpha,\beta)).
\tilde{\pi}(\phi;0)$$
 and
$$\tilde{\pi}(\phi;0)(z,r,\zeta)=(\phi(z),r\sqrt{\phi'(z)}, \zeta-{1\over 4}
{\phi''(z)\over
\phi'(z)}r^2).$$
Then $\tilde{\pi}$ is a representation of $SV$.
\item
The infinitesimal representation of $\tilde{\pi}$ is equal to $d\tilde{\pi}$.
\end{enumerate} }
{\bf Proof.}

Point (a) may be checked by direct verification (note that the formulas
were originally derived by exponentiating the vector fields in the
realization $d\tilde{\pi}$).

For (b), it is plainly enough to show that, for
any $f\in C^{\infty}(S^1)$ and $g,h\in C^{\infty}(\R)$,
$${d\over du}|_{u=0} \tilde{\pi}(\exp uL_f)=d\tilde{\pi}(L_f), {d\over
du}|_{u=0} \tilde{\pi}(\exp
uY_g)=d\tilde{\pi}(Y_g),
{d\over du}|_{u=0} \tilde{\pi}(\exp uM_h)=d\tilde{\pi}(M_h).$$

Put $\phi_u=\exp uL_f$, so that ${d\over du}|_{u=0} \phi_u(z)=f(z)$.
Then
$${d\over du} r(\phi'_u)^{\half}=\half r(\phi'_u)^{-\half} {d\over du}
\phi'_u\to_{u\to 0}
\half r f'(z),$$
$${d\over du} (r^2{\phi''_u\over \phi'_u})=r^2\left( {{d\over du}
\phi''_u\over \phi'_u}
-{\phi''_u\over (\phi'_u)^2} {d\over du} \phi'_u \right)\to_{u\to 0}
r^2 f''(z)$$
so  the equality ${d\over du}|_{u=0} \tilde{\pi}(\exp
uL_f)=d\tilde{\pi}(L_f)$ holds. The
two other
equalities can be proved in a similar way. \eop

Let us introduce another related representation, using the 'triangular'
structure of
the representation $\tilde{\pi}$. The action $\tilde{\pi}:SV\to
\Diff(S^1\times \R^2)$ can
be projected onto
an action $\bar{\pi}:SV\to \Diff(S^1\times \R)$ by 'forgetting' the
coordinate $\zeta$, since
the way coordinates $(t,r)$ are transformed does not depend on $\zeta$.
Note also
that $\tilde{\pi}$ acts by (time- and space- dependent) translations on the
coordinate $\zeta$, so one
 may define a
function $\Phi:SV\to C^{\infty}(\R^2)$ with coordinates $(t,r)$
by
$$\tilde{\pi}(g)(t,r,\zeta)=(\bar{\pi}(g)(t,r),\zeta+\Phi_g(t,r))$$
(independently of $\zeta\in\R$).  This action may be further projected onto
$\bar{\pi}_{S^1}:
SV\to Diff(S^1)$ by 'forgetting' the second coordinate $r$ this time, so
$$\bar{\pi}_{S^1}(\phi;(\alpha,\beta))=\phi.$$

{\bf Proposition 1.6.}

{\it
\begin{enumerate}
\item
One has the relation
$$\Phi_{g_2\circ
g_1}(t,r)=\Phi_{g_1}(t,r)+\Phi_{g_2}(\bar{\pi}(g_1)(t,r)).$$
In other words, $\Phi$ is a trivial $\pi$-cocyle: $\Phi\in
Z^1(G,C^{\infty}(\R^2))$.

\item
The application ${\pi}_{\lambda}:SV\to Hom(C^{\infty}(S^1\times\R),
C^{\infty}(S^1\times\R))$ defined by
$$\pi_{\lambda}(g)(\phi)(t,r)=\left(\bar{\pi}_{S^1}'\circ\bar{\pi}_{S^1}^{-1}(t
)
\right)^{\lambda}
e^{{\cal M}\Phi_g(\bar{\pi}(g)^{-1}.(t,r))} \
\phi(\bar{\pi}(g)^{-1}.(t,r))$$
defines a representation of $SV$ in $C^{\infty}(S^1\times\R)$.

\end{enumerate}
}

{\bf Proof.}

Straightforward. \eop

Note that the function $\Phi$ comes up naturally when considering
projective representations of the Schr\"odinger group in one space
dimension
$Sch^1\simeq SL(2,\R)\ltimes  \Gal_d$, where $\Gal_d$ is the Lie group naturally
associated to $\gal_d$
(see \cite{Per}).

 Let us look at the associated infinitesimal representation. Introduce the
function $\Phi'$ defined by $\Phi'(X)={d\over
du}|_{u=0} \Phi(\exp
uX),$ $X\in\sv$. If now $g=\exp X, \ X\in\sv$, then
\BEQ
{d\over du}|_{u=0} {\pi}_{\lambda}(\exp uX)(\phi)(t,r) =\left({\cal
M}\Phi'(X)+\lambda
 (d\bar{\pi}_{S^1}(X))'(t)+d\bar{\pi}(X)\right)\phi(t,r)
\EEQ
so $d{\pi}_{\lambda}(X)$ may be represented as the differential
operator of order one
$$d\bar{\pi}(X)+{\cal M}\Phi'(X)+\lambda (d\bar{\pi}_{S^1}(X))'(t).$$

So all this amounts to replacing formally $\partial_{\zeta}$ by ${\cal M}$
in the
formulas of Proposition 1.3 in the case $\lambda=0$. Then, for any  $\lambda$,
 \BEQ
d{\pi}_{\lambda}(Lf)=d{\pi}_0(L_f)-\lambda f',
\EEQ
while $d\pi_{\lambda}(Y_g)=d\pi_0(Y_g)$ and
$d\pi_{\lambda}(M_h)=d\pi_{0}(M_h).$

Let us write explicitly the action of all generators, both  for
completeness and for future reference :

$$
d{\pi}_{\lambda}(L_f)=-f(t)\partial_t-\half f'(t)
r\partial_{r}-
{1\over 4} f''(t) {\cal M} r^2 - \lambda f'(t)
$$
$$
d{\pi}_{\lambda}(Y_{g})=-g(t)\partial_{r}- {\cal M}
g'(t)  r
$$

\BEQ
d{\pi}_{\lambda}(M_h)=-{\cal M} h(t)  \label{gl:repsv}
\EEQ

{\bf Remarks:}

\begin{enumerate}
\item

One may check easily that one gets by restriction
 a representation $\pi_{\lambda}$ of the Schr\"odinger group in one
dimension
$\Sch^1$ whose infinitesimal representation coincides with
(\ref{gl:sch}). In particular,
$\pi_{1/4}|_{\Sch^1}$ acts on the space of solutions of the free
Schr\"odinger equation in one space dimension
(see Proposition 1.1).
\item
Taking the Laplace transform of (\ref{gl:repsv}) with respect to $\cal M$,
one gets a realization
$d\tilde{\pi}_{\lambda}$ of $\sv$ as differential operators of order one
acting on functions of $t,r$ and $\zeta$,
extending the formulas of Proposition 1.3 in the case $d=1$.
\end{enumerate}

\subsection{About graduations and deformations of the Lie algebra $\sv$}

We shall say in this paragraph a little more on the algebraic structure
of $\sv$ and introduce another related Lie algebra $\tsv$ ('{\it twisted
Schr\"odinger-Virasoro
algebra}').

The reader may wonder why we chose half-integer indices for the field $Y$.
The shift in the
indices in due to the fact that $Y$ behaves as a $(-\half)$-density, or,
in other words, $Y$
has conformal weight ${3\over 2}$ under the action of the Virasoro field
$L$ (see e.g. \cite{Kac} or \cite{Del}
 for
a mathematical introduction to conformal field theory and its terminology).

 Note in particular that,
although its weight is a half-integer, $Y$ is a bosonic field, which would
contradict spin-statistics
theorem, were it not for the fact that $Y$ is not meant to represent a
relativistic field (and also that we are in a one-dimensional context).

Nevertheless, as in the case of the double Ramond/Neveu-Schwarz
superalgebra (see \cite{KacRai}), one may
define a 'twisted' Schr\"odinger-Virasoro algebra $\tsv$ which is a priori
equally interesting, and
exhibits to some respects quite different properties (see Chapter 5).

{\bf Definition 1.5.}

{\it Let $\tsv$ be the Lie algebra generated by
$(L_n,Y_m,M_p)_{n,m,p\in\Z}$ with relations
\BEQ
[L_n,L_m]=(n-m)L_{n+m},\ [L_n,Y_m]=({n\over 2}-m)Y_{n+m},\ [L_n,M_m]=-mM_{n+m}
\EEQ
\BEQ
[Y_n,Y_{m}]=(n-m)M_{n+m},\ [Y_n,M_m]=0, [L_n,Y_m]=0, \label{gl:tsv}
\EEQ
where $n,m$ are integers.}

Notice that the relations are exactly the same as for $\sv$ (see
equations  (1.1)-(1.4)),
except for the values of the indices.

The simultaneous existence of two linearly independent graduations on
$\sv$ or $\tsv$ sheds some light
on this ambiguity in the definition.

{\bf Definition 1.6.}
{\it
Let $\del_1$, resp. $\del_2$, be the graduations on $\sv$ or $\tsv$ defined by
\BEQ
\del_1(L_n)=n,\ \del_1(Y_m)=m, \del_1(M_p)=p
\EEQ

\BEQ
\del_2(L_n)=n,\ \del_2(Y_m)=m-\half,\ \del_2(M_p)=p-1
\EEQ

with $n,p\in\Z$ and $m\in\Z$ or $\half+\Z$.}

One immediately checks that both $\del_1$ and $\del_2$ define graduations
and that they are linearly
independent.

{\bf Proposition 1.7.}
{\it
The graduation $\del_1$, defined either on $\sv$ or on $\tsv$,
 is given by the inner derivation $\del_1=\ad(-X_0)$, while
$\del_2$ is an outer derivation, $\del_2\in Z^1(\sv,\sv)\setminus
B^1(\sv,\sv)$ and
$\del_2\in Z^1(\tsv,\tsv)\setminus B^1(\tsv,\tsv)$.}

{\bf Remark.} As we shall see in Chapter 5, the space $H^1(\sv,\sv)$ or
$H^1(\tsv,\tsv)$
 of outer derivations modulo
inner derivations is three-dimensional, but only $\del_2$ defines a
graduation on the basis
$(L_n,Y_m,M_p)$.

{\bf Proof.}

The only non-trivial point is to prove that $\del_2$ is not an inner
derivation. Suppose (by absurd)
that $\del_2=\ad X$, $X\in\sv$ or $X\in\tsv$ (we treat both cases
simultaneously). Then
$\del_2(M_0)=0$ since $M_0$ is central in $\sv$ and in $\tsv$. Hence the
contradiction.\eop

Note that the graduation $\del_2$ is given by the Lie action of the Euler
vector field $t\partial_t
+r\partial_r+\zeta
\partial_{\zeta}$ in the representation $d\tilde{\pi}$ (see Proposition 1.3).

Let us introduce a natural deformation of $\sv$, anticipating on Chapter 5
(we shall need
the following definition in paragraph 3.5, see Theorem 3.10, and
chapter 5):

{\bf Definition 1.7}

{\it
Let $\sv_{\eps},\ \eps\in\R$ (resp. $\tsv_{\eps}$)
 be the Lie algebra generated by
$L_n,Y_m,M_p,\ n,p\in\Z, m\in\half+\Z$ (resp. $m\in\Z$), with
relations
$$
[L_n,L_m]=(n-m)L_{n+m},\ [L_n,Y_m]=({(1+\eps)n\over 2}-m)Y_{n+m},\
[L_n,M_m]=(\eps n-m)M_{n+m}
$$

\BEQ
[Y_n,Y_{m}]=(n-m)M_{n+m},\ [Y_n,M_m]=0, [L_n,Y_m]=0, \label{gl:sveps}
\EEQ

}

One checks immediately that this defines a Lie algebra, and that $\sv=\sv_0$.

All these Lie algebras may be extended by using the trivial extension of
the Virasoro cocycle of
Section 1.1, yielding Lie algebras denoted by $\widetilde{\sv}$,
$\widetilde{\tsv}$, $\widetilde{\sv}_{\eps}$, $\widetilde{\tsv}_{\eps}$.


\section{About the conformal embedding of the Schr\"odinger algebra}
\subsection{The conformal embedding}

The idea of embedding $\sch^d$ into $\conf(d+2)_{\C}$ comes naturally when
considering the wave equation
\BEQ
(2i{\cal M}\partial_t-\partial_r^2)\psi({\cal M};t,r)=0
\EEQ
where ${\cal M}$ is viewed no longer as a parameter, but as a coordinate.
Then the Fourier transform of the wave function with respect to the mass
\BEQ
\tilde{\psi}(\zeta;t,r)=\int_{\R} \psi({\cal M};t,r)e^{-i{\cal M}\zeta}\
d{\cal M}
\EEQ
satisfies the equation
\BEQ
(2\partial_{\zeta}\partial_t -\partial_r^2)\psi(\zeta;t,r)=0
\EEQ
which is none but a zero mass Klein-Gordon equation on
$(d+2)$-dimensional
space-time,
put into light-cone coordinates $(\zeta,t)=(x+y,x-y)$.

This simple idea has been developed in a previous article (see \cite{Unt1})
for $d=1$, in which case $\sch_1$ appears as a (non-trivial)
maximal parabolic subalgebra of $\conf(d+2)$ - which is no longer true
when $d>1$. Let us here give an explicit embedding for any dimension $d$.

We need first to fix some notations. Consider the conformal algebra in its
standard representation as infinitesimal conformal transformations on
$\R^{d+2}$ with coordinates $(\xi_1,\ldots,\xi_{d+2})$. Then there is
a natural basis of $\conf(d+2)$
 made of $(d+2)$ translations $P_{\mu}$, $\half(d+1)(d+2)$
rotations ${\cal M}_{\mu,\nu}$, $(d+2)$ inversions $K_{\mu}$ and the Euler
operator
$D$: in coordinates, one has
\BEA
P_{\mu}=\partial_{\xi_{\mu}}\\
{\cal M}_{\mu,\nu}=\xi_{\mu}\partial_{\nu}-\xi_{\nu}\partial_{\mu}\\
K_{\mu}=2\xi_{\mu}(\sum_{\nu=1}^{d+2} \xi_{\nu}\partial_{\nu})-
(\sum_{\nu=1}^{d+2} \xi_{\nu}^2) \partial_{\mu}\\
D=\sum_{\nu=1}^{d+2} \xi_{\nu}\partial_{\nu}.
\EEA

{\bf{Proposition 2.1.}}

{\it The formulas
\BEA
Y_{-\half}^j=-2^{\half}e^{-i\pi/4} P_j\\
Y_{\half}^j=-2^{-\half}e^{i\pi/4} ({\cal M}_{d+2,j}+i{\cal M}_{d+1,j}) \\
{\cal R}_{j,k}={\cal M}_{j,k}\\
X_{-1}=i(P_{d+2}-iP_{d+1})\\
X_0=-{D\over 2}+{i\over 2} {\cal M}_{d+2,d+1}\\
X_1=-{i\over 4} (K_{d+2}+iK_{d+1})
\EEA
give an embedding of $\sch^d$ into $\conf(d+2)_{\C}$.}

{\bf{Proof.}} Put
$$t=\half(-\xi_{d+1}+i\xi_{d+2}),\zeta=\half(\xi_{d+1}+i\xi_{d+2}),
r_j=2^{-\half} e^{i\pi/4} \xi_j\ (j=1,\ldots,d).$$
Then the previous definitions yield the representation
$d\tilde{\pi}^d_{d/4}$ of $\sch^d$
(see (\ref{gl:schzeta})). \hfill \eop

\subsection{\bf Relations betwen $\sv$ and the Poisson algebra on $T^\ast
S^1$ and 'no-go' theorem}
The relation betwen the Virasoro algebra and the Poisson algebra on $T^\ast S^1$
has been investigated in \cite{OvsRog2}. We shall consider more precisely the Lie
algebra ${\cal A} (S^1)$ of smooth functions on $\dot{T}^\ast S^1 = T^\ast
S^1  \setminus S^1$, the total space of the cotangent bundle with zero
section removed, which are Laurent series on the fibers. So ${\cal A}
(S^1) = C^{\infty} (S^1) \otimes \R [ \partial, \partial^{-1}]$ and $F \in
{\cal A} (S^1)$ is of the following form: \\
$$F (t, \partial, \partial^{-1}) = \displaystyle \sum_{k \in \Z} f_k (t)
\partial^k,$$
 with  $f_k = 0$ for large enough $k$. The Poisson bracket is
defined as usual, following: 
$$\{ F, G \} = \frac{\partial F}{\partial \partial} \frac{\partial
G}{\partial t} - \frac{\partial G}{\partial \partial} \frac{\partial
F}{\partial t}.$$
 (The reader should not be afraid by notation $\partial
\partial$ !). 
In terms of densities on the circle, one has the natural
decomposition: ${\cal A} (S^1) = \displaystyle \bigoplus_{k  0} {\cal
F}_k \bigoplus \left( \displaystyle \prod_{k \leq 0} {\cal F}_k \right)$.
The Poisson bracket turns out to be homogeneous with respect to that
decomposition: $\{ {\cal F}_k, {\cal F}_l \} \subset {\cal F}_{k + l-1}$,
and  more explicitly $\{ f (x) \partial^k, g (x) \partial^l \} = ( k f g'
- l f^{'} g) \partial^{k+l-1}$. One recovers the usual formulae for the
Lie bracket on ${\cal F}_1 = \Vect (S^1)$ and  its representations on
modules of densities; one has as well the embedding of the semi-direct product
$\Vect (S^1) \ltimes C^{\infty} (S^1) = {\cal F}_1 \ltimes {\cal F}_0$ as
a Lie subalgebra of ${\cal A} (S^1)$, representing differential operators
of order $\leq 1$.

\vspace{0.5cm}
\noindent {\bf Remark:} One can also consider the subalgebra of ${\cal A}
(S^1)$, defined as $\C [z, z^{-1} ] \otimes \C [\partial, \partial^{-1}]$;
it gives the usual description of the Poisson algebra on the torus $\T^2$,
sometimes denoted $SU(\infty)$. We also have to consider half densities
and developments into Laurent series in $z$ and $\sqrt{z}$. Geometrically
speaking, the half densities can be described as spinors: let $E$ be a
vector bundle over $S^1$, square root of $T^\ast S^1$; in other words one
has $E \otimes E = T^\ast S^1$. Then the space of Laurent polynomials on
the fibers of $E$  (minus the zero-section) is exactly the Poisson algebra
$\widetilde{{\cal A}} (S^1) = C^\infty (S^1) \otimes \C [ \partial^{1/2},
\partial^{-1/2} ]]$. Moreover, one also needs half-integer power series or
polynomials in $z$ as coefficients of the Laurent series in $\partial$; one
can obtain the corresponding algebra globally, using the pull-back though
the application $S^1 \longrightarrow S^1$ defined as $z \longrightarrow
z^2$.\\
 Finally one has obtained the subalgebra $\widehat{{\cal A}} (S^1) \subset
\widetilde{{\cal A}} (S^1)$ generated by terms $z^m \partial^n$ where $m$
and $n$ are either integers or half-integers. One can represent such
generators as the points with coordinates $(m, n)$ in the plane $\R^2$.\\
So our algebra $SV = \Vect (S^1) \ltimes \h$, with $\h \simeq {\cal F}_{1/2}
\ltimes {\cal F}_0$ as a $\Vect (S^1)$-module, can be naturally embedded
into $\widehat{{\cal A}} (S^1)$,\\

$\cdots \hspace{2cm} 
L_{-2} \hspace{2cm} L_{-1} \hspace{2cm} L_0 \hspace{0.5cm}
\hspace{2cm} L_1 \hspace{2cm} L_2 \hspace{2cm} \cdots$\\
\BEQ
\hspace{1.5cm} \cdots \hspace{2cm} Y_{-\frac{3}{2}} \hspace{2cm} Y_{-{\frac{1}{2}}}
\hspace{2cm} Y_{\frac{1}{2}} \hspace{2cm} Y_{\frac{3}{2}} \hspace{2cm}
\cdots  \hspace{2cm}
\EEQ
 \\
 $ \cdots \hspace{2cm}
M_{-2} \hspace{2cm} M_{-1} \hspace{2cm} M_0 \hspace{2cm}
 M_1 \hspace{2cm} M_2 \hspace{2cm} \cdots $

\vspace{1cm}
The above scheme represents pictorially the embedding. For the twisted
Schr\"odinger-Virasoro algebra, one considers the $Y_m$ field with integer
powers, or as described in the following scheme:

$$\cdots \hspace{2cm} L_{-2} \hspace{2cm} L_{-1} \hspace{2cm}
L_0 \hspace{2cm} L_1 \hspace{2cm} L_2 \hspace{2cm} \cdots$$

$$\cdots \hspace{2cm} Y_{-2} \hspace{2cm} Y_{-1} 
\hspace{2cm} Y_0 \hspace{2cm} Y_1 \hspace{2cm} Y_2 \hspace{2cm} \cdots$$

$$\cdots \hspace{2cm} M_{-2} \hspace{2cm} M_{-1} 
\hspace{2cm}   M_0 \hspace{2cm} M_1 \hspace{2cm} M_2 \hspace{2cm} \cdots
$$

\vspace{0.5cm}
One can naturally ask whether this defines a Lie algebra embedding, just
as in the case of $\Vect(S^1) \ltimes {\cal F}_0$. The answer is no:

\vspace{0.5cm}
\noindent {\bf Proposition 2.2.}\\
 {\it The natural vector space embedding $\sv \hookrightarrow
\widehat{\cal A} (S^1)$ is not a Lie algebra homomorphism.}

\vspace{0.5cm}
\noindent {\bf Proof:}
 One sees immediately that on the  one hand $[Y_n, M_m] =
0$ and $[M_n, M_m] = 0$, while on the other hand  $\{ {\cal F}_{1/2}, {\cal F}_0 \}
\subset {\cal F}_{-1/2}$ is in general  non trivial. The vanishing of $\{
{\cal F}_0, {\cal F}_0 \}$ which makes the embedding of $\Vect (S^1)
\ltimes {\cal F}_0$ as a Lie subalgebra possible was in some sense an
accident. In fact, one can show that from the image of the generators of
$\sv$ and computing successive Poisson brackets, one can generate all the
${\cal F}_\lambda$ with $\lambda \leq 0$.\\
Let $\widehat{\cal A} (S^1)_{(1)} = {\cal F}_1 \oplus {\cal F}_{1/2}
\oplus {\cal F}_0 \displaystyle \bigoplus_{\lambda \in 
\frac{\Z}{2} \ ,\ \lambda < 0} \Pi {\cal F}_\lambda$, it defines a Poisson
subalgebra of $\widehat{\cal A} (S^1)$, and it is in fact the smallest
possible Poisson algebra which contains the image of $\sv$. Now, let
$\widehat{{\cal A} (S^1)}_{(0)} = \displaystyle \prod_{\lambda \in 
\frac{\Z}{2},\ \  \lambda < 0} {\cal F}_\lambda$ the Poisson subalgebra of
$\widehat{\cal A} (S^1)$ which contains only negative powers of $\partial$
(its quantum analogue is known as Volterra algebra of integral operators,
see \cite{GuiRog05}, chap. X). One easily sees that it is an ideal of $\widehat{{\cal A}
(S^1)}_{(1)}$, as a Lie algebra, but of course not an associative ideal; if
one considers the quotient  $\widehat{{\cal A} (S^1)}_{(1)} /
\widehat{{\cal A}} (S^1)_{(0)}$, then all the obstruction for the
embedding to be a homomorphism disappears.
So one has:

\vspace{0.1cm}
\noindent {\bf Proposition 2.3.} \\
{\it There exists a natural Lie algebra embedding of $\sv$ into the
quotient $\widehat{{\cal A} (S^1)}_{(1)} / \widehat{{\cal A}}
(S^1)_{(0)}$. One can say briefly that $\sv$ is a subquotient of the
Poisson algebra $\widehat{{\cal A} (S^1)}$.
}

Now a natural question arises: the conformal embedding of Schr\"odinger
algebra described in paragraph 3.1 yields $\sch_1 \subset \conf (3)_\C$,
so one would like to extend the construction of $\sv$ as generalization of
$\sch_1$, in order that it contain $\conf (3)$: we are looking for an
hypothetic Lie algebra ${\cal G}$ making the following diagram of
embeddings complete:

\BEQ
\begin{array}{cccc}
\sch_1 & \hookrightarrow & \conf (3)\\
\downarrow & & \downarrow   \\
\sv & \hookrightarrow  & {\cal G}
 \end{array}
 \EEQ

In the category of abstract Lie algebras, one has an obvious solution to
this problem: simply take the amalgamated sum of $\sv$ and $\conf (3)$
over $\sch_1$. Such a Lie algebra is defined though generators and
relations, and is generally untractable.
We are looking here for a natural, geometrically defined construction of
such a $\cal G$; we shall give some evidence of its non-existence, a
kind of "no-go theorem", analogous to those well-known in gauge theory,
for example\footnote{Simply recall that this theorem states that there
doesn't exist a common non-trivial extension containing both the  Poincar\'e
group and the external gauge group.} (see  \cite{Kak}).

Let us consider the root diagram of $\conf (3)$ as drawn in \cite{Unt1}:

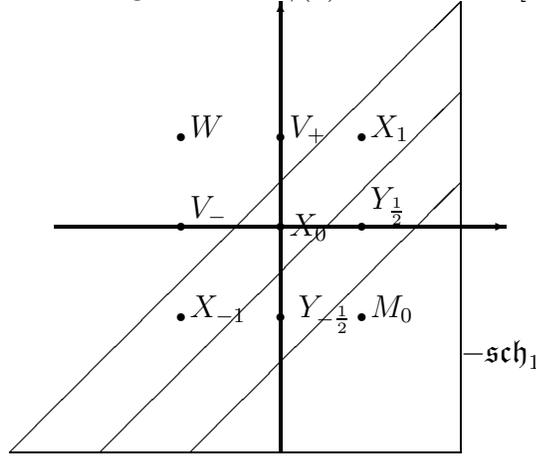
\vspace*{1cm}
\begin{picture}(50,100)(-200,-50)
\setlength{\unitlength}{1.2mm}
\linethickness{0.4mm}
\put(1,10){$V_+$}
\put(0,10){\circle*{1}}
\put(10,10){$X_1$}
\put(9,10){\circle*{1}}
\put(-10,10){$W$}
\put(-11,10){\circle*{1}}
\put(1,-1){$X_0$}
\put(-10,1){$V_-$}
\put(-11,0){\circle*{1}}
\put(10,2){$Y_{\frac12}$}
\put(9,0){\circle*{1}}
\put(2,-10){$Y_{-\frac12}$}
\put(-0,-10){\circle*{1}}
\put(-10,-10){$X_{-1}$}
\put(-11,-10){\circle*{1}}
\put(10,-10){$M_0$}
\put(20, -15){$-\sch_1$ \hspace{4.5cm}}
\put(9,-10){\circle*{1}}
\put(-25,0){\vector(1,0){50}}
\put(0,-25){\vector(0,1){50}}
\linethickness{0.1mm}
\put(-30,-25){\line(1,1){50}}
\put(-20,-25){\line(1,1){40}}
\put(-10,-25){\line(1,1){30}}
\put(-30,-25){\line(1,0){50}}
\put(20,-25){\line(0,1){50}}
\put(0,0){\circle*{1}}
\end{picture}

\vspace{2.5cm}
\noindent Comparing with $(2.14)$, one sees that the successive diagonal
strips are contained in ${\cal F}_1, {\cal F}_{1/2}, {\cal F}_0$
respectively. So the first idea might be to try to add ${\cal F}_{3/2}$
and ${\cal F}_2$, as an  infinite prolongation of the supplementary part to
$\sch_1$ in $\conf (3)$, so that $V_ - \longrightarrow t^{-1/2}
\partial^{3/2}, \ \ V_{+} \longrightarrow t^{1/2} \partial^{3/2}, \ \ W
\longrightarrow \partial^2$.\\

Unfortunately, this construction fails at once for two reasons: first, one
doesn't get the right brackets for $\conf (3)$ with such a choice, and
secondly the elements of ${\cal F}_\lambda, \  \lambda \in \{ 0,
\frac{1}{2}, 1, 3/2, 2 \}$, taken together with their successive brackets
generate the whole Poisson algebra $\widehat{{\cal A}} (S^1)$.

Another approach could be the following: take two copies of $\h$, say
$\h^+$ and $\h^-$ and consider the semi direct product ${\cal G} = \Vect
(S^1) \ltimes (\h^+ \oplus \h^-)$, so that $\h^+$ extends the $\{
Y_{-\frac{1}{2}}, Y_{\frac{1}{2}}, M_0 \}$ as in $\sv$ before, and $\h^-$
extends $\{ V_-, V_+, W \}$. Then ${\cal G}$ is obtained from density
modules, but doesn't extend $\conf (3)$, but only a contraction of it: all
the brackets between $\{ Y_{-\frac{1}{2}}, Y_{\frac{1}{2}}, M_0\}$ on  one
hand and $\{ V_-, V_+, W \}$ on the other are vanishing. Now, we can try
to deform ${\cal G}$ in order to obtain the right brackets for $\conf (3)$.
Let $Y^+_m, M^+_m$ and $Y^-_m, M^-_m$ be the generators of $\h^+$ and
$\h^-$; we want to find coefficients $a_{p,m}$ such that $[ Y^+_m, Y^-_p ]
= a_{p,m} L_{m+p}$ defines a Lie bracket. So let us check Jacobi identity
for $(L_n, Y^+_m, Y^-_p)$. One obtains $(m - \frac{n}{2}) a_{p, n+m} + (n
- m - p) a_{p, m} + ( p - \frac{n}{2}) a_{p+n, m} = 0$. If one tries $a_{p
m} = \lambda p + \mu m$, one deduces from this relation:
$n \lambda (p - \frac{n}{2}) + n \mu ( m - \frac{n}{2}) = 0$ for every $n
\in \Z, p, m \in \frac{1}{2} \Z$, so obviously $\lambda = \mu = 0$.

So our computations show there doesn't exist a geometrically defined
construction of ${\cal G}$ satisfying the conditions of diagram $(2.16)$.
The two possible extensions of $\sch_1$,  $\sv$ and $\conf (3)$ are shown
to be incompatible, and this is  our "no-go" theorem.

%

\section{On some natural representations of $\sv$}

We introduce in this chapter several natural representations of $\sv$ that
split into
two classes : the (centrally extended) coadjoint action on the one hand;
some apparently
unrelated representations on spaces of functions or differential operators
that can actually
all be obtained as particular cases of the general coinduction method for
$\sv$ (see
chapter 4).

It is interesting by itself that the coadjoint action should not belong to
the same family
of representations as the others. We shall come back to this later on in
this chapter.

\subsection{Coadjoint action of $\sv$}

Let us recall some facts about coadjoint actions of centrally extended Lie
groups and algebras, referring
to \cite{GuiRog05}, chapter 6, for details. So let $G$ be a Lie group with
Lie algebra $\g$, and let us
consider  central extensions of them, in the categories of groups and
algebras respectively:
\BEQ
(1)\longrightarrow \R \longrightarrow \tilde{G}\longrightarrow
G\longrightarrow (1) \label{gl:ext1}
\EEQ
\BEQ
(0)\longrightarrow \R \longrightarrow \tilde{\g} \longrightarrow \g
\longrightarrow (0) \label{gl:ext2}
\EEQ
\noindent with $\tilde{\g}=$Lie$(\tilde{G})$, the extension (\ref{gl:ext2})
representing the tangent spaces
at the identity of the extension (\ref{gl:ext1}) (see \cite{GuiRog05}, II 6.1.1.
for explicit formulas).
 Let $C\in Z^2_{{\mathrm{diff}}}(G,\R)$ and $c\in Z^2(\g,\R)$ the
respective cocycles. We want to study the coadjoint action on the dual
$\tilde{\g}^*=\g^*\times\R$. We
shall denote by $\Ad^*$ and $\widetilde{\Ad}^*$ the coadjoint actions of
$G$ and $\tilde{G}$ respectively,
and $\ad^*$ and $\widetilde{\ad}^*$ the coadjoint actions of $\g$ and
$\tilde{\g}$. One then has the following
formulas
\BEQ
\widetilde{\Ad}^*(g,\alpha)(u,\lambda)=(\Ad^*(g)u+\lambda\Theta(g),\lambda) 
 \label{gl:ad1}
\EEQ and
\BEQ
\widetilde{\ad}^*(\xi,\alpha)(u,\lambda)=(\Ad^*(\xi)u+\lambda
\theta(\xi),0) \label{gl:ad2}
\EEQ
where $\Theta:G\to\hat{\g^*}$ and $\theta:\g\to\hat{\g}^*$ are the Souriau
cocycles for differentiable
and Lie algebra cohomologies respectively; for $\theta$ one has the
following formula :
$\langle \theta(\xi),\eta\rangle=c(\xi,\eta).$ For details of the proof,
as well as 'dictionaries'
between the various cocycles, the reader is referred to \cite{GuiRog05}, chapter 6.

Note that formulas (\ref{gl:ad1}) and (\ref{gl:ad2}) define affine actions
of $G$ and $\g$
respectively, different from their coadjoint actions when $\lambda\not=0$.
The actions
on hyperplanes ${\g}^*_{\lambda}=\{(u,\lambda)\ |\ u\in\g^*\}\subset
\tilde{\g}^*$ with fixed
second coordinate
will be denoted by $\ad^*_{\lambda}$ and $\Ad^*_{\lambda}$ respectively.

Here we shall consider the central extension $\tilde{\sv}$ of $\sv$
inherited from Virasoro algebra,
defined by the cocycle $c$ such that
$$
c(L_n,L_p)=\del_{n+p,0}\ n(n+1)(n-1)
$$
\BEQ
c(L_n,Y_m)=c(L_n,M_p)=c(Y_m,Y_{m'})=0 \EEQ
(with $n,p\in\Z$ and $m,m'\in \half+\Z$). Note that we shall prove in
chapter 5 that this central
extension is universal (a more 'pedestrian' proof was given in \cite{Henk94}).

As usual in infinite dimension, the algebraic dual of $\tilde{\sv}$
 is untractable, so let us  consider the regular
dual, consisting of sums of modules of densities of $\Vect(S^1)$ (see
Definition 1.3): the dual module  ${\cal F}_{\mu}^*$ is identified with
${\cal F}_{-1-\mu}$ through
\BEQ
\langle u(dx)^{1+\mu},f\ dx^{-\mu}\rangle=\int_{S^1} u(x)f(x)\ dx.
\EEQ
So, in particular, $\Vect(S^1)^*\simeq {\cal F}_{-2}$, and (as a
$\Vect(S^1)$-module)
\BEQ
\sv^*={\cal F}_{-2}\oplus{\cal F}_{-{3\over 2}}\oplus{\cal F}_{-1};
\EEQ
we shall identify the element $\Gamma=\gamma_0 dx^2+\gamma_1 dx^{{3\over
2}}+\gamma_2 dx\in\sv^*$
with the triple $\left(\begin{array}{c} \gamma_0 \\ \gamma_1 \\
\gamma_2\end{array}\right)\in
(C^{\infty}(S^1))^3.$ In other words,

\BEQ
\langle  \left( \begin{array}{c}\gamma_0 \\ \gamma_1  \\  \gamma_2
\end{array} \right),
L_{f_0}+Y_{f_1}+M_{f_2} \rangle=\sum_{i=0}^2 \int_{S^1} (\gamma_i f_i)(z)
 \ dz.
\EEQ

The following Lemma describes the coadjoint representation of a Lie
algebra that can be written as a semi-direct product.

{\bf Lemma 3.1.}
{\it
Let $\s=\s_0\ltimes \s_1$ be a semi-direct product of two Lie algebra
$\s_0$ and $\s_1$. Then the coadjoint action
of $\s$ on $\s^*$ is given by
$$\ad^*_{\s}(f_0,f_1).(\gamma_0,\gamma_1)=\langle \ad^*_{\s_0}(f_0)
\gamma_0-\tilde{g}_1.\gamma_1,
\tilde{f}_0^*(\gamma_1)+\ad^*_{\s_1}(f_1)\gamma_1\rangle$$
where by definition

$$\langle  \tilde{f}_1. \gamma_1,X_0\rangle_{\s_0^*\times\s_0}=\langle
\gamma_1,[X_0,f_1]\rangle_{\s_1^*\times\s_1}$$
and
$$\langle  \tilde{f}_0^*(\gamma_1),X_1\rangle_{\s_1^*\times\s_1}=\langle
\gamma_1,[f_0,X_1]\rangle_{\s_1^*\times\s_1}.$$
}

{\bf Proof.} Straightforward.

{\bf Theorem 3.2.}
{\it

The coadjoint action of $\sv$ on the affine hyperplane ${\sv}^*_{\lambda}$
is given by the
following formulas:
\BEA
\ad^*(L_{f_0}) \left( \begin{array}{c}\gamma_0 \\ \gamma_1  \\  \gamma_2
\end{array}
\right)=
\left(\begin{array}{c} c f'''_0+2f'_0 \gamma_0+f_0 \gamma'_0 \\
   f_0 \gamma'_1+{3\over 2} f'_0 \gamma_1 \\     f_0 \gamma'_2+f'_0
\gamma_2 \end{array}
\right)\\
\ad^*(Y_{f_1})  \left( \begin{array}{c}\gamma_0 \\ \gamma_1  \\  \gamma_2
\end{array}
\right)=
 \left(\begin{array}{c} {3\over 2}\gamma_1 f'_1+\half \gamma'_1 f_1 \\
2\gamma_2
f'_1+\gamma'_2 f_1 \\ 0 \end{array}\right) \\
\ad^*(M_{f_2})  \left( \begin{array}{c}\gamma_0 \\ \gamma_1  \\  \gamma_2
\end{array}
\right)=
 \left(\begin{array}{c} -\gamma_2 f'_2\\ 0\\ 0 \end{array}\right).
\EEA

}

{\bf Proof.}

The action of $\Vect(S^1)\subset\sv$ follows from the identification of
${\sv}^*_{\lambda}$
with $\vir^*_{\lambda}\oplus{\cal F}_{-{3\over 2}}\oplus{\cal F}_{-1}.$

Applying the preceding Lemma, one gets now
\begin{eqnarray*}
\langle \ad^*(Y_{f_1}).  \left( \begin{array}{c}\gamma_0 \\ \gamma_1  \\
\gamma_2
\end{array} \right),L_{h_0}\rangle &=&
    -\langle \tilde{Y}_{f_1}.  \left( \begin{array}{c}0 \\ \gamma_1  \\
\gamma_2
\end{array} \right) , L_{h_0}\rangle \\
&=& \langle  \left( \begin{array}{c}0 \\ \gamma_1  \\  \gamma_2 \end{array}
\right),Y_{\half h'_0 f_1-h_0 f'_1}\rangle \\
&=& \int_{S^1} \gamma_1(\half h'_0 f_1-h_0 f'_1)\  dz\\
&=& \int_{S^1} h_0(-{3\over 2} \gamma_1 f'_1-\half \gamma'_1 f_1)\ dz;
\end{eqnarray*}
\begin{eqnarray*}
\langle  \ad^*(Y_{f_1}).  \left( \begin{array}{c}\gamma_0 \\ \gamma_1  \\
\gamma_2
\end{array} \right),Y_{h_1}\rangle &=&
\langle \ad^*_{\h}(Y_{f_1}).
 \left( \begin{array}{c} 0 \\ \gamma_1  \\  \gamma_2 \end{array} \right),
Y_{h_1}
\rangle \\
&=& -\langle   \left( \begin{array}{c} 0 \\ \gamma_1  \\  \gamma_2 \end{array}
\right), M_{f'_1 h_1-f_1 h'_1} \rangle\\
&=& -\int_{S^1} \gamma_2(f'_1 h_1-f_1 h'_1) \ dz\\
&=& \int_{S^1} h_1(-2\gamma_2 f'_1-\gamma'_2 f_1)\ dz
\end{eqnarray*}
and
$$ \langle \ad^*(Y_{f_1})  \left( \begin{array}{c}\gamma_0 \\ \gamma_1  \\
\gamma_2
\end{array} \right),M_{h_2}\rangle=0.$$
Hence the result for $\ad^*(Y_{f_1}).$

For the action of $\ad^*(M_{f_2})$, one gets similarly

\begin{eqnarray*}
\langle \ad^*(M_{f_2}) .  \left( \begin{array}{c}\gamma_0 \\ \gamma_1  \\
\gamma_2
\end{array} \right),L_{h_0}\rangle &=&
- \langle  \left( \begin{array}{c}0 \\ \gamma_1  \\  \gamma_2 \end{array}
\right),
M_{f'_2 h_0}\rangle \\
&=& -\int_{S^1} \gamma_2 f'_2 h_0\ dz
\end{eqnarray*}
and
$$
\langle \ad^*(M_{f_2}) .  \left( \begin{array}{c}\gamma_0 \\ \gamma_1  \\
\gamma_2
\end{array} \right),Y_{h_1}\rangle =
\langle \ad^*(M_{f_2}).  \left( \begin{array}{c}\gamma_0 \\ \gamma_1  \\
\gamma_2
\end{array} \right),M_{h_2}\rangle=0.
$$
Hence the result for $\ad^*(M_{f_2}).$ \eop

One can now easily construct the coadjoint action of the group $SV$, which
"integrates" the above defined coadjoint action of $\sv$; as usual in
infinite dimension, such an action should not be taken for granted and one has to construct
it explicitly case by case. The result is given by the following.

\noindent {\bf Theorem 3.3.}\\
{\it The coadjoint action of $SV$ on the affine hyperplane
$\tilde{\sv}^\ast_\lambda$ is given by the following formulas:

Let $(\varphi, \alpha, \beta) \in SV$, then:\\
\BEQ
  \ Ad^*(\varphi)  \left( \begin{array}{c}\gamma_0 \\ \gamma_1  \\  \gamma_2
\end{array}
\right)=
\left(\begin{array}{l} \lambda \Theta (\varphi) + (\gamma_0 \circ \varphi)
(\varphi')^2 \\
(\gamma_1\circ \varphi) (\varphi')^{\frac32} \\
(\gamma_2 \circ \varphi) \varphi'
\end{array}
\right) \label{coadj1}
\EEQ

 \BEQ
 Ad^*(\alpha, \beta)  \left( \begin{array}{c}\gamma_0 \\ \gamma_1  \\
\gamma_2
\end{array}
\right)=
 \left( \begin{array}{l}\gamma_0 + \frac{3}{2} \gamma_1 \alpha^{'} +
\frac{\gamma^{'}_{1}}{2} \alpha + \gamma_2 \beta^{'} -
\frac{\gamma_{2}}{2} ( 3 \alpha^{'2} + \alpha \alpha^{''}) - \frac{3}{2}
\gamma^{'}_{2} \alpha \alpha^{'} - \frac{\gamma^{''}_{2}}{4} \alpha^2 \\
\gamma_1 + 2 \gamma_2 \alpha^{'} + \gamma^{'}_2 \alpha \\  \gamma_2
\end{array}
\right)  \label{coadj2}
\EEQ

}

\noindent {\bf Proof:}

 The first part (\ref{coadj1}) is easily deduced from the
natural action of $\Diff (S^1)$ on $\tilde{\sv}^\ast_\lambda =
\vir^\ast_\lambda \oplus {\cal F}_{- 3/2} \oplus {\cal F}_{-1}$. Here $\Theta
(\varphi)$ denotes the Schwarzian derivative of $\varphi$. Let's only
recall that it is the Souriau cocycle in $H^1 ( \Vect (S^1), \vir^\ast)$
associated to Bott-Virasoro cocycle in $H^2 (\Diff (S^1) \ , \ \R)$,
referring to \cite{GuiRog05}, Chap. IV , VI for details.\\

 The problem of computing the coadjoint action of $(\alpha, \beta) \in H$
can be split into two pieces; the coadjoint action of $H$ on $\h^\ast$ is
readily computed and one finds:\\
$$ Ad^*(\alpha, \beta)   \left( \begin{array}{c} \gamma_1  \\  \gamma_2
\end{array}
\right)=
\left(\begin{array}{l} \gamma_1 + 2 \gamma_2 \alpha^{'} + \gamma^{'}_{2}
\alpha \\
\gamma_2
\end{array}
\right)$$

The most delicate part is to compute the part of coadjoint action of
$(\alpha, \beta) \in H$ coming from the adjoint action on $\Vect (S^1)$, by using:\\
$$\langle Ad^*(\alpha, \beta)  \left( \begin{array}{c}\gamma_0 \\ \gamma_1  \\
\gamma_2
\end{array} \right), f \partial
\rangle =  \langle  \left( \begin{array}{c}\gamma_0 \\ \gamma_1  \\
\gamma_2
\end{array} \right),Ad (\alpha, \beta)^{-1} (f \partial , 0, 0)\rangle.$$

One can now use conjugation in the group $SV$ and one finds\\
$$Ad (\alpha, \beta)^{-1} (f \partial, 0, 0) = \left( f \partial, f
\alpha^{'} - \frac{1}{2} \alpha f^{'}, f \beta^{'} + \frac{1}{2} (f
\alpha{''} \alpha + \frac{f^{'}}{2} \alpha \alpha^{'} -
\frac{\alpha^{2}}{2} f^{''} - f \alpha^{'2}  + \frac{f'}{2} \alpha
\alpha^{''})\right).$$

Now, using integration by part, one finds easily the formula (\ref{coadj2})
 above. \hfill \eop

\subsection{Action of $\sv$ on the affine space of Schr\"odinger operators}

The next three sections aim at generalizing an idea that appeared at a
crossroads between projective
geometry, integrable systems and the theory of representations of
$\Diff(S^1)$.  We shall, to our own
regret, give some new insights on $SV$ from the latter point of view
exclusively,
 leaving aside other aspects of a figure that will hopefully soon emerge.

Let $\partial={\partial \over \partial x}$ be the derivation operator on
the torus $\T=[0,2\pi]$. A {\it
Hill operator} is by definition a second order operator on $\T$ of the
form ${\cal L}_u:=\partial^2+u$,
$u\in C^{\infty}(\T)$. Let $\pi_{\lambda}$ be the representation of
$\Diff(S^1)$ on the space of $(-\lambda)$-densities ${\cal F}_{\lambda}$
(see Definition 1.3). One identifies the vector spaces $C^{\infty}(\T)$
and
${\cal F}_{\lambda}$ in the natural way, by associating to $f\in
C^{\infty}(\T)$ the density $f dx^{-\lambda}$.
Then, for any couple $(\lambda,\mu)\in\R^2$, one has an action
$\Pi_{\lambda,\mu}$ of $\Diff(S^1)$
on the space of differential operators on $\T$ through the left-and-right
action
$$\Pi_{\lambda,\mu}(\phi): D\to \pi_{\lambda}(\phi)\circ D\circ
\pi_{\mu}(\phi)^{-1},$$
with corresponding infinitesimal action
$$d\Pi_{\lambda,\mu}(\phi): D\to d\pi_{\lambda}(\phi)\circ D-D\circ
d\pi_{\mu}(\phi).$$

For a particular choice of $\lambda,\mu$, namely, $\lambda=-{3\over 2},
\mu={1\over 2}$, this representation
preserves the affine space  of Hill operators; more precisely,
\BEQ
\pi_{-3/2}(\phi)\circ (\partial^2+u)\circ
\pi_{1/2}(\phi)^{-1}=\partial^2+(\phi')^2 (u\circ \phi')+\half
\Theta(\phi)
\EEQ
where $\Theta$ stands for the Schwarzian derivative. In other words, $u$
transforms
as an element of $\vir_{\half}^*$ (see section 3.1). One may also - taking
an opposite point of view - say
that Hill operators define a $\Diff(S^1)$-equivariant morphism from ${\cal
F}_{\half}$ into ${\cal F}_{-{3\over
2}}$.

This program may be completed for actions of $SV$ on several affine spaces
of differential operators. This
will lead us to introduce several representations of $SV$ that may all be
obtained by the general method
of coinduction (see Chapter 4). Quite remarkably, when one thinks of the
analogy with the case of the
action of $\Diff(S^1)$ on Hill operators, the coadjoint action of $SV$ on
$\sv^*$ does not appear in this
context, and morerover cannot be obtained by the coinduction method, as one
concludes easily from the formulas of Chapter 4 (see Theorem 4.2).

{\bf Definition 3.1.} Let ${\cal S}^{lin}$ be the vector space of second order
operators on $\R^2$ defined by
$$D\in{\cal S}^{lin}\Leftrightarrow D=h(2{\cal M}\partial_t-\partial_r^2)
+V(r,t),
\quad h,V\in C^{\infty}(\R^2)$$
and ${\cal S}^{aff}\subset {\cal S}^{lin}$ the affine subspace of
'Schr\"odinger operators' given by the hyperplane $h=1$.

In other words, an element of ${\cal S}^{aff}$ is the sum of the free
Schr\"odinger
operator
${\Delta}_0=2{\cal M}\partial_t-\partial_r^2$ and of a potential $V$.

The following theorem proves that there is a natural family of  actions of
the group
$SV$
on the space ${\cal S}^{lin}$ : more precisely, for every $\lambda\in\R$,
and  $g\in SV$,  there is
a 'scaling function' $F_{g,\lambda}\in C^{\infty}(S^1)$ such that
\BEQ
\pi_{\lambda}(g)( {\Delta}_0+V)\pi_{\lambda}(g)^{-1}=F_{g,\lambda}(t)
({\Delta}_0
+V_{g,\lambda})
\EEQ
where $V_{g,\lambda}\in C^{\infty}(\R^2)$ is a 'transformed potential'
depending on
$g$ and on $\lambda$ (see Section 1.2, Proposition 1.6 and commentaries
thereafter for the definition
of $\pi_{\lambda}$).
Taking the infinitesimal representation of $\sv$ instead, this is
equivalent
to demanding that the 'adjoint' action of $d\pi_{\lambda}(\sv)$ preserve
${\cal S}^{lin}$, namely
\BEQ
[d\pi_{\lambda}(X),{\Delta}_0+V](t,r)=f_{X,\lambda}(t)({\Delta}_0+V_{X,\lambda}
),\quad X\in\sv
\EEQ
for a certain infinitesimal 'scaling' function $f_{X,\lambda}$ and with a
transformed potential $V_{X,\lambda}$.

We shall actually prove that this last property even characterizes in some
sense the
differential operators of order one that belong to $d\pi_{\lambda}(\sv)$.

{\bf Theorem 3.4. }

{\it
\begin{enumerate}
\item
The Lie algebra of differential operators of order one  $\cal X$ on $\R^2$
preserving
the space ${\cal S}^{lin}$, i.e., such that
$$[{\cal X},{\cal S}^{lin}]\subset {\cal S}^{lin}$$
is equal to the image of $\sv$ by the representation $d\pi_{\lambda}$ (modulo
the addition to $\cal X$ of  operators of
multiplication by an arbitrary function of $t$).
\item
The action of $d\pi_{\lambda+1/4}(\sv)$ on the free Schr\"odinger operator
${\Delta}_0$ is given by
\BEQ
[d\pi_{\lambda+1/4}(L_f),{\Delta}_0]=f'{\Delta}_0+{{\cal M}^2\over 2}f'''
r^2+2{\cal M}\lambda f''
\EEQ
\BEQ
[d\pi_{\lambda+1/4}(Y_g),{\Delta}_0]=2{\cal M}^2  rg''
\EEQ
\BEQ
[d\pi_{\lambda+1/4}(M_h),{\Delta}_0]= 2{\cal M}^2  h'
\EEQ

\end{enumerate}
}

{\bf Proof.}

Let ${\cal X}=f\partial_t+g\partial_r+h$ preserving the space ${\cal
S}^{lin}$:
this is equivalent to the existence of two functions $\phi(t,r),V(t,r)$
 such that
$[{\cal X},{\Delta}_0]=\phi({\Delta}_0+V).$ It is clear that $[h,{\cal
S}_{lin}]\subset {\cal S}_{lin}$
if $h$ is a function of $t$ only.

By considerations of degree, one must then have $[{\cal
X},\partial_r]=a(t,r)
\partial_r+b(t,r)$, hence $f$ is a function of $t$ only. Then
\BEQ
[f\partial_t, 2{\cal M}\partial_t-\partial_r^2]=-2{\cal M}f'\partial_t
\EEQ
\BEQ
[g\partial_r,2{\cal M}\partial_t-\partial_r^2]=-2{\cal M}\partial_t
g\partial_r+2\partial_r g
\partial_r^2+\partial_r^2 g\partial_r
\EEQ
\BEQ
[h,-\partial_r^2]=2\partial_r h \partial_r+\partial_r^2 h
\EEQ
so, necessarily, $$f'=2\partial_r g=-\phi$$ and
$$(2{\cal M}\partial_t-\partial_r^2)g=-2\partial_r h.$$

By putting together these relations, one gets points 1 and 2
simultaneously.
\eop

Using a left-and-right action of $\sv$ that combines $d\pi_{\lambda}$ and
$d\pi_{1+\lambda}$, one gets a new
family of
representations
$d\sigma_{\lambda}$ of $\sv$ which map the affine space ${\cal S}^{aff}$
into differential operators of order zero
 (that is to say, into functions) :

{\bf Proposition 3.5.}

{\it
Let $d\sigma_{\lambda}: \sv\to Hom({\cal S}^{lin},{\cal S}^{lin})$ defined
by the left-and-right infinitesimal action
$$d\sigma_{\lambda}(X): D\to d\pi_{1+\lambda}(X)\circ D-D\circ
d\pi_{\lambda}(X).$$
Then $d\sigma_{\lambda}$ is a representation of $\sv$ and
$d\sigma_{\lambda}(\sv)({\cal
S}^{aff})\subset C^{\infty}(\R^2).$
}

{\bf Proof.}

Let $X_1,X_2\in \sv$, and put $d\bar{\pi}_{S^1}(X_i)=f_i(t)$, $i=1,2$:
then, with a slight abuse of notations,
$d\sigma_{\lambda}(X_i)=\ad\ d\pi_{\lambda}(X_i)+f'_i$, so
\BEA
[d\sigma_{\lambda}(X_1),d\sigma_{\lambda}(X_2)]&=&
[\ad(d\pi_{\lambda}(X_1))+f'_1,
\ad(d\pi_{\lambda}(X_2))+f'_2] \\
&=& \ad\  d\pi_{\lambda}([X_1,X_2])+\left( [
d\pi_{\lambda}(X_1),f'_2(t)]-  [
d\pi_{\lambda}(X_2),f'_1(t)]\right).
\EEA
Now $\ad \ d\pi_{\lambda}$ commutes with operators of multiplication
by any function of time $g(t)$ if $X\in\h$, and
$$[d\pi_{\lambda}(L_f),g(t)]=[f(t)\partial_t,g(t)]=f(t)g'(t)$$
so as a general rule
$$[d\pi_{\lambda}(X_i),g(t)]=f_i(t)g'(t).$$
Hence
\BEA
[d\sigma_{\lambda}(X_1),d\sigma_{\lambda}(X_2)] &=&  \ad \
d\pi_{\lambda}([X_1,X_2])+
(f_1(t)f''_2(t)-f_2(t)f''_1(t))\\
 &=&  \ad \ d\pi_{\lambda}([X_1,X_2])+ (f_1f'_2-f_2 f'_1)'(t) \\
&=& d\sigma_{\lambda}([X_1,X_2]).
\EEA

By the preceding Theorem, it is now clear  that $d\sigma_{\lambda}(\sv)$
sends ${\cal
S}^{aff}$ into differential operators of order zero. \eop

{\bf Remark:} choosing $\lambda={1\over 4}$ leads to a representation of
$\Sch^1$ preserving the kernel of $\Del_0$
(as already known from Proposition 1.1). So, in some sense,
$\lambda={1\over 4}$ is the 'best' choice.

Clearly, the (affine) subspace ${\cal S}_{\le 2}^{aff}\subset {\cal
S}^{aff}$ of Schr\"odinger operators with potentials
that are at most quadratic in $r$, that is,
$$D\in {\cal S}_{\le 2}^{aff}\Leftrightarrow
D=2{\cal M}\partial_t-\partial_r^2+g_0(t)r^2+g_1(t)r+g_2(t)$$
is mapped into potentials of the same form under $d\sigma_{\lambda}(SV)$.

Let us use the same vector notation for elements of ${\cal S}_{\le
2}^{aff}$ and for potentials
that are at most quadratic in $r$ (what is precisely meant  will be
clearly seen from the context): set
$D=\left(\begin{array}{c} g_0\\ g_1\\ g_2
\end{array}\right)$, respectively $V=\left(\begin{array}{c} g_0\\ g_1\\ g_2
\end{array}\right)$  for
$D={\Delta}_0+g_0(t)r^2+g_1(t)r+g_2(t)\in{\cal S}_{\le
2}^{aff}$, respectively $V(t,r)=g_0(t)r^2+g_1(t)r+g_2(t)\in
C^{\infty}(\R^2)$. Then one can give
an explicit formula for the action of $d\sigma_{\lambda}$ on ${\cal
S}_2^{aff}$.

{\bf Proposition 3.6.}

{\it
\begin{enumerate}
\item
Let $D=\left(\begin{array}{c} g_0\\ g_1\\ g_2\end{array}\right)\in{\cal
S}_{\le 2}^{aff}$  and $f_0,f_1,f_2\in C^{\infty}(\R)$.
Then the following formulas hold:
\BEA
d\sigma_{\lambda+1/4}(L_{f_0})(D)=-\left(\begin{array}{c} {-{\cal
M}^2\over 2} f'''_0+2f'_0 g_0+f_0
g'_0 \\
   f_0 g'_1+{3\over 2} f'_0 g_1 \\     f_0 g'_2+f'_0 g_2-2{\cal M}\lambda
f''_0 \end{array}
\right) \\
d\sigma_{\lambda+1/4}(Y_{f_1})(D)=-\left(\begin{array}{c} 0 \\ 2f_1
g_0-2{\cal M}^2 f''_1
\\ f_1 g_1 \end{array}\right) \\
d\sigma_{\lambda+1/4}(M_{f_2})(D)=\left(\begin{array}{c}  0 \\ 0 \\ 2{\cal
M}^2 f'_2
\end{array} \right)
\EEA

\item
Consider the restriction of $d\sigma_{1/4}$ to $\Vect(S^1)\subset\sv$. Then
$d\sigma_{1/4}|_{\Vect(S^1)}$ acts diagonally on the
$3$-vectors $\left(\begin{array}{c} g_0\\ g_1\\ g_2\end{array}\right)$ and
its restriction to the subspaces
${\cal S}^{aff}_i:=\{{\Delta}_0+g(t) r^i\ |\ g\in C^{\infty}(\R)\}$,
$i=0,1,2$, is equal to the coadjoint action of
$\Vect(S^1)$ on the affine hyperplane ${\vir}^*_{1/4}$ $(i=2)$, and to the
usual action of $\Vect(S^1)$ on
${\cal F}_{-3/2}\simeq {\cal F}_{1/2}^*$ (when $i=1$), respectively on
${\cal F}_{-1}\simeq{\cal F}_0^*$ (when $i=0$). Taking $\lambda\not=0$
leads to an affine term proportional
to $f''_0$  on the third coordinate, corresponding to the non-trivial
affine cocycle in $H^1(\Vect(S^1),
{\cal F}_{-1})$.

\end{enumerate}
}

In other words, if one identifies ${\cal S}_{\le 2}^{aff}$ with
$\sv^*_{1\over 4}$ by
$$\langle \left( \begin{array}{c} g_0\\ g_1\\ g_2 \end{array}\right),
\left(
\begin{array}{ccc} f_0 & f_1 & f_2 \end{array} \right) \rangle_{{\cal
S}_{\le 2}^{aff}\times \sv}=\sum_{i=0}^2
\int_{S^1} (g_i f_i)(z) \ dz,$$
then the restriction of $d\sigma_{1/4}$ to $\Vect(S^1)$ is equal to the
restriction of the coadjoint action of $\sv$ on
$\sv^*_{1\over 4}$.

But  mind that $d\sigma_{1/4}$ is {\it not} equal to the coadjoint action
of $\sv$.

{\bf Proof.}
Point 2 is more or less obvious, and we shall only give some of the
computations for the first one. One has
$[d\pi_{\lambda+1/4}(L_{f_0}),{\Delta}_0]=f'_0 {\Delta}_0-{{\cal M}\over
2} f'''_0 r^2+2{\cal M}\lambda f''_0$,
$[d\pi_{\lambda+1/4}(L_{f_0}),g_2(t)]=-f_0(t) g'_2(t)$,
$[d\pi_{\lambda+1/4}(L_{f_0}),g_1(t) r]=
-(f_0(t) g'_1(t)+{\half} f'_0(t) g_1(t))r$,
$[d\pi_{\lambda+1/4}(L_{f_0}),g_0(t)r^2]=-(f_0(t)
g'_0(t)+f'_0(t) g_0(t))r^2$, so
\begin{eqnarray*}
d\sigma_{\lambda+1/4}(L_{f_0})(D) &=& - f'_0. D+
[d\pi_{\lambda+1/4}(L_{f_0}),D] \\
&=& -(f_0 g'_2+f'_0 g_2-2{\cal M}\lambda f''_0)- (f_0 g'_1+{3\over 2} f'_0
g_1)r-({{\cal M}^2\over
2}f'''_0 + 2f'_0 g_0 +f_0 g'_0)r^2.
\end{eqnarray*}
Hence the result for $d\sigma_{\lambda+1/4}(L_{f_0})$. The other
computations are
similar though somewhat simpler. \eop

This representation is easily integrated to a representation $\sigma$ of
the group $SV$.
We let $\Theta(\phi)={\phi'''\over \phi'}-{3\over 2} \left({\phi''\over
\phi'}\right)^2$
$(\phi\in\Diff(S^1))$ be the Schwarzian of the function $\phi$.

{\bf Proposition 3.7.}\\
{\it
Let $D=\left(\begin{array}{c} g_0\\g_1\\g_2 \end{array}\right)\in{\cal
S}_{\le 2}^{aff}$, then
\BEQ
\sigma_{\lambda+1/4}(\phi;(a,b))D=\sigma_{\lambda+1/4}(1;(a,b))\sigma_{\lambda+
1/4}(\phi;0)D
\EEQ
\BEQ
\sigma_{\lambda+1/4}(\phi;1).D=\left(\begin{array}{c}
(\phi')^2.(g_0\circ\phi)-{{\cal M}^2\over 2} \Theta(\phi)\\
(\phi')^{3\over 2} (g_1\circ\phi)\\ \phi' (g_2\circ\phi) -2{\cal M}\lambda
{\phi''\over \phi} \end{array} \right)
\EEQ
\BEQ
\sigma_{\lambda+1/4}(1;(a,b)).D=\left(\begin{array}{c} g_0\\
g_1-2ag_0+2{\cal M}^2 a'' \\
g_2-a g_1+a^2 g_0+{\cal M}^2 (2b'-a a'') \end{array}\right)
\EEQ
defines a representation of $SV$ that integrates $d\sigma$, and maps the
affine space ${\cal S}^{aff}_{\le 2}$
into itself.

In other words, elements of ${\cal S}^{aff}_{\le 2}$ define an
$SV$-equivariant morphism from ${\cal H}_{\lambda}$
into ${\cal H}_{\lambda+1}$, where ${\cal H}_{\lambda}$, respectively
${\cal H}_{\lambda+1}$,
 is the space $C^{\infty}(\R^2)$
of functions of $t,r$ that are at most quadratic in $r$, equipped with the
action $\pi_{\lambda}$, respectively
$\pi_{\lambda+1}$ (see \ref{gl:repsv}).

}

{\bf Proof.}

Put $SV=G\ltimes H$. Then the restrictions $\sigma|_G$ and $\sigma|_H$
define representations
(this is a classical result for the first action, and may be checked by
direct computation
for the second one). The associated infinitesimal representation of $\sv$
is easily seen to be
equal to $d\sigma$. \eop

In particular, the orbit of the free Schr\"odinger operator ${\Delta}_0$
is given by the
remarkable formula
\BEQ
\sigma_{\lambda+1/4}(\phi;(a,b)){\Delta}_0=\left(\begin{array}{c} -{{\cal
M}^2\over 2} \Theta(\phi) \\
 2{\cal M}^2 a''  \\  {\cal M}^2 (2b'- a a'') 
 \end{array}\right) +\lambda
\left(\begin{array}{c} 0\\ 0\\ -2{\cal M}{\phi''\over \phi}\end{array}\right),
\EEQ
 mixing a third-order cocycle with coefficient ${\cal M}^2$ which
extends the Schwarzian cocycle $\phi\to\Theta(\phi)$ with a second-order cocyle
with coefficient $-2{\cal M}\lambda$ which extends the well-known cocycle
 $\phi\to
{\phi''\over \phi}$ in
 $H^1(\Vect(S^1),{\cal F}_{-1})$. The following paragraph shows that all affine cocycles
of $\sv$ with coefficients in the (linear) representation space ${\cal S}^{aff}_{\le 2}$
are of this form.

\vspace{1cm}

\subsection{Affine cocycles of $\sv$ and $SV$ on the space of Schr\"odinger operators}

The representations of groups and Lie algebras described above are of
affine type; if one has linear representations of the group $G$ and Lie
algebra ${\cal G}$ on a module $M$, one can deform these representations
into affine ones, using the following construction. 

Let $C : G \rightarrow M$ (resp $c : {\cal G}\longrightarrow M$) be a
$1$-cocycle in $Z^1_{diff} (G, M)$ (resp $Z^1 ({\cal G}, M)$); it defines
an affine action of $G$ (resp ${\cal G}$) by deforming the linear action
as follows:
$$g \ast m = g . m + C (g)$$
$$\xi \ast m = \xi . m + c ( \xi)$$ respectively.
Here the dot indicates the original linear action and $\ast$ the affine
action. One deduces from the formulas given in propositions 3.5 and 3.6
that the above representations are of this type; the first cohomology of
$SV$ (resp $\sv$) with coefficients in the module ${\cal S}^{aff}_{\le 2}$ 
(equipped with the linear action) classifies all the
affine deformations of the action, up to isomorphism. They are given by
the following theorem:\\

\vspace{0.5cm}
\noindent {\bf Theorem 3.8.}\\
{\it The degree-one cohomology of the group $SV$ (resp Lie algebra $\sv$)
with coefficients in the module ${\cal S}^{aff}_{\le 2}$ (equipped with the linear
action) is two-dimensional and can be
represented by the following cocycles:\\
- for
SV: \hspace{4cm} \ $C_1 (\phi, (a, b)) = \left( \begin{array}{c}-
\frac{1}{2} \Theta ( \phi ) \\ 2a''  \\  2 b' - a a''
\end{array}
\right)$\\

\hspace{5.2cm}
$C_2(\phi,(a, b)) =  \left( \begin{array}{c}0 \\ 0  \\  \phi''
\end{array}
\right)$
\\
 - for
$\sv:$ \hspace{4.2cm} $c_1 ( L_{f_{0}}+ Y_{f_{1}}+ M_{f_{2}}) =  \left(
\begin{array}{c}\frac{1}{2} f^{'''}_0 \\ -2 f^{''}_1  \\  -2 f^{'}_2
\end{array}
\right)$\\

\hspace{5.2cm}
$ c_2 (L_{f_{0}}+ Y_{f_{1}}+ M_{f_{2}}) = \left(\begin{array}{c} 0\\ 0\\
f^{''}_0 \end{array}\right)$\\
(one \ easily  \ recognizes \ that $C_1$ and $c_1$ correspond to the
representations given in prop.3.7 and prop.3.6 respectively.)
}

\vspace{0.5cm}
\noindent {\bf Proof:} One shall first make the computations  for the Lie algebra and
then try to integrate explicitly; here,  the "heuristical" version of
Van-Est theorem, generalized to the infinite-dimensional case, guarantees
the isomorphism between the $H^1$ groups for $SV$ and $\sv$ (see \cite{GuiRog05},
chapter  IV).

So let us compute  $H^1(\sv,M)\simeq H^1({\cal G} \ltimes \h,M)$
 for $M = {\cal S}^{aff}_{\le 2}$ (equipped with the linear action). Let $c : {\cal G} \times \h
\longrightarrow M$ be a cocycle and set $c = c' + c''$ where $c' =
 c|_{\cal G}$ and $c'' = c|_{\h}$. One has $c' \in Z^1 ( {\cal G} , M)$ and
$c'' \in Z^1 (\h, M)$, and these two cocycles are linked
together by the  compatibility relation 
\BEQ
c'' ( [ X, \alpha]) - X. (c'' (\alpha)) + \alpha . (c' (X)) = 0 \label{compat}
\EEQ
As a ${\cal G}$-module, $M = {\cal F}_{-2} \oplus {\cal F}_{- 3/2} \oplus
{\cal F}_{-1}$, so one determines easily that $H^1 ( {\cal G}, {\cal
F}_{-2})$ and $H^1 ( {\cal G}, {\cal F}_{-1})$ are one-dimensional,
generated by $L_{f_{0}} \longrightarrow f^{'''}_0 dx^2$ and $L_{f_{0}}
\longrightarrow f^{''}_0 dx$ respectively, and $H^1 ({\cal G}, {\cal F}_{-
3/2} ) = 0$ (see \cite{GuiRog05},  chapter IV). One  can now readily compute the
1-cohomology of the nilpotent part $\h$; one easily remarks that the linear
action on ${\cal S}^{aff}_{\le 2}$ is defined as follows:
$$
(Y_{f_{1}}+ M_{f_{2}}) \  . \  \left( \begin{array}{c}\gamma_0 \\ \gamma_1  \\
\gamma_2
\end{array}
\right)=
 \left( \begin{array}{c}0 \\ 2 f_1 \gamma_0  \\  f_1 \gamma_1
\end{array}
\right)
$$
Direct computation shows that there are several cocycles, but
compatibility condition (\ref{compat}) destroys all of them but one:
$$
c{''} (Y_{f_{1}}+ M_{f_{2}})  = \left( \begin{array}{c} 0 \\ f^{''}_1  \\
f'_2
\end{array}
\right)
$$
The compatibility condition gives
$$c{''} ( [ L_{f_{0}}, Y_{f_{1}}+ M_{f_{2}} ] ) - L_{f_{0}} \ (c{''}
(Y_{f_{1}}+ M_{f_{2}})) = - \frac{1}{2} f_1 f^{'''}_0$$
On the other hand one finds:
$$
(Y_{f_{1}} +M_{f_{2}}). (c^{'}(L_{f_{0}})) = (Y_{f_{1}}+ M_{f_{2}})\ .\  
 \left(
\begin{array}{c} f^{'''}_0 \\ 0  \\  f^{''}_0
\end{array}
\right) = \left( \begin{array}{c} 0 \\ 2 f_1 f^{'''}_0  \\  0
\end{array}
\right)
$$
Hence the result, with the right proportionality coefficients, and one
obtains the formula for $c_1$.\\
One also remarks that the term with $f^{''}_0 dx$ disappears through the
action of $\h$, so it will induce an independent generator in $H^1 ( \sv,
{\cal S}^{aff})$, precisely $c_2$. \\
Finally the cocycles $C_1$ and $C_2$ in $H^1 ( SV, {\cal S}^{aff}_{\le 2})$ are
not so hard to compute, once we have determined the action of $H$ on
${\cal S}^{aff}_{\le 2}$, which is unipotent as follows:
$$
(a, b) . \left( \begin{array}{c} \gamma_0 \\ \gamma_1  \\  \gamma_2
\end{array}
\right)= \left( \begin{array}{l} \gamma_0 \\ \gamma_1 + 2 a \gamma_0  \\
\gamma_2 + a \gamma_1 - a^2 \gamma_0
\end{array}
\right)
$$

\eop

\subsection{Action on Dirac-L\'evy-Leblond operators}

L\'evy-Leblond introduced in \cite{LL}  a matrix differential operator ${\cal
D}_0$ on $\R^{d+1}$ (with
coordinates $t,r_1,\ldots,r_d$) of order one,
similar to the Dirac operator, whose square is equal to $-\Del_0\otimes
\Id=-\left(\begin{array}{ccc}
\Del_0 & & \\ & \ddots & \\ & & \Del_0\end{array}\right)$ for
$\Del_0=2{\cal M}\partial_t-\sum_{i=1}^d
\partial_{r_i}^2.$ So, in some sense, ${\cal D}_0$ is a square-root of the
free Schr\"odinger operator,
just as the Dirac operator is a square-root of the D'Alembertian. The
group of Lie invariance of ${\cal D}_0$
has been studied in \cite{Fus93}, it is isomorphic to $\Sch^d$ with a
realization different but close to that
of Proposition 1.1.

Let us restrict to the case $d=1$ (see \cite{Unt2} for details). Then ${\cal
D}_0$ acts on spinors, or couples of
functions $\left(\begin{array}{c} \phi_1\\ \phi_2\end{array}\right)$ of
two variables $t,r$, and
may be written as
\BEQ
{\cal D}_0=\left(\begin{array}{cc} \partial_r & -2{\cal M}\\ \partial_t &
-\partial_r \end{array}\right).
\EEQ
One checks immediately that ${\cal D}_0^2=-\Del_0\otimes \Id.$

From the explicit realization of $\sch$ on spinors (see \cite{Unt2}), one
may easily guess a realization of
$\sv$ that extends the action of $\sch$, and, more interestingly perhaps,
acts on an affine space ${\cal D}^{aff}$ of Dirac-L\'evy-Leblond operators
with potential, in the same spirit as in the previous section. More
precisely, one has the following theorem (we need to introduce some
notations first).

{\bf Definition 3.2.}

{\it Let ${\cal D}^{lin}$ be the vector space of first order matrix
operators on $\R^2$ defined by
$$D\in {\cal D}^{lin}\Leftrightarrow D=h(r,t){\cal
D}_0+\left(\begin{array}{cc} 0 & 0\\ V(r,t) & 0\end{array}
\right),\quad h,V\in C^{\infty}(\R^2)$$
and ${\cal D}^{aff}$, ${\cal D}^{aff}_{\le 2}$ be the affine subspaces of
${\cal D}^{lin}$ such that
$$D\in {\cal D}^{aff}\Leftrightarrow D={\cal D}_0+\left(\begin{array}{cc}
0 & 0\\ V(r,t) & 0\end{array}
\right),$$
$$D\in {\cal D}^{aff}_{\le 2} \Leftrightarrow D={\cal
D}_0+\left(\begin{array}{cc} 0 & 0\\
g_0(t) r^2+g_1(t)r+g_2(t) & 0\end{array}
\right).$$ }

We shall call {\it Dirac potential} a matrix of the form
$\left(\begin{array}{cc} 0 & 0\\ V(r,t) & 0\end{array}
\right),$ with  $V\in C^{\infty}(\R^2)$.

{\bf Definition 3.3.}

{\it
Let $d\pi_{\lambda}^{\sigma}$  $(\lambda\in\C)$ be the infinitesimal
representation of $\sv$ on the
space $\widetilde{\cal H}_{\lambda}^{\sigma}\simeq (C^{\infty}(\R^2))^2$
with coordinates $t,r$, defined by
\BEA
d\pi_{\lambda}^{\sigma}(L_f) &=& (-f(t)\partial_t-\half
f'(t)r\partial_r-{1\over 4} {\cal M} f''(t) r^2
)\otimes \Id \nonumber  \\
 & & -f'(t)\otimes \left(\begin{array}{cc} \lambda-{1\over 4} & \\
& \lambda+{1\over 4} \end{array}
\right)-\half f''(t) r\otimes \left(\begin{array}{cc} 0& 0\\ 1 &
0\end{array}\right);  \\
d\pi_{\lambda}^{\sigma}(Y_g) &=& (-g(t)\partial_r-{\cal M}f'(t)
r)\otimes\Id-f'(t)\otimes
 \left(\begin{array}{cc} 0& 0\\ 1 & 0\end{array}\right); \\
d\pi_{\lambda}^{\sigma}(M_h) &=&  -{\cal M}f(t)\otimes\Id. 
\EEA
}

{\bf Theorem 3.9.}
{\it
\begin{enumerate}

\item
Let $d\sigma: \sv\to {\mathrm{Hom}}({\cal D}^{lin},{\cal D}^{lin})$ defined
by the  left-and-right infinitesimal
action
$$d\sigma(X): D\to d\pi_1^{\sigma}(X)\circ D-D\circ
d\pi_{\half}^{\sigma}(X).$$
Then $d\sigma$ maps ${\cal D}_{\le 2}^{aff}$ into the vector space of
Dirac potentials.

\item
If one represents the Dirac potential $V=\left(\begin{array}{cc} 0 & 0\\
g_0(t)r^2+g_1(t)r+g_2(t) & 0\end{array}
\right) $ or, indifferently, the Dirac operator ${\cal D}_0+V$, by the
vector $\left(\begin{array}{c}
g_0\\ g_1\\ g_2\end{array}\right)$, then the action of $d\sigma$ on ${\cal
D}_{\le 2}^{aff}$ is given by the same
formula as in Proposition 3.6, except for the affine terms (with
coefficient proportional to ${\cal M}$ or
${\cal M}^2$) that should all be divided by $2{\cal M}$.
\end{enumerate}
}

We shall skip the proof (in the same spirit as Theorem 3.4, Proposition
3.5 and
Proposition 3.6) which presents no difficulty, partly for lack of space,
partly because the action on Dirac
operators doesn't give anything new by comparison with the case of
Schr\"odinger operators.

Note that, as in the previous section, one may define a 'shifted' action
\BEQ
d\sigma_{\lambda}(X): D\to d\pi_{\lambda+1}^{\sigma}(X)\circ D-D\circ
d\pi_{\lambda+\half}^{\sigma}(X)
\EEQ
which will only modify the coefficients of the affine cocycles.

As a concluding remark of these two sections, let us emphasize two points:

--contrary to the case of the Hill operators, there is a free parameter
$\lambda$ in the left-and-right actions on
the affine space of Schr\"odinger or Dirac operators;

--looking at the differences of indices between the left action and the
right action, one may consider somehow
that Schr\"odinger operators are of order one, while Dirac operators are
of order $\half$! (recall the
difference of indices was $2={3\over 2}-(-\half)$ in the case of the Hill
operators, which was the signature
of operators of order 2 -- see \cite{DiF}).

So Schr\"odinger operators are somehow reminiscent of the operators
$\partial+u$ of order one on the line, which
intertwine ${\cal F}_{\lambda}$ with ${\cal F}_{1+\lambda}$ for any value
of $\lambda$. The case of the Dirac
operators, on the other hand, has no counterpart whatsoever for
differential operators on the line.

\subsection{About multi-diagonal differential operators and some
Virasoro-solvable Lie algebras}

The original remark that prompted the introduction of multi-diagonal
differential operators in our context
(see below for a definition) was the following. Consider  the space $\R^3$
with coordinates
$r,t,\zeta$ as in Chapter 1. We introduce the two-dimensional Dirac operator
\BEQ
\widetilde{\cal D}_0=\left(\begin{array}{cc} \partial_r &
-2\partial_{\zeta}\\ \partial_t & -\partial_r
\end{array}\right),
\EEQ
acting on spinors $\left(\begin{array}{c} \phi_1 \\
\phi_2\end{array}\right)\in(C^{\infty}(\R^3))^2$ -
the reader will have noticed that $\widetilde{\cal D}_0$ can be obtained
from the Dirac-L\'evy-Leblond
operator ${\cal D}_0$ of section 3.3 by taking a formal Laplace transform
with respect to the mass.
The kernel of $\widetilde{\cal D}_0$ is given by the equations of motion
obtained from the Lagrangian
density
$$ \left(\bar{\phi}_2(\partial_r \phi_1-2\partial_{\zeta}
\phi_2)-\bar{\phi}_1(\partial_t \phi_1-\partial_r
\phi_2)\right) \ dt \ dr\ d\zeta.$$
Let $d\tilde{\pi}_{\half}^{\sigma}$ be the Laplace transform with respect
to $\cal M$ of the infinitesimal
representation of $\sv$ given in Definition 3.3. Then
$d\tilde{\pi}_{\half}^{\sigma}(\sch)$ preserves the
space of solutions of the equation $\widetilde{\cal D}_0
\left(\begin{array}{c} \phi_1\\ \phi_2 \end{array}
\right)=0,$ $\phi_1,\phi_2\in C^{\infty}(\R^3).$ Now, by computing
 $\widetilde{\cal D}_0 \left(d\tilde{\pi}_{\half}^{\sigma}
(X)\left(\begin{array}{c} \phi_1 \\ \phi_2\end{array}\right) \right)$ for
$X\in\sv$ and $\left(\begin{array}{c}
\phi_1\\ \phi_2 \end{array}\right)$ in the kernel of $\widetilde{\cal
D}_0$, it  clearly appears (do it!) that
if one adds the constraint $\partial_{\zeta}\phi_1=0$, then
 $\widetilde{\cal D}_0 \left(d\tilde{\pi}_{\half}^{\sigma}
(X)\left(\begin{array}{c} \phi_1 \\ \phi_2\end{array}\right) \right)=0$
for every $X\in\sv$, and, what is more,
the transformed spinor $\left(\begin{array}{c} \psi_1\\
\psi_2\end{array}\right)=d\tilde{\pi}_{\half}(X)
\left(\begin{array}{c} \phi_1\\ \phi_2\end{array}\right)$ also satisfies
the same constraint $\partial_{\zeta}
\psi_1=0$. One may realize this constraint by adding to the Lagrangian
density the Lagrange
multiplier term
$(h\partial_{\zeta}\bar{\phi}_1-\bar{h}\partial_{\zeta}\phi_1)\ dt\ dr\
d\zeta$.
The new equations of motion read then $\nabla\left(\begin{array}{c} -h/2
\\ \phi_2 \\
     -\phi_1 \end{array}\right)=0$, with
\BEQ
\nabla=\left(\begin{array}{ccc} 2\partial_{\zeta} & \partial_r &
\partial_t \\ & 2\partial_{\zeta} & \partial_r\\
 & & 2\partial_{\zeta} \end{array}\right).
\EEQ

This is our main example of a multi-diagonal differential operator. Quite
generally, we shall call
{\it multi-diagonal} a function- or operator-valued matrix
$M=(M_{i,j})_{0\le i,j\le d-1}$ such that
$M_{i,j}=M_{i+k,j+k}$ for every admissible triple of indices $i,j,k$. So
$M$ is defined for instance
 by the $d$ independent
coefficients $M_{0,0},\ldots,M_{0,j},\ldots,M_{0,d-1}$, with $M_{0,j}$
located on the {\it $j$-shifted
diagonal}.

An obvious generalization in $d$ dimensions leads to the following definition.

\vspace{0.5cm}
\noindent {\bf Definition 3.4.}\\
{\it Let $\nabla^d$ be the $d\times d$ matrix differential operator of order
one, acting on $d$-uples of
functions $H=\left(\begin{array}{c} h_0 \\ \vdots \\ h_{d-1}
\end{array}\right)$ on $\R^d$ with coordinates
$t=(t_0,\ldots,t_{d-1})$, given by
\BEQ
\nabla^d=\left(\begin{array}{ccccc}  \partial_{t_{d-1}} &
\partial_{t_{d-2}} & \cdots & \partial_{t_1} &
 \partial_{t_0}\\
 0 & \ddots & &  & \partial_{t_1} \\
\vdots & \ddots & & & \vdots \\
& & & & \partial_{t_{d-2}} \\
0 & \cdots & & 0 & \partial_{t_{d-1}} \end{array}\right)
\EEQ }
So $\nabla^d$ is upper-triangular, with coefficients
$\nabla^d_{i,j}=\partial_{t_{i-j+d-1}}$, $i\le j$.
The kernel of $\nabla^d$ is defined by a system of equations linking
$h_0,\ldots,h_{d-1}$. The set of
differential operators of order one of the form
\BEQ
X=X_1+\Lambda= (\sum_{i=0}^{d-1} f_i(t)\partial_{t_i})\otimes \Id +
\Lambda,\quad \Lambda=(\lambda_{i,j})\in
Mat_{d\times d}(C^{\infty}(\R^d))
\EEQ
preserving the equation $\nabla^d H=0$ forms a Lie algebra, much too large
for our purpose.

Suppose now (this is a very restrictive condition) that
$\Lambda={\mathrm{diag}}(\lambda_0,\ldots,\lambda_{d-1})$
is diagonal. Since $\nabla^d$ is an operator with constant coefficients,
$[X_1,\nabla^d]$ has no term of
zero order, whereas $[\Lambda,\nabla^d]_{i,j}=\lambda_i
\partial_{t_{i-j+d-1}}-\partial_{t_{i-j+d-1}}
\lambda_j$ $(i\le j)$ does have terms of zero order in general. One
possibility to solve this constraint,
motivated by the preceding examples (see for instance the representation
$d\pi_{\lambda}$ of (\ref{gl:repsv}), with
${\cal M}$ replaced by $\partial_{\zeta}$), but also by the theory of
scaling in statistical physics
 (see commentary following
Proposition 1.1.),
is to impose $\lambda_i=\lambda_i(t_0)$, $i=0,\ldots,d-2$, and
$\lambda_{d-1}=0$.  Since $[X,\nabla^d]$ is
of first order, preserving $\Ker\nabla^d$ is equivalent to a relation of
the type $[X,\nabla^d]=A\nabla^d$,
with $A=A(X)\in Mat_{d\times d}(C^{\infty}(\R^d)).$ Then the matrix operator
 $[X_1,\nabla^d]$ is upper-triangular, and
multi-diagonal, so this must also hold for $A\nabla^d-[\Lambda,\nabla^d]$.
By looking successively at the
coefficients of $\partial_{t_{d-1-l}}$ on the $l$-shifted diagonals,
$l=0,\ldots,d-1$, one sees easily
that $A$ must also be upper-triangular and multi-diagonal, and that one
must have $\lambda_i(t_0)
-\lambda_{i+1}(t_0)=\lambda(t_0)$ for a certain function $\lambda$
independent of $i$, so
$\Lambda=\left(\begin{array}{cccc} (d-1)\lambda & & & \\ & \ddots & & \\ &
& \lambda & \\ & & & 0
\end{array}\right).$ Also, denoting by
$a_0=A_{0,0},\ldots,a_{d-1}=A_{0,d-1}$ the
 coefficients of the first line of the matrix $A$, one
obtains :
\BEQ
\partial_{t_i}f_j=0\quad (i>j);
\EEQ
\BEQ
a_0=\partial_{t_0}f_0+(d-1)\lambda=\partial_{t_1}f_1+(d-2)\lambda=\ldots=
\partial_{t_{d-1}}f_{d-1};
\EEQ
\BEQ
a_i=\partial_{t_0}f_i=\partial_{t_1}f_{i+1}=\ldots=\partial_{t_{d-i-1}}f_{d-1}
\quad (i=1,\ldots,d-1).
\EEQ
In particular, $f_0$ depends only on $t_0$.

From all these considerations follows quite naturally the following
definition. We let $\Lambda_0\in Mat_{
d\times d}(\R)$ be the diagonal matrix $\Lambda_0=\left(\begin{array}{ccc}
d-1 & &\\ & \ddots & \\ & & 0
\end{array}\right)$.

\vspace{0.5cm}
\noindent {\bf Lemma 3.10.}

{\it Let $\md_{\eps}^d$ $(\eps\in\R)$ be the set of differential operators
of order one of the type
\BEQ
X=\left(f_0(t_0)\partial_{t_0}+\sum_{i=1}^{d-1}
f_i(t)\partial_{t_i}\right)\otimes\Id-\eps f'_0(t_0)\otimes
\Lambda_0
\EEQ
preserving $\Ker\nabla^d$.

Then $\md_{\eps}$ forms a Lie algebra.
}

\noindent {\bf Proof.}

Let $\cal X$ be the Lie algebra of vector fields $X$ of the form
$$X=X_1+X_0=\left(f_0(t_0)\partial_{t_0}+\sum_{i=1}^{d-1}
f_i(t)\partial_{t_i}\right)\otimes\Id+\Lambda, \quad
\Lambda=\diag(\lambda_i)_{i=0,\ldots,d-1}\in Mat_{d\times d}
(C^{\infty}(\R^d))$$
preserving $\Ker \nabla^d$. Then the set $\{Y=\sum_{i=0}^{d-1}
f_i(t)\partial_{t_i}\ |\ \exists \lambda\in
Mat_{d\times d}(C^{\infty}(\R^d)),\ Y+\Lambda\in{\cal X}\}$ of the
differential parts of order one
of the elements of $\cal X$ forms a Lie algebra, say ${\cal X}_1$. Define
$${\cal X}_1^{\eps}:=\{ \sum_{i=0}^{d-1} f_i(t)\partial_{t_i}-\eps
f'_0(t)\otimes \Lambda_0\ |\
\sum_{i=0}^{d-1} f_i(t)\partial_{t_i}\in {\cal X}_1\}.$$
Let $Y=(\sum f_i(t)\partial_{t_i})\otimes \Id-\eps f'_0(t)\otimes \Lambda_0,
Z=(\sum g_i(t)\partial_{t_i})\otimes \Id-\eps g'_0(t)\otimes \Lambda_0$ be
two elements of ${\cal X}_1^{\eps}$:
then $$[Y,Z]=\left((f_0 g'_0-f'_0
g_0)(t_0)\partial_{t_0}+\ldots\right)\otimes\Id-\eps (f_0 g'_0 - f'_0
g_0)'(t_0)\otimes \Lambda_0$$ belongs to ${\cal X}_1^{\eps}$, so ${\cal
X}_1^{\eps}$ forms a Lie algebra.
Finally, $\md_{\eps}^d$ is the Lie subalgebra of ${\cal X}_1^{\eps}$
consisting of all differential operators
preserving $\Ker \nabla^d$.\\
\eop

It is quite possible to give a family of generators and relations for
$\md_{\eps}^d$. The surprising fact,
though, is
the following :  for $d\ge 4$, one finds by solving the equations that
$f''_0$ is necessarily zero
if $\eps\not=0$ (see proof of Theorem 3.11). So in any case, the only Lie
algebra that deserves to
 be considered for $d\ge 4$ is $\md_0^d$.

The algebras $\md_{\eps}^d$ $(d=2,3)$, $\md_0^d$ $(d\ge 4)$ are
semi-direct products of a Lie subalgebra
isomorphic to $\Vect(S^1)$, with generators
$$ L_{f_0}^{(0)}=(-f_0(t_0)\partial_{t_0}+\ldots)\otimes\Id-\eps
f'_0(t_0)\otimes \Lambda_0$$
and commutators $[L_{f_0}^{(0)},L_{g_0}^{(0)}]=L^{(0)}_{f'_0 g_0-f_0
g'_0},$ with a nilpotent Lie algebra
consisting of all generators with coefficient of $\partial_{t_0}$
vanishing. When $d=2,3$, one retrieves
realizations of the familiar Lie algebras $\Vect(S^1)\ltimes {\cal
F}_{1+\eps}$ and $\sv_{\eps}$.

{\bf Theorem 3.11. (structure of $\md_{\eps}^d$).}

{\it

\begin{enumerate}
\item
(case $d=2$). Put $t=t_0,r=t_1$: then $\md_{\eps}^2=\langle L_f^{(0)},
L_g^{(1)}\rangle_{f,g\in C^{\infty}(S^1)}$
with
\BEQ
L_f^{(0)}=\left(-f(t)\partial_t-(1+\eps)f'(t)r\partial_r\right)\otimes \Id
 +\eps f'(t)\otimes \left(\begin{array}{cc}
1 & \\ & 0\end{array}\right),
\EEQ
\BEQ
L_g^{(1)}= -g(t)\partial_r.
\EEQ
It is isomorphic to $\Vect(S^1)\ltimes {\cal F}_{1+\eps}.$
\item
(case $d=3$). Put $t=t_0,r=t_1, \zeta=t_2$: then $\md_{\eps}^3=\langle
L_f^{(0)}, L_g^{(1)}, L_h^{(2)}\rangle_{f,g,h\in C^{\infty}(S^1)}$ with
\BEQ
L_f^{(0)}=\left(-f(t)\partial_t-(1+\eps)f'(t)r\partial_r-\left[
(1+2\eps)f'(t)\zeta+{1+\eps\over 2}f''(t)
r^2\right] \partial_{\zeta}\right)\otimes \Id +\eps f'(t)\otimes
\left(\begin{array}{ccc} 2 & & \\
& 1 & \\ & & 0\end{array}\right),
\EEQ
\BEQ
L_g^{(1)}= -g(t)\partial_r-g'(t)r\partial_{\zeta},
\EEQ
\BEQ
L_h^{(2)}=-h(t)\partial_{\zeta}.
\EEQ
The Lie algebra obtained by taking the modes 
\BEQ
L_n=L^{(0)}_{t^{n+1}},\ Y_m=L^{(1)}_{t^{m+1+\eps}},\
M_p=L^{(2)}_{t^{p+1+2\eps}}
\EEQ
 is isomorphic to $\sv_{1+2\eps}$ {\em (see Definition 1.7)}.  In particular, the differential
parts give three independent copies of the
representation $d\tilde{\pi}$ of $\sv$ when $\eps=-\half.$

\item
(case $\eps=0, d\ge 2$) Then $\md^d_{\eps}\simeq \Vect(S^1)\otimes
\R[\eta]/\eta^d$ is generated by the
\BEQ
L_g^{(k)}=-g(t_0)\partial_{t_k}-\sum_{i=1}^{d-1-k} g^{(i)}(t_0)t_1^{i-1}
\left( {1\over i!} t_1 \partial_{t_{i+k}}
+{1\over (i-1)!} \sum_{j=2}^{d-i-k} t_j \partial_{t_{i+j+k-1}}\right),
\quad g\in C^{\infty}(S^1)
\EEQ
$k=0,\ldots,d-1$, with commutators
$[L_g^{(i)},L_h^{(j)}]=L_{g'h-gh'}^{(i+j)} $ if $i+j\le d-1$, $0$ else.

\end{enumerate}

}

{\bf Proof.}

Let $X=-\left(f_0(t_0)\partial_{t_0}+\sum_{i=1}^{d-1}
f_i(t)\partial_{t_i}\right)\otimes \Id+\eps f'_0(t)\otimes
\Lambda_0$ : a set of necessary and sufficient conditions for $X$ to be in
$\md_{\eps}^d$ has been given before
Lemma 3.10, namely
$$\partial_{t_i}f_j=0\ {\mathrm{if}}\  i>j,$$
$$(1+\eps(d-1))f'_0(t_0)=\partial_{t_1}f_1(t_0,t_1)+\eps(d-2)f'_0(t_0)=\ldots=
\partial_{t_{d-1}}
f_{d-1}(t_0,\ldots,t_{d-1})$$ and
$$\partial_{t_0}f_i=\partial_{t_1}f_{i+1}=\ldots=\partial_{t_{d-i-1}}f_{d-1}
\quad (i=1,\ldots,d-1).$$

Solving successively these equations yields
\BEQ
f_i(t_0,\ldots,t_i)=(1+\eps i)f'_0(t_0) \ .\
t_i+f_i^{[1]}(t_0,\ldots,t_{i-1}), \quad i\ge 1;
\EEQ
\BEQ
f_i^{[1]}(t_0,\ldots,t_{i-1})=\partial_{t_0}f_1^{[1]}(t_0)\ .\
t_{i-1}+(1+\eps)f''_0(t_0)\int_0^{t_{i-1}}
t_1 dt_{i-1}+f_i^{[2]}(t_0,\ldots,t_{i-2});
\EEQ
At the next step, the relation $\partial_{t_0}f_2=\partial_{t_1}f_3$
yields the equation
$$(1+2\eps)f''_0(t_0)\ .\ t_2+(f_1^{[1]})''(t_0) \ .\
t_1+2(1+\eps)f'''_0(t_0)\ .\ t_1+
(f_2^{[2]})'(t_0)=(1+\eps)f''_0(t_0)\ .\
t_2+\partial_{t_1}f_3^{[2]}(t_0,t_1)$$
which has no solution as soon as $\eps\not=0$ and $f''_0\not=0$. So, as we
mentioned without proof before
the theorem, the most interesting case is $\eps=0$ when $d\ge 4$.

 The previous computations completely solve
the cases $d=2$ and $d=3$. So let us suppose that $d\ge 4$ and $\eps=0$.

Then, by solving the next equations, one sees by induction that
$f_0,\ldots,f_{d-1}$ may be expressed
in terms of $d$ arbitrary functions of $t_0$, namely, $f_0=f_0^{[0]},
f_1^{[1]}, f_2^{[2]},\ldots, f_{d-1}^{[d-1]}$, and that generators
satisfying $f_i^{[i]}=0$ for every $i\not=k$, $k$ fixed, are necessarily
of the form
$$f_k^{[k]}(t_0)\partial_{t_k}+\sum_{j=1}^{d-1-k}
g_{k+j}(t_0,\ldots,t_j)\partial_{t_{k+j}}$$
for functions $g_{k+j}$ that may be expressed in terms of $f_k^{[k]}$ and
its derivatives.

One may then easily check that $L_{-f_k^{[k]}}^{(k)}$ is of this form and
satisfies the conditions for
being in $\md_{\eps}^d$, so  we have proved that the $L_f^{(k)}$,
$k=0,\ldots, d-1$, $f\in C^{\infty}(S^1)$,
 generate $\md_{\eps}^d$.

All there remains to be done is to check for commutators. Since
$L_f^{(i)}$ is homogeneous of degree $-i$
for the Euler-type operator $\sum_{k=0}^{d-1} kt_k \partial_{t_k}$, one
necessarily has $[L_f^{(i)},
L_g^{(j)}]=L_{C(f,g)}^{(i+j)}$ for a certain function $C$ (depending on
$f$ and $g$) of the time-coordinate
$t_0$. One gets immediately $[L_f^{(0)},L_g^{(0)}]=L_{f'g-fg'}^{(0)}$.
Next (supposing $l>0$),  since
$$L_g^{(0)}=-\sum_{i=0}^{l-1} E_i^0(g)\partial_{t_i} -
(g'(t_0)t_l+F_l^0(t_0,\ldots,t_{l-1}))\partial_{t_l}+\ldots$$ where
$E_i^0(g)$, $i=0,\ldots,l-1$ do not depend on $t_l$, and
$$L_h^{(l)}=-h(t_0)\partial_{t_l}+\ldots,$$
one gets $[L_g^{(0)},L_h^{(l)}]=(gh'-g'h)(t_0) \partial_{t_l}+\ldots$, so
$[L_g^{(0)},L_h^{(l)}]=L_{g'h-gh'}^{(l)}.$
Considering now $k,l>0$, then one has
$$L_g^{(k)}=-\sum_{i=0}^{l-1}
E_i^k(g)\partial_{t_{i+k}}-(h'(t_0)t_l+F_l^k(t_0,\ldots,t_{l-1})\partial_{t_{l+
k
}}$$
where $E_i^k(g)$, $i=0,\ldots,l-1$, do not depend on $t_l$, and  a similar
formula for $L_h^{(l)}$, which give
together the right formula for $[L_g^{(k)},L_h^{(l)}].$ \\
\eop

Let us come back to the original motivation, that is, finding new
representations of $\sv$ arising in a geometric
context. Denote by $d\pi^{(3,0)}$ the realization of $\sv$ given in
Theorem 3.11.

{\bf Definition 3.5}

{\it
Let $d\pi^{\nabla}$ be the infinitesimal representation of $\sv$ on the
space $\widetilde{\cal H}^{\nabla}\simeq
(C^{\infty}(\R^2))^3$ with coordinates $t,r$, defined by

\BEA
d\tilde{\rho}(L_f)&=&\left( -f(t)\partial_t-\half f'(t)r\partial_r
\right)\otimes {\mathrm{Id}}
+f'(t) \otimes \left(\begin{array}{ccc} -1 & & \\ & -\half & \\ & & 0
\end{array}\right) \\
&+& {1\over 2} f''(t) r \otimes \left(\begin{array}{ccc} 0& 1& 0\\ & 0&1\\
& & 0\end{array}\right)
+{1\over 4} f'''(t) r^2 \otimes \left(\begin{array}{ccc} 0 &0 & 1\\ &
0&0\\ & & 0\end{array}\right);
\EEA
\BEQ
d\tilde{\rho}(Y_f)=-f(t)\partial_r
\otimes{\mathrm{Id}}
+f'(t) \otimes  \left(\begin{array}{ccc} 0& 1& 0\\ & 0&1\\ & &
0\end{array}\right)
+f''(t) r\otimes    \left(\begin{array}{ccc} 0 &0 & 1\\ & 0&0\\ & &
0\end{array}\right);
\EEQ
\BEQ
d\tilde{\rho}(M_f)=
    f'(t) \otimes \left(\begin{array}{ccc} 0 &0 & 1\\ & 0&0\\ & &
0\end{array}\right) .
\EEQ
}

{\bf Proposition 3.12.}

{\it
For every $X\in\sv$, $d\pi^{\nabla}(X)\circ\nabla-\nabla\circ
d\pi^{(3,0)}(X)=0.$
}

{\bf Proof.}

Let $X\in\sv$; put
$d\pi^{(3,0)}(X)=-(f_0(t)\partial_t+f_1(t,r)\partial_r+f_2(t,r,\zeta)
\partial_{\zeta})\otimes
\Id-f'_0(t)\otimes \left(\begin{array}{ccc} 1&& \\ & \half &\\ & &
0\end{array}\right).$

The computations preceding Lemma 3.10 prove that
$[d\pi^{(3,0)}(X),\nabla^d]=A(X)\nabla^d,$ $A(X)$ being
the upper-triangular, multi-diagonal matrix defined by
$$A(X)_{0,0} =\partial_{\zeta}f_2, \ A(X)_{0,1}=\partial_r f_2,\
A(X)_{0,2}=\partial_t f_2.$$
Hence one has
$$A(L_f)={r\over 2} f''(t) \left(\begin{array}{ccc} 0& 1&0\\ & 0& 1\\ & &
0\end{array}\right)+{r^2\over 4}
f'''(t) \left(\begin{array}{ccc} 0&0&1\\ & 0&0\\ & & 0\end{array}\right),$$
$$A(Y_g)=g'(t)  \left(\begin{array}{ccc} 0& 1&0\\ & 0& 1\\ & &
0\end{array}\right) + rg''(t)
 \left(\begin{array}{ccc} 0&0&1\\ & 0&0\\ & & 0\end{array}\right),$$
and
$$A(M_h)= f'(t) \otimes \left(\begin{array}{ccc} 0 &0 & 1\\ & 0&0\\ & &
0\end{array}\right).$$
Hence the result. \\
\eop

{\bf Remark.}

Consider the affine space
$${\cal H}_{\nabla}^{aff}=\{ \nabla+\left(\begin{array}{ccc} g_0 & g_1&
g_2\\ & g_0 & g_1\\ & & g_0
\end{array}\right)\ |\ g_0,g_1,g_2\in C^{\infty}(S^1\times\R^2)\}.$$
Then one may define an infinitesimal left-and-right action $d\sigma$ of
$\sv$ on ${\cal H}_{\nabla}^{aff}$
by putting
$$d\sigma(X)(\nabla+V)=d\pi^{\nabla}(X)\circ(\nabla+V)-(\nabla+V)\circ
d\pi^{(3,0)}(X),$$
but the action is simply linear this time, since
$d\pi^{\nabla}\circ\nabla=\nabla\circ d\pi^{(3,0)}(X).$
So this action is not very interesting and doesn't give anything new.

%

\section{Cartan's prolongation and generalized modules of tensor
densities.}

\subsection{The Lie algebra $\sv$ as a Cartan prolongation}

As in the case of vector fields on the circle, it is natural, starting
from the  representation $d\tilde{\pi}$ of $\sv$ given by formula
(\ref{gl:repsvzeta}), to
consider
the subalgebra $\fsv\subset\sv$ made up of the vector fields with
polynomial
coefficients. Recall from Definition 1.6 that  the outer derivation
$\del_2$ of $\sv$ is defined by
\BEQ
\del_2(L_n)=n,\ \del_2(Y_m)=m-\half,\ \del_2(M_n)=n-1\quad
(n\in\Z,m\in\half+\Z).
\EEQ
and  that $\del_2$  is simply
obtained from the Lie action of the Euler operator
$t\partial_t+r\partial_r
+\zeta \partial_{\zeta}$ in the  representation $d\tilde{\pi}$.
The Lie subalgebra $\fsv$ is given more abstractly, using $\del_2$, as
\BEQ
\fsv=\oplus_{k=-1}^{+\infty} \sv_k
\EEQ
where $\sv_k=\{X\in\sv\ |\ \del_2(X)=kX\}=\langle
L_k,Y_{k+\half},M_{k+1}\rangle$
 is the eigenspace of $\del_2$
corresponding to the eigenvalue $k\in\Z$.

Note in particular that $\sv_{-1}=\langle L_{-1},Y_{-\half},M_0\rangle$
is commutative, generated by the infinitesimal translations $\partial_t,
\partial_r,\partial_{\zeta}$ in the vector field representation,
and that $\g_0=\langle L_0,Y_{\half},M_1\rangle=\langle L_0\rangle\ltimes
\langle Y_{\half},M_1\rangle$ is solvable.

{\bf Theorem 4.1.}

{\it The Lie algebra $\fsv$ is isomorphic to the Cartan prolongation of
$\sv_{-1}\oplus\sv_0$
where $\sv_{-1}\simeq\langle X_{-1},Y_{-\half},M_0\rangle$ and
$\sv_0\simeq\langle
X_0,Y_{\half},M_1\rangle.$}

{\bf Proof.}

Let $\sv_n$ ($n=1,2,\ldots$) the $n$-th level vector space obtained from
Cartan's construction, so that the Cartan prolongation of
$\sv_{-1}\oplus\sv_0$
is equal to the Lie algebra $\sv_{-1}\oplus\sv_0\oplus\ \oplus_{n\ge 1}
\g_n$. It will be enough, to establish the required isomorphism,
to prove the following. Consider the  representation $d\tilde{\pi}$ of
$\sv$.
 Then the space $\h_n$  defined through induction on $n$ by
\BEA
\h_{-1}=\pi(\sv_{-1})= \langle
\partial_t,\partial_r,\partial_{\zeta}\rangle\\
\h_0=\pi(\sv_0)=\langle t\partial_t+\half r\partial_r,
t\partial_r+r\partial_{\zeta},t\partial_{\zeta} \rangle  \\
\h_{k+1}=\{X\in {\cal X}_{k+1}\ | \ [X,\h_{-1}]\subset \h_k \}, \ (k\ge 0)
\EEA
(where ${\cal X}_{k}$ is the space of vector fields with polynomial
coefficients of degree $k$)
is equal to $\pi(\sv_n)$ for any $n\ge 1$.

So assume that
$X=f(t,r,\zeta)\partial_t+g(t,r,\zeta)\partial_r+h(t,r,\zeta)
\partial_{\zeta}$ satisfies
\BEQ
[X,\h_{-1}]\subset\pi(\sv_n)=\langle
t^{n+1}\partial_t+\half(n+1)t^n r\partial_r+{1\over 4} (n+1)nt^{n-1}r^2
\partial_{\zeta}, t^{n+1}\partial_r+(n+1)t^n r\partial_{\zeta},
t^{n+1}\partial_{\zeta}\rangle.
\EEQ

In the following lines, $C_1,C_2,C_3$ are undetermined constants.
Then (by comparing the coefficients of $\partial_t$) $$f(t,r,\zeta)=C_1
t^{n+2}.$$
By inspection of the coefficients of $\partial_r$, one gets then
$$\partial_t g(t,r,\zeta)={C_1\over 2} (n+1)(n+2)t^n r+C_2 (n+2)t^{n+1}$$
so
$$g(t,r,\zeta)={C_1\over 2}(n+2)t^{n+1}r+C_2 t^{n+2}+G(r,\zeta)$$
with an unknown polynomial $G(r,\zeta)$. But
$$[X,Y_{-\half}]=\left({C_1\over 2}(n+2)t^{n+1}+\partial_r
G(r,\zeta)\right)\partial_r
\ {\mathrm{mod}}\ \partial_{\zeta}$$
so $\partial_r G(r,\zeta)=0$.

Finally, by comparing the coefficients of $\partial_{\zeta}$, one gets
$$[X,L_{-1}]=(n+2)C_1[t^{n+1}\partial_t+\half(n+1)t^n r\partial_r]+
C_2 (n+2)t^{n+1}\partial_r+\partial_t h\ \partial_{\zeta}$$
so
$$\partial_t h(t,r,\zeta)={C_1\over 4}(n+2)(n+1)nt^{n-1}r^2+C_2 (n+2)(n+1)
t^n r+C_3 (n+2)t^{n+1}$$
whence
$$h(t,r,\zeta)= {C_1\over 4}(n+2)(n+1)t^{n}r^2+C_2 (n+2)
t^{n+1} r+C_3 t^{n+2}+H(r,\zeta)$$
where $H(r,\zeta)$ is an unknown polynomial. Also
$$[X,Y_{-\half}]={C_1\over 2} (n+2)t^{n+1}\partial_r+{C_1\over
2}(n+2)(n+1)t^n
r\partial_{\zeta}+C_2(n+2)t^{n+1}\partial_{\zeta}+\partial_r
H(r,\zeta)\partial_{\zeta},$$
so $H=H(\zeta)$ does not depend on $r$; finally
$$[X,M_0]={dG(\zeta)\over d\zeta}\partial_r+{dH(\zeta)\over
d\zeta}\partial_{\zeta}$$
so $G=H=0$. \\
\eop

{\bf Remark :} by modifying slightly the definition of $\sv_0$, one gets related Lie algebras. For instance, substituting $L_0^{\eps}:=-t\partial_t
-(1+\eps) r\partial_r-(1+2\eps)\zeta \partial_{\zeta}$ for $L_0$ 
leads to the 'polynomial part' of $\sv_{1+2\eps}$ (see Theorem 3.11 for
an explicit realization of $\sv_{1+2\eps}$). 

\subsection{Coinduced representations of $\sv$}

In order to classify 'reasonable' representations of the Virasoro algebra,
V. G. Kac made the following
conjecture: the Harish-Chandra representations, those for which $\ell_0$
acts semi-simply with
finite-dimensional eigenspaces, are either higher- (or lower-) weight
modules, or tensor density
modules. As proved in \cite{MarPia} and \cite{Mat}, 
 one has essentially two  types
of Harish-Chandra representations of the Virasoro algebra :

- Verma modules which are {\it induced} to $\vir$  from a character of
$\vir_+=\langle L_0,L_1,\ldots\rangle$,
zero on the subalgebra $\vir_{\ge 1}=\langle L_1,\ldots\rangle$, and
quotients of degenerate Verma modules (see Section
6 for a generalization in our case);

- tensor modules of formal densities which are {\it coinduced} to the
subalgebra of formal or polynomial vector fields
$\Vect(S^1)_{\ge -1}=\langle L_{-1},L_0,\ldots\rangle$ from a character of
$\Vect(S^1)_{\ge 0}$ that is zero on the subalgebra
$\Vect(S^1)_{\ge 1}$. These modules extend naturally to representations of
$\Vect(S^1)$.

We shall generalize in this paragraph this second type of representations
to the case of $\sv$. Note that although
 we have
two natural graduations on $\sv$, the one given by the structure of Cartan
prolongation is most adapted here since $\sv_{-1}$
is commutative (see \cite{Alb}).

Let $d\rho$ be a representation of $\sv_0=\langle
L_0,Y_{\half},M_1\rangle$ into a vector space ${\cal H}_{\rho}$. Then
$d\rho$ can be trivially extended to $\sv_+=\oplus_{i\ge 0}\sv_i$ by
setting $d\rho(\sum_{i>0}\sv_i)=0$. Let
$\fsv=\oplus_{i\ge -1}\sv_i\subset\sv$ be the subalgebra of 'formal'
vector fields: in the representation $d\pi$,
the image of $\fsv$ is the subset of  vector fields that are polynomial in
the time coordinate.

Let us now define the  representation of $\fsv$ coinduced from $d\rho$.

{\bf Definition 4.1.}

{\it The {\it $\rho$-formal density module} $(\tilde{\cal
H}_{\rho},d\tilde{\rho})$ is the coinduced module
\BEA
\tilde{\cal H}_{\rho} &=& {\mathrm{Hom}}_{{\cal U}(\sv_+)} ({\cal
U}(\fsv),{\cal H}_{\rho}) \\
&=& \{ \phi:\ {\cal U}(\fsv)\to {\cal H}_{\rho} \ {\mathrm{linear}}\ |\
\phi(U_0 V)=d\rho(U_0).\phi(V),
\ \ U_0\in{\cal U}_{\sv_+},V\in{\cal U}(\fsv) \}
\}
\EEA
with the natural action of ${\cal U}(\fsv)$ on the right
\BEQ
(d\tilde{\rho}(U).\phi)(V)=\phi(VU),\quad U,V\in{\cal U}(\fsv).
\EEQ}

By Poincar\'e-Birkhoff-Witt's theorem, this space can be identified with
$$ {\mathrm{Hom}}({\cal U}(\sv_+) \setminus {\cal U}(\fsv),{\cal
H}_{\rho})\simeq {\mathrm{Hom}}({\mathrm{Sym}}(\sv_{-1}),{\cal H}_{\rho})$$
(linear applications from the symmetric algebra on $\sv_{-1}$ into ${\cal
H}_{\rho}$), and this last space is in turn
isomorphic with the space ${\cal H}_{\rho}\otimes\R[[t,r,\zeta]]$ of ${\cal
H}_{\rho}$-valued functions of $t,r,\zeta$,
through the application
\BEA
{\cal H}_{\rho}\otimes\R[[t,r,\zeta]] &\longrightarrow&
{\mathrm{Hom}}({\mathrm{Sym}}(\sv_{-1}),{\cal H}_{\rho}) \\
F(t,r,\zeta)&\longrightarrow& \phi_{F}: (U\to \partial_U
F|_{t=0,r=0,\zeta=0})
\EEA
where $\partial_U$ stands for the product derivative $\partial_{L_{-1}^j
Y_{-\half}^k M_0^l}=(-\partial_t)^j
(-\partial_r)^k(-\partial_{\zeta})^l$ (note our choice of signs!).

We shall really be interested in the action of $\fsv$ on functions
$F(t,r,\zeta)$ that we shall denote by $d\sigma_{\rho}$, or $d\sigma$ for
short.

The above morphisms allow one to compute the action of $\fsv$ on monomials
through the equality

\BEA
\left( {\partial_t^j\over j!}\  {\partial_r^k\over k!}\
{\partial_{\zeta}^l\over l!}\right)|_{t=0,r=0,\zeta=0}
(d\sigma(X).F) &=& {(-1)^{j+k+l} \over j!k!l!}
(d\tilde{\rho}(X).\phi_F)(L_{-1}^j Y_{-\half}^k M_0^l) \\
&=&  {(-1)^{j+k+l} \over j!k!l!} \phi_F(L_{-1}^j Y_{-\half}^k M_0^l X),
\quad X\in\fsv.
\EEA

In particular,

$$\partial_t^j \partial_r^k \partial_{\zeta}^l|_{t=0,r=0,\zeta=0}
(d\sigma(L_{-1}).F) =
-\partial_t^{j+1} \partial_r^k \partial_{\zeta}^l|_{t=0,r=0,\zeta=0} F$$
so $d\sigma(L_{-1}).F=-\partial_t F$; similarly,
$d\sigma(Y_{-\half}).F=-\partial_r F$ and
$d\sigma(M_0).F=-\partial_{\zeta} F.$

So one may assume that $X\in\sv_+$ : by Poincar\'e-Birkhoff-Witt's
theorem, $L_{-1}^j Y_{-\half}^k M_0^l X$ can be
rewritten as $U+V$ with
$$U\in \sv_{>0} {\cal U}(\fsv)$$
and
$$V=V_1 V_2, \quad V_1\in {\cal U}(\sv_0), V_2\in{\cal U}(\sv_{-1}).$$

Then $\phi_F(U)=0$ by definition of $\tilde{\cal H}_{\rho}$, and
$\phi_F(V)$ may easily be computed as
$\phi_F(V)=d\rho(V_1)\otimes \partial_{V_2}|_{t=0,r=0,\zeta=0} F$.

{\bf Theorem 4.2.}

{\it
Let $f\in\R[t]$, the coinduced representation $d\tilde{\rho}$ is given by
the action of the
following matrix differential operators on functions:

\BEQ
d\tilde{\rho}(L_f)=\left( -f(t)\partial_t-\half f'(t)r\partial_r-{1\over 4}
f''(t) r^2 \partial_{\zeta}\right)\otimes {\mathrm{Id}}_{{\cal H}_{\rho}}
+f'(t) d\rho(L_0)
+{1\over 2} f''(t) r d\rho(Y_{\half})+{1\over 4} f'''(t) r^2 d\rho(M_1);
\EEQ
\BEQ
d\tilde{\rho}(Y_f)=\left(-f(t)\partial_r-f'(t) r\partial_{\zeta}\right)
\otimes{\mathrm{Id}}_{{\cal H}_{\rho}}
+f'(t)
d\rho(Y_{\half})+f''(t) r\  d\rho(M_1);
\EEQ
\BEQ
d\tilde{\rho}(M_f)=-f(t)\partial_{\zeta}\otimes{\mathrm{Id}}_{{\cal H}_{\rho}}
    +f'(t) \ d\rho(M_1).
\EEQ
}

{\bf Proof.}

One easily checks that these formulas define a representation of $\fsv$.
Since $(L_{-1},Y_{-\half},M_0,L_0,L_1,L_2)$
generated $\fsv$ as a Lie algebra, it is sufficient to check the above
formulas for $L_0,L_1,L_2$ (they
are obviously correct for $L_{-1},Y_{-\half},M_0$).

Note first that $M_0$ is central in $\fsv$, so $$\partial_{\zeta}^l
(d\sigma(X).F)=d\sigma(X). (\partial_{\zeta}^l F).$$
Hence it will be enough to compute the action on monomials of the form
$ t^j r^l\otimes v$, $v\in {\cal H}_{\rho}$.

We shall give a detailed proof since the computations in ${\cal U}(\fsv)$
are rather involved.

Let us first compute $d\sigma(L_0)$ : one has
\begin{eqnarray*}
(-\partial_t)^j (-\partial_r)^k|_{t=0,r=0} (d\sigma(L_0).F) &=&
\phi_F(L_{-1}^j Y_{-\half}^k L_0) \\
&=& \phi_F(L_{-1}^j L_0 Y_{-\half}^k -{k\over 2} L_{-1}^j Y_{-\half}^k) \\
&=& \phi_F(L_0 L_{-1}^j Y_{-\half}^k - (j+{k\over 2}) L_{-1}^j
Y_{-\half}^k) \\
&=& \left[ d\rho(L_0) (-\partial_t)^j (-\partial_r)^k - (j+{k\over
2})(-\partial_t)^j (-\partial_r)^k\right]F(0)
\end{eqnarray*}
 so $$d\sigma(L_0)=-t\partial_t-\half r\partial_r+d\rho(L_0).$$

Next,
\begin{eqnarray*}
\phi(L_{-1}^j Y_{-\half}^k L_1) &=& \phi(L_{-1}^j L_1 Y_{-\half}^k - k
L_{-1}^j Y_{\half} Y_{-\half}^{k-1}+
{k(k-1)\over 2} L_{-1}^j Y_{-\half}^{k-2} M_0) \\
&=& \phi((-2jL_0 L_{-1}^{j-1}+j(j-1) L_{-1}^{j-1})Y_{-\half}^k) -k
\phi(Y_{\half}L_{-1}^jY_{-\half}^{k-1})\\
&+&
jk \phi(L_{-1}^{j-1}Y_{-\half}^k)+{k(k-1)\over 2} \phi(L_{-1}^j
Y_{-\half}^{k-2}M_0)\\
&=& \big[ (-2j) (-\partial_r)^k (-\partial_t)^{j-1} d\rho(L_0)+j(j-1)
(-\partial_r)^k (-\partial_t)^{j-1}-k
(-\partial_t)^j (-\partial_r)^{k-1} d\rho(Y_{\half}) \nonumber
\\ &+& jk (-\partial_t)^{j-1}(-\partial_r)^k - {k(k-1)
\over 2} \partial_{\zeta} (-\partial_t)^j (-\partial_r)^{k-2}\big]F(0)
\end{eqnarray*}
hence the result for $d\sigma(L_1)$.

Finally,
\begin{eqnarray*}
\phi(L_{-1}^j Y_{-\half}^k L_2) &=& \phi(L_{-1}^j L_2 Y_{-\half}^k-{3\over
2} k L_{-1}^j Y_{3\over 2} Y_{-\half}^{k-1}
+3 {k(k-1)\over 2} L_{-1}^j M_1 Y_{-\half}^{k-2}) \\
&=& \phi((-3j L_1 L_{-1}^{j-1}+3 j(j-1) L_0 L_{-1}^{j-2}-j(j-1)(j-2)
L_{-1}^{j-3})Y_{-\half}^k) \\
&-& {3\over 2} k\phi(-2j Y_{\half} L_{-1}^{j-1} Y_{-\half}^{k-1}+j(j-1)
L_{-1}^{j-2} Y_{-\half}^k)
+{3\over 2} k(k-1) \phi(M_1 L_{-1}^j Y_{-\half}^{k-2} -jM_0
L_{-1}^{j-1}Y_{-\half}^{k-2})\nonumber\\
&=& \big[ \big( 3j(j-1) d\rho(L_0) (-\partial_t)^{j-2}-j(j-1)(j-2)
(-\partial_t)^{j-2}\big) (-\partial_r)^k\\
&-&{3\over 2} k \big( -2j d\rho(Y_{\half}) (-\partial_t)^{j-1}
(-\partial_r)^{k-1}+j(j-1) (-\partial_t)^{j-2}
(-\partial_r)^k\big) \\ &+& {3\over 2} k(k-1) \big( d\rho(M_1)
(-\partial_t)^j (-\partial_r)^{k-2}+j\partial_{\zeta}
(-\partial_t)^{j-1}(-\partial_r)^{k-2}\big)\big]F(0).
\end{eqnarray*}
Hence  $$d\sigma(L_2)=-t^3 \partial_t-{3\over 2} t^2 r\partial_r-{3\over
2} tr^2\partial_{\zeta}+3t^2 d\rho(L_0)
+{3\over 2} t^2 r\ d\rho(Y_{\half})+{3\over 2} r^2\ d\rho(M_1).$$\\
\eop

Let us see how all actions defined in Chapter 3 (except for the coadjoint
action!) derive from this construction.

{\bf Example 1.} Take ${\cal H}_{\rho_{\lambda}}=\R, \
d\rho_{\lambda}(L_0)=-\lambda,\ d\rho_{\lambda}(Y_{\half})
=d\rho_{\lambda}(M_1)=0$ $(\lambda\in\R)$. Then
$d\tilde{\rho}_{\lambda}=d\tilde{\pi}_{\lambda}$
(see second remark following Proposition 1.6 for a definition of
$d\tilde{\pi}_{\lambda}$).

{\bf Example 2.} The linear part of the infinitesimal action on the affine
space of Schr\"odinger operators
(see Proposition 3.4) is given by the restriction of $d\tilde{\rho}_{-1}$
to functions of the type
$g_0(t)r^2+g_1(t)r+g_2(t).$

{\bf Example 3.} Take ${\cal H}_{\rho_{\lambda}}=\R^2$,
$d\rho_{\lambda}(L_0)=\left(\begin{array}{cc}
1/4 &\\ & -1/4 \end{array}\right)-\lambda \Id,$
$d\rho_{\lambda}(Y_{\half})=-\left(\begin{array}{cc}
0 & 0\\ 1 & 0\end{array}\right),$ $d\rho_{\lambda}(M_1)=0$. Then the
infinitesimal representation
$d\pi_{\lambda}^{\sigma}$ of Definition 3.3 (associated with the action on
Dirac operators) is equal
to $d\tilde{\rho}_{\lambda}$ (up to a Laplace transform in the mass).

{\bf Example 4.} (action on multi-diagonal matrix differential operators)
Take ${\cal H}_{\rho}=\R^3,$
$d\rho(L_0)=\left(\begin{array}{ccc} -1 & &\\ & -\half &\\ & &
0\end{array}\right),$
$d\rho(Y_{\half})=d\rho(M_1)=0$ on the one hand;
$${\cal H}_{\sigma}=\R^3,\ d\rho(L_0)=\left(\begin{array}{ccc} -1 & &\\ &
-\half &\\ & & 0\end{array}\right),
\ d\rho(Y_{\half})=\left(\begin{array}{ccc} 0& 1& 0\\ & 0& 1\\ & &
0\end{array}\right)$$
and $d\rho(M_1)=\left(\begin{array}{ccc} 0&0&1\\ & 0&0\\ & &
0\end{array}\right)$ on the other.
Then $d\pi^{(3,0)}=d\tilde{\rho}$ and $d\pi^{\nabla}=d\tilde{\sigma}$ (see
Proposition 3.11 in Section 3.4).

The fact that the coadjoint action cannot be obtained by this construction
follows easily by comparing the
formula for the action of the $Y$ and $M$ generators of Theorem 3.2 and
Theorem 4.2: the second derivative
$f''$ does not appear in $\ad^*(Y_f)$, while it does in
$d\tilde{\rho}(Y_f)$ for any representation $\rho$
such that $d\rho(M_1)\not=0$; if $d\rho(M_1)=0$, then, on the contrary,
there's no way to account for the
first derivative $f'$ in $\ad^*(M_f)$.

{\bf Remark:} The problem of classifying all coinduced representations
 is hence reduced to the problem
of classifying the representations $d\rho$ of the Lie algebra $\langle X_0,
 Y_{\half},M_1\rangle$.
This is {\it a priori} an untractable problem (due to the non-semi-simplicity
 of
this Lie algebra), even if one is satisfied with finite-dimensional
 representations. An interesting
class of examples (to which examples 1 through 4 belong) is provided
 by extending a 
(finite-dimensional, say) representation $d\rho$ of the $(ax+b)$-type Lie
 algebra $\langle X_0,Y_{\half}
\rangle$ to $\langle X_0,Y_{\half},M_1\rangle$ by putting 
$d\rho(M_1)=d\rho(Y_{\half})^2$.
In particular, one may consider the spin $s$-representation $d\sigma$
of $\slin(2,\R)$, restrict it to
the Borel subalgebra considered as $\langle X_0,Y_{\half}\rangle$, 'twist' it by putting
$d\sigma^{\lambda}:=d\sigma+\lambda\Id$ and extend it to $\langle X_0,Y_{\half},M_1\rangle$ as we just explained.

%

\section{Cohomology of
$\sv$ and $\tsv$  and applications to central extensions and deformations}

Cohomological computations for Lie algebras are mainly motivated by the
search for deformations and central extensions. We concentrate on $\tsv$ in
the first three paragraphs of this section, because the generators of $\tsv$
bear integer indices, which is more natural for computations. The main
theorem is Theorem 5.1 in paragraph 5.1, which classifies all deformations
of $\tsv$; Theorem 5.5 shows that all the infinitesimal deformations obtained
in paragraph 5.1 give rise to genuine deformations. One particularly
interesting family of deformations is provided by the Lie algebras
$\tsv_{\lambda}$ ($\lambda\in\R$), which
were introduced in Definition 1.7. We compute their central extensions in
paragraph 5.2, and compute in paragraph 5.3 their deformations in the particular case $\lambda=1$, for which $\tsv_1$ is the tensor product of
$\Vect(S^1)$ with a nilpotent associative and commutative algebra. Finally,
in paragraph 5.4, we come back to the original Schr\"odinger-Virasoro algebra
and compute its deformations, as well as the central extensions of the
family of deformed Lie algebras $\sv_{\lambda}$.

\subsection{Classifying deformations of $\tsv$}

We shall be interested in the classification of all formal deformations of $\tsv$,
following the now classical scheme of Nijenhuis and Richardson:
deformation of a Lie algebra ${\cal G}$ means that one has a formal family
of Lie brackets on ${\cal G}$, denoted $[\ , \ ]_t$, inducing a Lie algebra
structure on the extended  Lie algebra ${\cal G} \displaystyle
\bigotimes_{k} k [[ t ]] = {\cal G} [[t]]$. As well-known, one has to
study the cohomology of ${\cal G}$ with coefficients in the adjoint
representation; degree-two cohomology $H^2 ({\cal G}, {\cal G})$
classifies the infinitesimal deformations (the terms of order one in the
expected formal deformations) and $H^3 ({\cal G}, {\cal G})$ contains the
potential obstructions to a  further prolongation of the deformations. So we
shall naturally begin with the computation of $H^2 (\tsv, \tsv)$ (as usual,
we shall consider only local cochains, equivalently given by differential
operators, or polynomial in the modes):

\vspace{0.5cm}
\noindent {\bf Theorem 5.1} \\
{\it One has $dim \ H^2 ( \tsv, \tsv) = 3$. A set of generators is provided
by the cohomology
classes of the cocycles $c_1, c_2$ and $c_3$, defined as follows in terms of
modes (the missing components of the cocycles are meant to vanish):\\
$c_1 (L_n, Y_m) = -\frac{n}{2} Y_{n+m}, \hspace{1cm} c_1 (L_n, M_n) = - n M_{n+m}$\\
$c_2 (L_n, Y_m) = Y_{n+m} \hspace{1.4cm} c_2 (L_n, M_m) = 2 M_{n+m}$\\
$c_3 (L_n, L_m) = (m-n) M_{n+m}$ }

\vspace{0.5cm}
\noindent {\bf Remarks:}

1.\ The cocycle $c_1$ gives rise to the family of Lie algebras $\tsv_{\eps}$
described in Definition 1.7.

2.\  The cocycle $c_3$ can be described globally as  $c_3:
\Vect (S^1) \times \Vect (S^1) \longrightarrow {\cal F}_0$ given by
$$c_3 (f \partial, g \partial ) = \left| \begin{array}{ccc} f & g \\ f'&
g' \end{array} \right|$$
This cocycle appeared in \cite{Fuk}  and has been used in a
different context in \cite{GuiRog05}.  

\bigskip

Before entering the technicalities of the proof, we shall indicate
precisely, for the comfort of the reader, some cohomological results on
${\cal G} = \Vect (S^1)$ which will be extensively used in the sequel.

\vspace{0.5cm}
\noindent {\bf Proposition 5.2.} {\em 
{\em (see \cite{Fuk}, or \cite{GuiRog05},
chap. IV for a more elementary approach)}
\begin{enumerate}
\item[(1)] $Inv_{\cal G} ({\cal F}_\lambda \otimes {\cal F}_\mu) = 0$
unless $\mu = -1 - \lambda$ and ${\cal F}_\mu = {\cal F}^\ast_\lambda$;
then $Inv_{\cal G}  ({\cal F}_\lambda \otimes {\cal F}^\ast_\lambda )$ is
one-dimensional, generated by the identity mapping.
\item[(2)] $H^i ( {\cal G}, {\cal F}_\lambda \otimes {\cal F}_\mu) \equiv
0$ if $\lambda \neq 1 - \mu$ and $\lambda$ or $\mu$ are not integers.\\
{\em $(1)$ and $(2)$ can be immediately deduced from \cite{Fuk},
theorem 2.3.5 p. 136-137.}
\item[(3)] Let $W_1$ be the Lie algebra of formal vector fields on the
line, its cohomology represents the algebraic part of the cohomology of
${\cal G} = \Vect (S^1)$ {\em (see again \cite{Fuk},  theorem 2.4.12)}.
 Then  $H^1 (W_1,
Hom ({\cal F}_\lambda, {\cal F}_\lambda ))$ is  one-dimensional, generated
by the cocycle $(f \partial, a dx^{- \lambda}) \longrightarrow f^{'} a
dx^{- \lambda}$ {\em (cocycle $I_\lambda$ in \cite{Fuk}, p. 138)}.
\item[(4)] Invariant antisymmetric bilinear operators  ${\cal F}_\lambda \times {\cal F}_\mu
\longrightarrow {\cal F}_\nu$
between densities
have been determined by P. Grozman {\em (see \cite{Gro},  p. 280)}.\\
\noindent They are of the following type:
\begin{enumerate}
\item[(a)]  the Poisson bracket for  $\nu = \lambda + \mu -1$, defined by 
$$\{ f dx ^{- \lambda}, g dx^{- \mu} \} = (\lambda f g' - \mu f' g) dx^{-
(\lambda + \mu - 1)}$$
\item[(b)] the following three exceptional brackets : \\
${\cal F}_{1/2} \times {\cal F}_{1/2} \rightarrow {\cal F}_{-1} \ given \
by \ ( f \partial^{1/2}, g \partial^{1/2}) \rightarrow \frac{1}{2} (f
g^{''} - gf^{''}) dx$;\\
${\cal F}_0 \times {\cal F}_0 \rightarrow {\cal F}_{-3} \ given \ by \ (
f, g) \rightarrow (f^{''}g^{'} - g^{''}f^{'}) dx^3$;\\
and an operator ${\cal F}_{2/3} \times {\cal F}_{2/3} \rightarrow {\cal F}_{-\frac{5}{3}}$
called the Grozman bracket {\em (see \cite{Gro}, p 274)}.
\end{enumerate}
\end{enumerate} }

\vspace{0.5cm}
\noindent {\bf Proof of Theorem  5.1.}

 We shall use standard
techniques in Lie algebra cohomology; the proof will be rather technical,
but without specific difficulties. Let us fix the notations: set $\tsv =
{\cal G} \ltimes \h$ where ${\cal G} = \Vect (S^1)$ and $\h$ is  the nilpotent
part of $\tsv$.\\*One can consider the exact sequence

\BEQ
 0 \longrightarrow \h \longrightarrow {\cal G} \ltimes \h \longrightarrow
{\cal G} \longrightarrow 0 \ \
 \EEQ
as a short exact sequence of ${\cal G} \ltimes \h$ modules, thus inducing
a long exact sequence in cohomology:
\BEQ
\cdots \longrightarrow H^1 ( \tsv, {\cal G}) \longrightarrow H^2 (\tsv, \h)
\longrightarrow H^2 (\tsv, \tsv) \longrightarrow H^2 ( \tsv, {\cal G})
\longrightarrow H^3 (\tsv, \h) \longrightarrow \cdots
\EEQ
So we shall consider $H^\ast (\tsv, {\cal G})$ and $H^\ast (\tsv, \h)$
separately.

\vspace{0.5cm}
\noindent{\bf Lemma 5.3:}\\
{\it $H^\ast (\tsv, {\cal G}) = 0$ for $\ast = 0, 1, 2$.}

\vspace{0.5cm}
\noindent{\bf Proof of  Lemma 5.3:}

One uses the Hochschild-Serre spectral sequence associated with the exact
sequence $(5.1)$. Let us remark first that $H^\ast ({\cal G}, H^\ast (\h,
{\cal G})) = H^\ast ({\cal G}, H^\ast (\h) \otimes {\cal G})$ since $\h$
acts trivially on ${\cal G}$. So one has to understand $H^\ast (\h)$ in
low dimensions; let us consider the exact sequence
$0 \longrightarrow \n \longrightarrow \h  \longrightarrow y
\longrightarrow 0$, where $\n = [ \h, \h]$. As ${\cal G}$-modules, these
algebras are density modules, more precisely $\n = {\cal F}_0$ and $y =
{\cal F}_{ 1/2}$. So $H^1 (\h) = y^\ast = {\cal F}_{-3/2}$ as a ${\cal
G}$-module. Let us recall that, as a module on itself, ${\cal G} = {\cal
F}_{1}$. One gets: $E^{p, 0}_2 = H^p ({\cal G}, {\cal G}) = 0$ as 
well-known (see \cite{Fuk}),
$$E^{1, 1}_2 = H^1 ({\cal G}, H^1 (\h) \otimes {\cal G}) = H^1 ({\cal G},
{\cal F}_{-3/2} \otimes {\cal F}_1)$$
The determination of cohomologies of $\Vect (S^1)$ with coefficients  in
tensor products of modules of densities has been done by Fuks (see \cite{Fuk},
chap. 2, thm 2.3.5, or Proposition 5.2 (2) above), in  this  case everything vanishes and $E^{1,1}_2 = 0$.

One has now to compute $H^2(\h)$ in order to get $E^{0, 2}_2 = Inv_{\cal
G} (H^2 (\h) \otimes {\cal G})$. We shall use the decomposition of
cochains on $\h$ induced by its splitting into vector subspaces: $\h = \n
\oplus y$. So $C^1 (\h) = \n^\ast \oplus y^\ast$ and $C^2 (\h) = \Lambda^2
\n^\ast \oplus \Lambda^2 y^\ast \oplus y^\ast \wedge \n^\ast$. The
coboundary $\partial$ is induced by the only non-vanishing part $\partial
: \n^\ast \longrightarrow \Lambda^2 y^\ast$ which is dual to the bracket
$\Lambda^2 y \longrightarrow \n$. So the cohomological complex splits
into three subcomplexes and one deduces the following exact sequences:
$$ 0 \longrightarrow \n^\ast \stackrel{\partial}{\longrightarrow}
\Lambda^2 y^\ast \longrightarrow M_1 \longrightarrow 0$$
$$ 0 \longrightarrow M_2 \longrightarrow  \Lambda^2 \n^\ast
\stackrel{\partial}{\longrightarrow}  \Lambda^2 y^\ast \otimes \n^\ast$$
$$ 0 \longrightarrow M_3 \longrightarrow  y^\ast \wedge \n^\ast
\stackrel{\partial}{\longrightarrow}  \Lambda^3 y^\ast $$
and $H^2 (\h) = M_1 \oplus M_2 \oplus M_3$. One can then easily deduce the
invariants 
$Inv_{\cal G} (H^2 (\h) \otimes {\cal G}) = \displaystyle \bigoplus^{3}_{i
= 1} Inv_{\cal G} (M_i \otimes {\cal G})$ from the cohomological exact
sequences associated with the above short exact sequences.
 One has: 
$$0 \longrightarrow Inv_{\cal G} (M_2 \otimes {\cal G}) \longrightarrow
Inv_{\cal G} (\Lambda^2 \n^\ast \otimes {\cal G}) = 0$$
 and
$$0 \longrightarrow Inv_{\cal G} (M_3 \otimes {\cal G}) \longrightarrow
Inv_{\cal G} (y^\ast \wedge \n^\ast \otimes {\cal G}) = 0$$
 from  Proposition 5.2; \\
$$\cdots \longrightarrow Inv_{\cal G} (\Lambda^2 y^\ast \otimes {\cal G})
\longrightarrow Inv_{\cal G} (M_1 \otimes {\cal G}) \longrightarrow H^1
({\cal G}, \n^\ast \otimes {\cal G}) \stackrel{\partial_
\ast}{\longrightarrow} H^1 ({\cal G} , \Lambda^2 y^\ast \otimes {\cal G})
\longrightarrow \cdots$$

From the same proposition, one gets  $Inv_{\cal G} (\Lambda^2 y^\ast
\otimes {\cal G}) = 0$  and we shall see later (see the last part of the proof)
that $\partial_\ast$ is an isomorphism. So $Inv_{\cal G} (H^2 (\h)
\otimes {\cal G}) = 0$ and $E^{0, 2}_2 = 0$. The same argument shows that
$E^{0, 1}_2 = 0$, which ends the proof of the lemma.\\
 \eop

\vspace{0.5cm}
From the long exact sequence $(5.2)$ one has now: $H^\ast (\tsv, \tsv) =
H^\ast (\tsv, \h)$ for $\ast = 0, 1, 2$. We shall compute $H^\ast (\tsv,
\h)$ by  using the Hochschild-Serre spectral sequence once more; there are three
terms to compute. 

\medskip

1.\ \ \ First $E^{2, 0}_2 = H^2 ({\cal G}, H^0 (\h, \h))$, but
$H^0 (\h, \h) = Z (\h) = \n = {\cal F}_0$ as ${\cal G}$-module. So $E^{2,
0}_2 = H^2 ({\cal G}, {\cal F}_0)$ which is one-dimensional, given by
$c_3 ( f \partial, g \partial) = \left| \begin{array}{ccc} f & g \\ f'
& g'
\end{array} \right|$, or in terms of modes $c_3 (L_n, L_m) = (m-n) M_{n+m}$.
Hence  we have found one of the classes announced in the theorem.

\medskip

2.\ \ \  One must now
compute  $E^{1, 1}_2 = H^1 ( {\cal G}, H^1 (\h, \h))$. The following
lemma will be useful for this purpose, and also for the last part of the
proof.

\vspace{0.5cm}
\noindent{\bf Lemma 5.4 
(identification of $H^1 (\h, \h)$ as a ${\cal G}$-module).}\\
{\it The space $H^1 (\h, \h)$ splits into the direct sum of  two ${\cal
G}$-modules $H^1 (\h, \h) = {\cal H}_1 \oplus {\cal H}_2$ such that
\begin{enumerate}
\item
$Inv_{\cal G} {\cal H}_2 = 0$, $H^1 ({\cal G}, {\cal H}_2) = 0$;
\item

$Inv_{\cal G} {\cal H}_1$ is one-dimensional, generated by the
'constant multiplication' cocycle $l$ defined by
\BEQ l(Y_n)=Y_n, \quad l(M_n)=2M_n \EEQ
\item
 $H^1 ({\cal G}, {\cal H}_1)$ is two-dimensional, generated by two cocycles $c_1,c_2$ defined by 
$$c_1 (f \partial, g \partial^{1/2}) = f^{'} g \partial^{1/2},\quad 
 c_1 (f \partial, g) = 2 f' g$$ and $$c_2 (f \partial, g \partial^{1/2}) = f
g \partial^{1/2}, \quad c_2 (f \partial, g) = 2 fg.$$
\item
 $H^2 ({\cal G}, {\cal H}_1)$ is one-dimensional, generated by the cocycle
$c_{12}$ defined by 
 $$c_{12} (f \partial, g \partial, h \partial^{1/2}) = \left|
\begin{array}{ccc} f & g\\
 f' & g' \end{array} \right| h \partial^{1/2}, \quad    c_{12} (f
\partial, g \partial, h) = 2 \left| \begin{array}{ccc} f & g\\
 f' & g' \end{array} \right| h$$
\end{enumerate}
}

\vspace{0.5cm}
\noindent{\bf Proof of lemma 5.4:}\\
We shall split the  cochains according to the decomposition
$\h = y \oplus \n$. Set $C^1 (\h, \h) = C_1 \oplus C_2$, where:
$$ C_1 = (\n^\ast \otimes \n ) \oplus (y^\ast \otimes y) \ \ \ \ C_2 =
(\n^\ast \otimes y) \oplus (y^\ast \otimes \n).$$
So one readily obtains the splitting $H^1 (\h, \h)= {\cal H}_1 \oplus
{\cal H}_2$ where
$$0 \longrightarrow {\cal H}_1 \longrightarrow (\n^\ast \otimes \n) \oplus
(y^\ast \otimes y) \stackrel{\partial}{\longrightarrow} \Lambda^2 y^\ast
\otimes \n$$
$$ 0 \longrightarrow y \stackrel{\partial}{\longrightarrow} y^\ast \otimes
\n \longrightarrow {\cal H}_2 \longrightarrow 0$$
$\partial$ being the coboundary  on the space of cochains on $\h$ with
coefficients into itself. Its non vanishing pieces in degrees $0, 1$ and
$2$ are the following ones: $y \stackrel{\partial}{\longrightarrow} y^\ast
\otimes \n, \ \ \n^\ast \otimes \n \stackrel{\partial}{\longrightarrow}
\Lambda^2 y^\ast \otimes \n, \ \  y^\ast \otimes y
\stackrel{\partial}{\longrightarrow} \Lambda^2 y^\ast \otimes \n$.
 We can now describe the second exact sequence in terms of densities as
follows:
\BEQ
0 \longrightarrow {\cal F}_{1/2} \longrightarrow {\cal F}_{-3/2} \otimes
{\cal F}_0 \longrightarrow {\cal H}_2 \longrightarrow 0 \ \ \ \ \ \ \ \ \
\ \
\EEQ

From Proposition 5.2, one has $Inv_{\cal G} ({\cal F}_{-3/2}
\otimes {\cal F}_0) = 0$ as well as $H^i ({\cal G}, {\cal F}_{1/2}) = 0$,
for $i = 0, 1, 2$, and $H^1 ({\cal G}, {\cal F}_{-3/2} \otimes {\cal
F}_0) = 0$. So the long exact sequence in cohomology associated with
$(5.3)$ gives $Inv_{\cal G} ({\cal H}_2) = 0$ and $H^1 ({\cal G}, {\cal
H}_2) = 0$.

 For ${\cal H}_1$, one has to analyse the cocycles by direct computation.
So let $l \in C_1$ given by $l (Y_n) = a_n (k) Y_{n+k}, \ l (M_n) = b_n
(k) M_{n+k}$. The cocycle conditions are given by:
 $$\partial l (Y_n, Y_m) = l (( m-n) M_{n+m} ) - Y_n \ \cdot \ l (Y_m) +
Y_m \ \cdot \ l (Y_n) = 0$$
 for all $(n, m) \in \Z^2$.
 So identifying the term in $M_{n+m+k}$, one  obtains:
 $$(m-n) b_{n+m} (k) = (m-n +k) a_m (k) - (n - m +k) a_n (k)$$
 so $b_{n+m} (k) = a_m (k) + a_n (k) + \frac{k}{m-n} (a_m (k) - a_n (k))
= f (n, m, k)$. \\
 One can now determine the $a_n(k)$, remarking that the function $f (n, m,
k)$ depends only on $k$ and $(n +m)$. One then obtains that $a_n (k)$ must
be affine in $n$ so:
 $$a_n (k) = n \lambda (k) + \mu (k)$$
 $$b_n (k) = n \lambda (k) + k \lambda (k) + 2 \mu (k)$$
 So, as a vector space ${\cal H}_1$ is isomorphic to $\displaystyle
\bigoplus_k \C (\lambda (k)) \bigoplus_k \C (\mu (k))$, two copies of an
infinite direct sum of a numerable family of one-dimensional vector
spaces.

 Now we have to compute the action of ${\cal G}$ on ${\cal H}_1$;
let $L_p \in {\cal G}$, one has
 $$\begin{array}{lll} (L_p \cdot l) (Y_n) &=& ((n - \frac{p}{2}) a_{n+p}
(k) - (n +k - \frac{p}{2}) a_n (k)) Y_{n+p+k}\\
 &=&\left(  n (p-k) \lambda (k) - (\frac{p^{2}}{2} \lambda (k) + k \mu (k))
\right) Y_{n+p+k}
 \end{array}$$
 So if one sets $(L_p \cdot l) (Y_n) = (n (L_p \cdot \lambda)(k+p) + (L_p
\cdot \mu)(k+p)) Y_{n+p+k}$ \\
 one obtains:
 $$(L_p \cdot \lambda) (k+p) = (p - k) \lambda (k)$$
 $$(L_p \cdot \mu) (k + p) = - \frac{p^{2}}{2} \lambda (k) + k \mu (k)$$

 \noindent Finally, ${\cal H}_1$ appears as an extension of modules of
densities of the following type:\\
 $0 \longrightarrow {\cal F}_0 \longrightarrow {\cal H}_1 \longrightarrow
{\cal F}_1 \longrightarrow 0$, in which ${\cal F}_0$ corresponds to
$\displaystyle \bigoplus_k \C (\mu (k))$ and ${\cal F}_1$ to
$\displaystyle \bigoplus_k \C (\lambda (k))$.

 There is a non-trivial extension cocycle $\gamma$ in
 $Ext^1_{\cal G} ({\cal F}_1, {\cal F}_0) = H^1 ({\cal G}, Hom ({\cal
F}_1, {\cal F}_0))$, given by $\gamma (f \partial) (g \partial) = f^{''}
g$; this cocycle corresponds to the term in $p^2$ in the above formula. In
any case one has a long exact sequence in cohomology
 $$\cdots \longrightarrow H^i ({\cal G}, {\cal F}_0) \longrightarrow H^i
({\cal G}, {\cal H}_1) \longrightarrow H^i ({\cal G}, {\cal F}_1)
\longrightarrow H^{i+1} ({\cal G}, {\cal F}_0) \longrightarrow \cdots$$
 As well-known, $H^\ast ({\cal G}, {\cal F}_1) = H^\ast ({\cal G}, {\cal
G})$ is trivial, and finally $H^i ({\cal G}, {\cal F}_0)$ is isomorphic to
$H^i ({\cal G}, {\cal H}_1)$. In particular $H^0 ({\cal G}, {\cal H}_1) =
H^0 ({\cal G}, {\cal F}_0)$ is one-dimensional, given by the constants; a
scalar $\mu$ induces an invariant cocycle as $l (Y_n) = \mu Y_n, \ \
l(M_n) = 2 \mu M_n$.

Moreover, $H^1 ({\cal G}, {\cal F}_0)$ has  dimension $2$: it is  generated
by the cocycles $\overline{c}_1$ and $\overline{c}_2$, defined by
$\overline{c}_1 (f \partial)= f'$ and $\overline{c}_2 (f \partial) = f$
respectively.
So one obtains two generators of $H^1 ({\cal G}, {\cal H}_1)$ given by
$$c_1 (f \partial, g \partial^{1/2}) = f^{'} g \partial^{1/2},\quad
 c_1 (f \partial, g) = 2 f' g$$ and $$c_2 (f \partial, g \partial^{1/2}) = f
g \partial^{1/2}, \quad c_2 (f \partial, g) = 2 fg$$ respectively.

Finally $H^2 ({\cal G}, {\cal F}_0)$ is one-dimensional, with the
cup-product $\overline{c}_{12}$ of $\overline{c}_1$ and $\overline{c}_2$
as generator (see \cite{Fuk},  p. 177),
 so $\overline{c}_{12} (f \partial, g \partial) = \left|
\begin{array}{ccc} f & g\\
 f' & g' \end{array} \right|$, and  one deduces the
formula for the corresponding cocycle 
 $c_{12}$ in $H^2 ( {\cal G}, {\cal H}_1)$:
 $$c_{12} (f \partial, g \partial, h \partial^{1/2}) = \left|
\begin{array}{ccc} f & g\\
 f' & g' \end{array} \right| h \partial^{1/2}, \quad    c_{12} (f
\partial, g \partial, h) = 2 \left| \begin{array}{ccc} f & g\\
 f' & g' \end{array} \right| h$$
 This finishes the proof of Lemma 5.4. \hfill\eop

\vspace{0.1cm}
 So, from Lemma 5.4, we have computed $E^{1,1}_2 = H^1 ( {\cal G}, H^1
(\h, \h))$; it is two-dimensional, generated by $c_1$ and $c_2$, while
earlier we had $H^2 ({\cal G}, H^0 (\h, \h)) = E^{2,0}_2$, a one-dimensional
vector space generated by $c_3$. We have to check now that these cohomology classes
shall not disappear in the spectral sequence; the only potentially
 non-vanishing differentials are $E^{0,1}_2 \longrightarrow E^{2,0}_2$ and
$E^{1,1}_2 \longrightarrow E^{3,0}_2$.
One has $E^{3,0}_2 = H^3 ({\cal G}, \h)= H^3 ({\cal G}, \n) = H^3 ({\cal
G}, {\cal F}_0) = 0$ (see \cite{Fuk} p. 177); here we consider only local
cohomology), then $E^{0,1}_2$ is one-dimensional determined by the
constant multiplication (see above) and direct verification shows that
$E^{0,1}_2 \longrightarrow E^{2,0}_2$ vanishes. So we have just proved that
the cocycles $c_1$, $c_2$ and $c_3$ defined in the theorem represent
genuinely non-trivial cohomology classes in $H^2 (\tsv, \tsv)$.

\medskip

3. \ \ \ 
In order to finish the proof, we still have to prove that there does not
exist any other non-trivial class in the last piece of the
Hochschild-Serre spectral sequence. We shall thus prove that $E^{0,2}_2 =
Inv_{\cal G} H^2 (\h, \h)= 0$ As in the proofs of the previous lemmas, we
shall use decompositions of the cohomological complex of $\h$ with
coefficients into itself as sums of ${\cal G}$-modules.

The space of adjoint cochains $C^2 (\h, \h)$ will split into six subspaces
according to the vector space decomposition $\h = y \oplus \n$. So we can
as well split the cohomological complex
 $$C^1 (\h, \h) \stackrel{\partial}{\longrightarrow} C^2 (\h, \h)
\stackrel{\partial}{\longrightarrow} C^3 (\h, \h)$$
into its components, and the coboundary operators 
will as well split into different
components, as we already explained. So one obtains  the following
families of exact sequences of ${\cal G}$-modules:
\BEQ
(\n^\ast \otimes \n) \oplus (y^\ast \otimes y)
\stackrel{\partial}{\longrightarrow} \Lambda^2 y^\ast \otimes \n
\longrightarrow A_1 \longrightarrow 0
\EEQ
 $$0 \longrightarrow K \longrightarrow (\Lambda^2 y^\ast \otimes y) \oplus
(\n^\ast \wedge y^\ast \otimes y) \stackrel{\partial}{\longrightarrow}
\Lambda^3 y^\ast \otimes \n$$
\BEQ
0 \longrightarrow \n^\ast \otimes y \stackrel{\partial}{\longrightarrow} K
\longrightarrow A_2 \longrightarrow 0
\EEQ
 $$0 \longrightarrow A_3 \longrightarrow \n^\ast \wedge y^\ast \otimes y
\stackrel{\partial}{\longrightarrow} (\Lambda^3 y^\ast \otimes y) \oplus
(\n^\ast \wedge \Lambda^2 y^\ast) \otimes \n$$
 $$0 \longrightarrow A_4  \longrightarrow \Lambda^2 \n^\ast \otimes \n
\stackrel{\partial}{\longrightarrow} (\n^\ast \wedge \Lambda^2 y^\ast)
\otimes \n$$
 $$0 \longrightarrow A_5 \longrightarrow \Lambda^2 \n^\ast \otimes y
\stackrel{\partial}{\longrightarrow} (\n^\ast \wedge \Lambda^2 y^\ast)
\otimes y \oplus (\Lambda^2 \n^\ast \wedge y^\ast ) \otimes \n$$
 The restrictions of coboundary operators are still denoted by $\partial$, and the
other arrows are either inclusions of subspaces or projections onto
quotients. So one has $H^2 (\h, \h) = \displaystyle \bigoplus^5_{i=1}
A_i$, and our result will follow from $Inv_{\cal G} A_i = 0, \  i = 1, ...
5$. For the last three sequences, the result  follows immediately from
the cohomology long exact sequence by using
 $Inv_{\cal G} (\n^\ast \wedge
y^\ast \otimes y) = 0,$ \ $Inv_{\cal G} \Lambda^2 \n^\ast \otimes \n = 0$,
$Inv_{\cal G} \Lambda^2 \n^\ast \otimes y = 0$: there results are deduced
from those of Grozman, recalled in Proposition 5.2.
(Note that  the obviously ${\cal G}$-invariant maps $\n
\otimes \n \longrightarrow \n$ and $\n \otimes y \longrightarrow y$ are
not antisymmetric!) 
So one has $Inv_{\cal G} A_i = 0$ for $i = 3, 4, 5$.

An analogous argument will work for $K$, since $Inv_{\cal G} \  \Lambda^2
y^\ast \otimes y = 0$ and $Inv_{\cal G} (\n^\ast \wedge y^\ast) \otimes y
= 0$ from the same results. So the long exact sequence associated with the
short sequence $(5.5)$ above will give:
$$0 \longrightarrow Inv_{\cal G} (K) \longrightarrow Inv_{\cal G} (A_2)
\longrightarrow H^1 ({\cal G}, \n^\ast \otimes y)$$
One has $H^1 ({\cal G}, \n^\ast \otimes y) = H^1 ({\cal G}, {\cal F}_{-1}
\otimes {\cal F}_{1/2}) = 0$ (see Proposition 5.2). So $Inv_{\cal G} (A_2) =
0$.

For $A_1$, we  shall require  a much more subtle argument. First of all, the
sequence $(5.4)$ can be split into two short exact sequences:
$$0 \longrightarrow {\cal H}_1 \longrightarrow (\n^\ast \otimes \n) \oplus
(y^\ast \otimes y) \longrightarrow B \longrightarrow 0$$
$$0 \longrightarrow B \longrightarrow \Lambda^2 y^\ast \otimes \n
\longrightarrow A_1 \longrightarrow 0.$$

Let us consider the long exact sequence associated with the first one:
$$0 \longrightarrow Inv_{\cal G} {\cal H}_1 \longrightarrow Inv_{\cal G}
(\n^\ast \otimes \n) \oplus Inv_{\cal G} (y^\ast \otimes y)
\longrightarrow Inv_{\cal G} B \longrightarrow \cdots$$
$$\cdots \hookrightarrow H^1 ({\cal G}, {\cal H}_1) \longrightarrow H^1 ({\cal G},
\n^\ast \otimes \n) \oplus H^1 ({\cal G}, y^\ast \otimes y)
\longrightarrow H^1 ({\cal G}, B) \longrightarrow \cdots$$
$$\cdots\longrightarrow H^2 ({\cal G}, {\cal H}_1) \longrightarrow H^2 ({\cal G},
\n^\ast \otimes \n) \oplus H^2 ({\cal G}, y^\ast \otimes y)
\longrightarrow H^2 ({\cal G}, B) \longrightarrow \cdots$$

The case of $H^i ({\cal G}, {\cal H}_1)$, $i = 0, 1, 2$ has been treated in
Lemma 5.4, and analogous techniques can be used to study $H^i ({\cal G},
\n^\ast \otimes \n)$ and $H^i ({\cal G}, y^\ast \otimes y)$ for $i = 0, 1,
2$. The cohomology classes come from the inclusion ${\cal F}_0 \subset \n^\ast
\otimes \n$, $y^\ast \otimes y$ or ${\cal H}_1$, and from the well-known
computation of $H^\ast ({\cal G}, {\cal F}_0)$ (Remark:  using the results of
Fuks \cite{Fuk}, chap 2, one should keep in mind the fact that he computes
cohomologies for $W_1$, the formal part of ${\cal G} = \Vect (S^1)$. To get
the cohomologies for $\Vect (S^1)$ one has to add the classes of
differentiable order $0$ (or "topological" classes), this is the reason
for the occurrence  of $c_2$ in Lemma 5.4).

So $H^i ({\cal G}, {\cal H}_1) = H^i ({\cal G}, \n^\ast \otimes \n) = H^i
({\cal G}, y^\ast \otimes y), \ i = 0, 1, 2$, and  the maps on the modules
are naturally defined  through the injection ${\cal H}_1 \longrightarrow
(\n^\ast \otimes \n) \oplus (y^\ast \otimes y)$: each generator of $H^i
({\cal G}, {\cal H}_1), \ i = 0, 1, 2,$, say $e$, will give $(e, -e)$ in
the corresponding component of $H^i ({\cal G}, (\n^\ast \otimes \n) \oplus
(y^\ast \otimes y))$. So $Inv_{\cal G} B$ and $H^2 ({\cal G}, B)$ are one-dimensional
 and  $H^1 ({\cal G}, B)$ is two-dimensional.

Now we can examine the long exact sequence associated with:\\
$$0 \longrightarrow B \stackrel{\partial}{\longrightarrow} \Lambda^2 y^\ast
\otimes \n \longrightarrow A_1 \longrightarrow 0,$$
 which is:
$$0 \longrightarrow Inv_{\cal G} B
\stackrel{\partial^{\ast}}{\longrightarrow} Inv_{\cal G} \Lambda^2 y^\ast
\otimes \n \longrightarrow Inv_{\cal G} A_1 \longrightarrow H^1 ({\cal G},
B)
\longrightarrow H^1 ({\cal G}, \Lambda^2 y^\ast \otimes \n)
\longrightarrow \cdots$$
The generator of $Inv_{\cal G} B$ comes from the identity map $\n
\longrightarrow \n$, and $Inv_{\cal G}\ \Lambda^2 y^\ast \otimes \n$ is
generated by the bracket $y \wedge y \longrightarrow \n$, so
$\partial^\ast$ is an isomorphism in this case. So one has
$$0 \longrightarrow Inv_{\cal G} A_1 \longrightarrow H^1 ({\cal G}, B)
\stackrel{\partial^{\ast}}{\longrightarrow} H^1 ({\cal G}, \Lambda^2
y^\ast \otimes \n)$$
The result will follow from the fact that this $\partial^*$ is also an
isomorphism. The two generators in $H^1 ({\cal G}, B)$ come from the
corresponding ones in $H^1 ({\cal G}, \n^\ast \otimes \n) \oplus H^1
({\cal G}, y^\ast \otimes y)$, modulo the classes coming from $H^1 ({\cal
G}, {\cal H}_1)$; so these generators can be described in terms of Yoneda
extensions, since $H^1 ({\cal G}, {\cal F}^\ast_0 \otimes {\cal F}_0) =
Ext^1_{\cal G} ({\cal F}_0, {\cal F}_0)$, as well as $H^1 ({\cal G}, {\cal
F}^\ast_{1/2} \otimes {\cal F}_{1/2}) = Ext^1_{\cal G} ( {\cal F}_{1/2},
{\cal F}_{1/2})$.

Let us write this extension as $0 \longrightarrow {\cal F}_0 \longrightarrow E
\longrightarrow {\cal F}_0 \longrightarrow 0;$  the action on $E$ can
be given in terms of modes as follows:
$$e^1_n (f_a, g_b) = (a f_{n+a} + ng_{b+n}, bg_{b+n})$$
or
$$e^2_n (f_a, g_b) = (af_{n+a} + g_{b+n}, bg_{b+n}).$$
The images of these classes in $H^1 ({\cal G}, \Lambda^2 y^\ast \otimes
\n)$ are represented by the extensions obtained through a pull-back
$$\begin{array}{ccccccccccc}
0 &\longrightarrow & {\cal F}_0 & \longrightarrow & E & \longrightarrow &
{\cal F}_0 & \longrightarrow  & 0\\
 &  & \| & & \uparrow & & \ \ \ \uparrow [\ ,\ ] & & \\
 0 &\longrightarrow & {\cal F}_0 & --->   & E' & --->
  & \Lambda^2{\cal F}_{1/2} & \longrightarrow  & 0
\end{array}$$
where $[\ ,\ ]$ denotes the mapping given by the Lie bracket $\Lambda^2 {\cal
F}_{1/2} \longrightarrow {\cal F}_0$. One can easily check that these
extensions are non-trivial, so finally $\partial^\ast$ is injective and
$Inv_{\cal G} (A_1) = 0$, which finishes the proof of $E^{0,2}_2 =
Inv_{\cal G} H^2 (\h, \h) = 0$ and the proof of Theorem 5.1. \\
\eop

\vspace{2cm}

 Theorem 5.1 implies that we have three independent
infinitesimal deformations of $\tsv$, defined by the cocycles $c_1, c_2$
and $c_3$, so the most general infinitesimal deformation of $\tsv$ is of
the following form:
$$[ \ \ , \ \ ]_{\lambda, \mu, \nu} = [ \ \ , \ \ ] + \lambda c_1 + \mu
c_2 + \nu c_3.$$
In order to study further deformations of this bracket, one has to compute
the Richardson-Nijenhuis brackets of $c_1$, $c_2$ and $c_3$ in $H^3 (\tsv,
\tsv)$. One can compute directly using our explicit formulas and  finds
$[c_i, c_j] = 0$ in
$H^3(\tsv, \tsv)$ for $i, j = 1, 2, 3$; and even better, the bracket of the
cocycles themselves  vanish,  not only their cohomology classes.
So one has the

\vspace{0.5cm}
\noindent {\bf Theorem 5.5.}\\
{\it The bracket $[\ ,\ ]_{\lambda, \mu, \nu} = [\ ,\ ] + \lambda c_1 + \mu c_2 +
\nu c_3$ where $[\ ,\ ]$ is the Lie bracket on $\tsv$ and $c_i$, $i = 1, 2, 3$
the cocycles given in Theorem 5.1, defines a three-parameter
family of Lie algebra brackets on $\tsv$.\\
For the sake of completeness, we  give below
the full formulas in terms of modes:

$$[L_n, L_m]_{\lambda, \mu, \nu} = (m-n) L_{n+m} + \nu (m-n) M_{n+m}$$

$$[L_n, Y_m]_{\lambda, \mu, \nu} = (m - \frac{n}{2} - \frac{\lambda n}{2}
+ \mu) Y_{n+m}$$

$$[L_n, M_m]_{\lambda, \mu, \nu} = (m - \lambda n + 2\mu) M_{n+m}$$

$$[Y_n, Y_m] = (n-m) M_{n+m}$$
All other terms are vanishing.}

\vspace{0.1cm}
\noindent
The term with cocycle $c_3$ has already been considered in a slightly
different context in \cite{GuiRog05}. The term with $c_2$ induces only a small
change in the action on $\h$: the modules ${\cal F}_{1/2}$ and ${\cal
F}_0$ are changed into ${\cal F}_{1/2, \mu}$ and ${\cal F}_{0, \mu}$ (see
\cite{Fuk}, p.127), the bracket on $\h$ being fixed. This is nothing but
a reparametrization of the generators in the module,and for integer values
of $\mu$, the Lie algebra given by $[\ ,  \ ]_{0, \mu, 0}$ is isomorphic to
the original one. 

We shall focus in the sequel on the term proportional to  $c_1$,
and denote by $\tsv_{\lambda}$ the one-parameter family of Lie algebra
structures on $\tsv$ given by $[\ ,\ ]_\lambda = [\ , \ ]_{\lambda, 0, 0}$,
in coherence with Definition 1.7.
Inspection of the above formulas shows that $\tsv_\lambda$ is a semi-direct
product $\Vect (S^1) \ltimes \h_\lambda$ where $\h_\lambda$ is a
deformation of $\h$ as a $\Vect(S^1)$-module; one has $\h_\lambda = {\cal
F}_{\frac{1 + \lambda}{2}} \oplus {\cal F}_{ \lambda}$, and the bracket
${\cal F}_{\frac{1+\lambda}{2}} \times {\cal F}_{\frac{1+\lambda}{2}}
\longrightarrow {\cal F}_{\lambda}$ is the usual one, induced by the
Poisson bracket on the torus.

Now, as a by-product of the above computations, we shall determine
explicitly $H^1 (\tsv, \tsv)$.

\vspace{0.1cm}
\noindent
{\bf Theorem 5.6.}\\
{\it $H^1 (\tsv, \tsv)$ is three-dimensional, generated by the following
cocycles, given in terms of modes by:
$$c_1(L_n) = M_n \ \ \ \ \ c_2 (L_n) = n M_n$$
$$l (Y_n) = Y_n \ \ \ \ \ \ \ l(M_n) = 2 M_n.$$
The cocycle $l$ already appeared in Chapter one, when we discussed the
derivations of $\tsv$; with the notations of Definition 1.6 one has $l = 2
(\delta_2 - \delta_1)$.}

\vspace{0.5cm}
\noindent
{\bf Proof:}

 From Lemma 5.3 above, one has $H^1 (\tsv, {\cal G}) = 0$, and so
$H^1 (\tsv, \tsv) = H^1 (\tsv, \h)$. One is led to compute the $H^1$  of a
semi-direct product, as already done in paragraph 3.3. The space
$H^1 (\tsv, \h)$ is  made from two parts $H^1 ({\cal G}, \h)$ and $H^1
(\h, \h)$ satisfying the compatibility condition as in Theorem 3.8:
$$c ([ X, \alpha ]) - [X, c (\alpha)] = - [ \alpha, c(X)]$$
for $X \in {\cal G}$ and $\alpha \in \h$.

The result is then easily deduced from the previous computations :  $H^1
({\cal G}, \h) = H^1 ({\cal G}, \n)$ is generated by $f \partial
\longrightarrow f$ and $f \partial \longrightarrow f'$, which correspond in the
mode decomposition to the cocycles $c_1$ and $c_2$. As a corollary, one has $[\alpha,
c(X)] = 0$ for $X \in {\cal G}$ and $\alpha \in \h$. Hence  the
compatibility condition reduces to $c([ X, \alpha]) = [X, c(\alpha)]$ and
thus  $c \in Inv_{\cal G} H^1 (\h, \h)$. It can now be deduced from Lemma 5.4
above, that the latter space is one-dimensional, generated by $l$.

We shall now determine the central charges of $\tsv_\lambda$; the
computation will shed light on some exceptional values of $\lambda$,
corresponding to interesting particular cases.

\noindent
\subsection{Computation of $H^2 (\tsv_\lambda, \R)$}

We shall again make use of the exact sequence decomposition $0
\longrightarrow \h_\lambda \longrightarrow \tsv_\lambda
\stackrel{\pi}{\longrightarrow}  {\cal G} \longrightarrow 0$, and classify
the cocycles with respect to their "type" along this decomposition;
trivial coefficients will make computations much easier than in the above
case. First of all,  $0 \longrightarrow H^2 ({\cal G}, \R)
\stackrel{\pi^{\ast}}{\longrightarrow} H^2 (\tsv_\lambda, \R)$ is an
injection.
So the Virasoro class $c \in H^2 (\Vect (S^1), \R)$ always survives in $H^2
(\tsv_\lambda, \R)$.

 For $\h_\lambda$, let us  use once again  the decomposition
$0\longrightarrow \n_\lambda \longrightarrow \h_\lambda \longrightarrow
y_\lambda \longrightarrow 0$ where $\n_\lambda = [\h_\lambda,
\h_\lambda]$. One has:  $H^1 ({\cal G}, H^1 (\h_\lambda)) = H^1 ({\cal G},
y^\ast_\lambda) = H^1 ({\cal G}, {\cal F}^*_{\frac{1+\lambda}{2}}) = H^1
({\cal G}, {\cal F}_{-(\frac{3+\lambda}{2})})$. \\
The cohomologies of degree one of $\Vect (S^1)$ with coefficients in
densities are known (see \cite{Fuk},Theorem 2.4.12):
 the space $H^1 ({\cal G}, {\cal
F}_{-(\frac{3+\lambda}{2})})$ is trivial, except for the three exceptional
cases $\lambda = - 3, -1, 1$:

$H^1 ({\cal G}, {\cal F}_0)$ is generated by the cocycles $f \partial
\longrightarrow f$ and $f \partial \longrightarrow f'$;

$H^1 ({\cal G}, {\cal F}_{-1})$ is generated by the cocycle $f \partial
\longrightarrow f^{''} dx$;

$H^1 ({\cal G}, {\cal F}_{-2})$ is generated by the cocycle $f \partial
\longrightarrow f^{'''}(dx)^2$,
corresponding to  the "Souriau cocycle" associated to the central charge of
the Virasoro algebra (see \cite{GuiRog05}, chapter  IV).

In terms of modes, the corresponding cocycles are given by:

$ \hspace{1cm} c_1 (L_n, Y_m) = \delta^0_{n+m}, \ \ \ \ c_2(L_n,
Y_m) = n \delta^0_{n+m}$ for $\lambda = -3;$

 $\hspace{1cm} c (L_n, Y_m) = n^2 \delta^0_{n+m}$ for $\lambda =-1;$

 $\hspace{1.3cm} c (L_n, Y_m) = n^3 \delta^0_{n+m}$ for $\lambda = 1.$

The most delicate part is the investigation of the term $E^{0,2}_2 =
Inv_{\cal G} H^2 (\h_\lambda)$ (this refers of course
to the Hochschild-Serre spectral sequence associated to the above
decomposition). For $H^2 (\h_\lambda)$ we shall use the same short exact
sequences as for $\h$ in the proof of  Lemma 5.3:

$$0 \longrightarrow Ker \partial \longrightarrow \Lambda^2 \n^\ast_\lambda
\stackrel{\partial}{\longrightarrow} \Lambda^2 y^\ast_\lambda \wedge
\n^\ast_\lambda$$
$$0 \longrightarrow Ker \partial \longrightarrow y^\ast_\lambda \wedge
\n^\ast_\lambda  \stackrel{\partial}{\longrightarrow} \Lambda^3
y^\ast_\lambda $$
$$0\longrightarrow \underline{\n}^\ast_\lambda \longrightarrow \Lambda^2
y^\ast_\lambda \longrightarrow Coker \partial \longrightarrow 0$$
(where $\underline{\n}_\lambda$ stands for $\n_\lambda$ divided out by the space
of constant functions).
One readily shows that for the first two sequences one has $Inv_{\cal G}
Ker \partial = 0$. The third one is more complicated; the cohomology exact
sequence yields:

$$0 \longrightarrow Inv_{\cal G} Coker \partial \longrightarrow H^1 ({\cal
G}, \underline{\n}^\ast_\lambda) \longrightarrow H^1 ({\cal G}, \lambda^2
y^\ast_\lambda) \longrightarrow \cdot \cdot \cdot$$

The same result as above (see \cite{Fuk}, Theorem 2.4.12) shows that:
$$H^1 ({\cal G}, \n^\ast_\lambda) = H^1 ({\cal G}, {\cal
F}_{(-1-\lambda)}) = 0 \ {\rm unless} \ \lambda = 1, -1, 0,$$
and one then has to investigate case by case; set $c (Y_p, Y_q) = a_p
\delta^0_{p+q}$ for the potential cochains on $y_\lambda$.
For each $n$ one has the relation:

$$(ad_{L_{n}} c) (Y_p, Y_q) - (q-p) \gamma (L_n) (M_{p+q}) = 0 \quad (5.6)$$
 for
some $1$-cocycle $\gamma: \ {\cal G} \longrightarrow \n^\ast_\lambda$, and
for all $(p, q)$ such that $n+ p + q = 0$; if $\gamma (L_n) (M_k) =
b_n \delta^0_{n+m+k}$, one obtains in terms of modes, using $a_p =
-a_{-p}$:
$$(p+q) \left( \frac{1+ \lambda}{2}\right) (a_p - a_q) + q a_q - p a_q -
(q - p) b_{-(p+q)} = 0$$

Let us now check  the different cases of non-vanishing terms in $H^1 ({\cal
G}, \n^\ast_\lambda)$.

For $\lambda = 1$ one has $b_n = n^3$, and one deduces $a_p = p^3$.

For $\lambda = -1$ there are two possible cases $b_n = n$ or $b_n = 1$,
the above equation gives
$$q a_p - p a_q = (q - p) ( \alpha (p+q) + \beta);$$
the only possible solution would be to set $a_p$ constant, but this is not
consistent with $a_p = - a_{-p}$. For $\lambda = 0$, one gets $b_n = n^2$
and the equation gives
$$\left(\frac{p+q}{2} \right) (a_p + a_q) + q a_p - p a_q - (q - p) (p +
q)^2 = 0$$
One easily checks that there are no solutions.

 Finally, one gets a new cocycle generating an independent class in $H^2
(\tsv_1, \R)$, given by the formulas:
$$ \begin{array}{lll}
c (Y_n, Y_m) = n^3 \delta_{n+m},\\
c (L_n, M_m) = n^3 \delta_{n+m}
\end{array} 
$$
Let us  summarize our results:

\vspace{0.5cm}
\noindent
{\bf Theorem 5.7.}\\
{\it For $\lambda \neq -3, -1, 1,$  $H^2 (\tsv_\lambda, \R) \simeq \R$
is generated by the Virasoro cocycle.\\
For $\lambda = -3, -1,$  $H^2 (\tsv_\lambda, \R) \simeq \R^2$ is generated by
the Virasoro cocycle and an independent cocycle of the form $c (L_n, Y_m) =
\delta^0_{n+m}$ for $\lambda = -3$ or $c (L_n, Y_m) = n^2 \delta^0_{n+m}$
for $\lambda = -1$.

\noindent For $\lambda = 1,$ $H^2 (\tsv_1, \R) \simeq \R^3$ is generated by the Virasoro
cocycle  and the two independent cocycles $c_1$ and $c_2$ defined by (all
other components vanishing) 
$$c_1 (L_n, Y_m) = n^3 \delta^0_{n+m};$$
$$c_2 (L_n, M_m) = c_2 (Y_n, Y_m) = n^3 \delta^0_{n+m}$$}

{\bf Remark :} The isomorphism $H^2(\sv_0,\R)\simeq \R$ has already been
proved in \cite{Henk94}. As we shall see in paragraph 5.4, generally
speaking, local cocycles
may be carried over from $\tsv$ to $\sv$ or from $\sv$ to $\tsv$
 without any difficulty.

Let us look more carefully at the $\lambda = 1$ case. One has that $\h_1 =
{\cal F}_1 \oplus {\cal F}_1$ with the obvious bracket ${\cal F}_1 \times
{\cal F}_1 \longrightarrow {\cal F}_1$; so, algebraically, $\h_1 = \Vect
(S^1) \otimes \varepsilon  \R [\varepsilon] / (\varepsilon^3=0)$. One
deduces immediately that $\tsv_1 = \Vect (S^1) \otimes \R
[\varepsilon]/(\varepsilon^3=0)$; so the cohomological result for $\tsv_1$
can be easily reinterpreted. Let $f_{\varepsilon} \partial$ and
$g_{\varepsilon} \partial$ be two elements in $\Vect (S^1) \otimes \R
[\varepsilon]/(\varepsilon^3=0)$, and compute the Virasoro cocycle $c
(f_\varepsilon \partial, g_{\varepsilon} \partial) = \int_{S^{1}}
f^{'''}_{\varepsilon} g_{\varepsilon} dt$ as a truncated polynominal in
$\varepsilon$; one has $f_{\varepsilon} = f_0 + \varepsilon f_1 +
\varepsilon^2 f_2$ and $g_{\varepsilon} = g_0 + \varepsilon g_1 +
\varepsilon^2 g_2$ so finally:
$$ c ( f_{\varepsilon} \partial, g_{\varepsilon} \partial) = \int_{S^{1}}
f^{'''}_0 g_0 dt + \varepsilon \int_{S^{1}} (f^{'''}_0 g_1 + f^{'''}_1
g_0) dt + \varepsilon^2 \int_{S^{1}} (f^{'''}_0 g_2 + f^{'''}_1 g_1 +
f^{'''}_2 g_0) dt.$$
In other terms: $c (f_{\varepsilon} \partial, g_{\varepsilon} \partial) = c_0
(f_{\varepsilon} \partial, g_{\varepsilon} \partial) + \varepsilon c_1
(f_{\varepsilon} \partial, g_{\varepsilon} \partial) + \varepsilon^2 c_2
(f_{\varepsilon} \partial, g_{\varepsilon} \partial)$. One can easily identify
the $c_i, \ i = 0, 1, 2$ with the cocycles defined in the above theorem,
using a decomposition into modes. This situation can be described by a
universal central extension
$$ 0 \longrightarrow \R^3 \longrightarrow \widehat{\tsv}_1 \longrightarrow
\Vect (S^1) \otimes \R [\varepsilon] / (\varepsilon^3 = 0) \longrightarrow
0$$ and the formulas of the cocycles show that  $\widehat{\tsv}_1$ is
isomorphic to Vir $\otimes \R [\varepsilon]/(\varepsilon^3 = 0)$

\vspace{0.2cm}
\noindent
{\bf Remarks:}
\begin{enumerate}
\item Cohomologies of Lie algebra of type $\Vect (S^1) \displaystyle
\bigotimes_{\R} A$, where $A$ is an associative and commutative algebra
(the Lie bracket being as usual given by [$f \partial \otimes a, g
\partial \otimes b] = (fg^{'} - g f^{'}) \partial \otimes ab)$, have been
studied by C. Sah and collaborators (see \cite{Ram}). Their result is:
 $H^2
\left(\Vect (S^1) \displaystyle \bigotimes_{\R} A \right) = A'$ where
$A' = Hom_\R (A, \R)$; all cocycles are given by the Virasoro cocycle
composed with the linear form on $A$. The isomorphism $H^2(\tsv_1,\R)\simeq
\R^3$ (see Theorem 5.7)  could have
been deduced from this general theorem.

\item One can obtain various generalisations of our algebra $\h$ as
nilpotent Lie  algebras with $\Vect (S^1)$-like brackets, such as 
\BEQ
[Y_n, Y_m]
=(m-n) M_{n+m} \label{nilpo}
\EEQ
by  using the same scheme. Let $A$ be an artinian ring
quotient of some polynomial ring $\R [t_1,\ldots,t_n]$ and $A_0 \subset A$
its maximal ideal; then $\Vect (S^1) \displaystyle  \bigotimes_\R A_0$ is
a nilpotent Lie algebra whose successive brackets are of the
same type (\ref{nilpo}). One
could speak of a "virasorization" of nilpotent Lie algebras. Explicit
examples are provided in the subsection 3.5  about
multi-diagonal operators of the present article.

\item It is interesting in itself  to look at  how  the dimension of $H^2
(\tsv_\lambda, \R)$ varies under deformations. For generic values of
$\lambda$, this dimension is equal to one, and it increases for some
exceptional values of $\lambda$; one can consider this as an example of so
called "Fuks  principle" in infinite dimension: deformations can decrease
the rank of cohomologies but never increase it.

\item Analogous Lie algebra structures, of the  "Virasoro-tensorized" kind,
have been considered in a quite different context in algebraic topology by
Tamanoi, see \cite{Tam}.
\end{enumerate}

\subsection{ About deformations of $\tsv_1$}

We must consider the local cochains $C^\ast_{loc} (\tsv_1, \tsv_1)$. The Lie
algebra $\tsv_1$ admits a graduation mod 3 by the degree of polynomial in
$\varepsilon$, the Lie bracket obviously respects this graduation; this
graduation induces on the space of local cochains a graduation by weight, and
$C^\ast_{loc} (\tsv_1, \tsv_1)$ splits into direct sum of subcomplexes of
homogeneous weight denoted by $C^\ast_{loc} (\tsv_1, \tsv_1)_{(p)}$.
Moreover, as classical in computations for Virasoro algebra, one can use
the adjoint action of the  zero mode $L_0$ (corresponding geometrically to the
Euler field $z \frac{\partial}{\partial z}$) to reduce cohomology
computations to the subcomplexes $C^\ast_{loc} (\tsv_1, \tsv_1)_{(p)(0)}$ of
cochains which are  homogeneous of weight 0 with respect to ad $L_0$ (see
e.g. \cite{GuiRog05}, chapter IV).

We can use the graduation in $\varepsilon$ and consider homogeneous
cochains with respect to that graduation.
 Here is what one gets, according to the weight: 
\begin{itemize}
\item weight 1: one has cocycles of the form
$$c(L_n, Y_m)= (m-n) M_{n+m}, \ c(L_n, L_m) = (m-n) Y_{n+m}$$
but if $b (L_n) = Y_n$, then $c = \partial b$.
\item weight 0: all cocycles are coboundaries, using the well-known result
$H^\ast (\Vect (S^1), \Vect (S^1))= 0.$
\item weight -1: one has to consider cochains of the following from \\
$$c(Y_n, M_m) = a (m-n) M_{n+m}$$
$$c (Y_n, Y_m) = b (m - n) Y_{n+m}$$
$$c(L_n, M_m) = e (m - n) Y_{n+m}$$
$$c (L_n, Y_m) = d (m - n) L_{n+m}$$
and check that $\partial c = 0$. It readily gives $e = d = 0$.\\
If one sets $\widetilde{c} (Y_n) = \alpha L_n$ and $\widetilde{c}( M_n)
=\beta Y_n$, then
$$\partial \widetilde{c} (Y_n, M_m) = (\alpha + \beta) (m-n) M_{n+m}$$
$$\partial \widetilde{c} (Y_n, M_m)= (2 \alpha - \beta) (m-n) Y_{n+m}$$
So all these cocycles are cohomologically trivial,
\item weight -2: set \\
$$c (Y_n, M_m) = \alpha (m-n) Y_{n+m}$$
$$c (M_n, M_m) = \beta (m - n) M_{n+m}$$
$$c (L_n, M_m) = \gamma (m-n) L_{n+m}$$
Coboundary conditions  give $\gamma = \alpha$ and $\beta = \gamma +
\alpha$, but if $\overline{c} (M_n) = L_n$, then \\
$$\partial \overline{c} (M_n, M_m) = (m-n) Y_{n+m}$$
$$\partial \overline{c} (Y_n, M_m) = 2 (m - n) M_{n+m}$$
$$\partial \overline{c} (L_n, M_m) = (m-n) L_{n+m}$$
\item weight -3: we  find for this case the only surviving cocycle.\\
One readily checks that $C\in C^*_{loc}(\tsv_1,\tsv_1)_{(-3)(0)}$ defined by
$$C (Y_n, M_m) = (m-n) L_{n+m}$$
$$C (M_n, M_m) = (m-n) Y_{n+m}$$
is a cocycle and cannot be a coboundary.
\end{itemize}
We can describe the cocycle $C$ above more pleasantly
by a global formula: let
$ \ f_{\varepsilon} = f_0 + \varepsilon f_1 + \varepsilon^2 f_2 $ and $
 g_{\varepsilon} = g_0 + \varepsilon g_1 + \varepsilon^2 g_2$,
with $f_i, g_i$ elements of $\Vect (S^1)$.
The bracket $[\ ,\  ]$ of $\tsv_1$ is then the following:
$$[f_{\varepsilon}, g_{\varepsilon} ] = \displaystyle \sum^2_{k=0} \sum_{i+j =
k} [ f_i, g_i] \ \ {\rm or} \ \ (f_{\varepsilon} g^{'}_{\varepsilon} -
g_{\varepsilon} f^{'}_{\varepsilon})|_{ \varepsilon^3 = 0},$$
\noindent and the deformed bracket $[\ ,\ ] + \mu C$ will be $[ f_{\varepsilon},
g_{\varepsilon}]_\mu = (f_{\varepsilon} g'_{\varepsilon} - g_{\varepsilon}
f'_{\varepsilon})|_{ \varepsilon^3 = \mu}$.
So we have found that $\dim \ H^2 (\tsv_1, \tsv_1) = 1$.

In order to construct deformations, we still have to check for the 
Nijenhuis-Richardson bracket $[C, C]$ in $C^3 (\tsv_1, \tsv_1)$. The only
possibly non-vanishing term is:
$$
\begin{array}{llll} [C, C] (M_n, M_n, M_p) = \displaystyle \sum_{(cycl)} C
(C(M_n, M_m,), M_p) &=& \displaystyle \sum_{(cycl)} (m-n) C (Y_{n+m}, M_p
)\\
&= & \displaystyle \sum_{(cycl)} (pm - pn + n^2-m^2) L_{n+m+p} = 0
\end{array}$$
So there does not exist any obstruction and we have obtained a genuine
deformation. We  summarize all these results in the following

\vspace{0.5cm}
\noindent
{\bf Proposition 5.8.} \\{\it There exists a one-parameter deformation of
the Lie algebra $\tsv_1$, as $\tsv_{1, \mu} = \Vect (S^1) \otimes \R
[\varepsilon] /_{(\varepsilon^3 = \mu)}$. This deformation in the only one
possible, up to isomorphism.}

If one is interested in central charges, the above-mentioned theorem of
C. Sah and al., see \cite{Ram}, shows that $\dim H^2 (\tsv_{1, \mu}, \R) = \R^3$ and the
universal central extension $\widehat{\tsv}_{1, \mu}$ is isomorphic to Vir
$\otimes \R [\varepsilon] /_{(\varepsilon^3 = \mu)}$. We did not do the
computations, but we conjecture that $\tsv_{1, \mu}$ is rigid, the ring $\R
[\varepsilon]/_{(\varepsilon^3 = \mu)}$ being more generic than $\R
[\varepsilon] /_{(\varepsilon^3 = 0)}$. More generally, it could be
interesting to study systematically Lie algebras of type $\Vect (S^1)
\otimes A$ where $A$ is a commutative ring, their geometric interpretation
being "Virasoro current algebras".

\subsection{Coming back to the original Schr\"odinger-Virasoro algebra}

The previous results concern the twisted Schr\"odinger-Virasoro algebra
generated by the modes $(L_n, Y_m, M_p)$ for $(n, m, p) \in \Z^3$, which make
computations easier and allows direct application of Fuks' techniques. The
"actual" Schr\"odinger-Virasoro algebra is generated by the modes $(L_n, Y_m,
M_p)$ for $(n, p) \in \Z^2$ but $m \in  \Z+\half$. Yet Theorem 5.1
and Theorem 5.5 on deformations of $\tsv$ are also valid for $\sv$: one
has dim $H^2(\sv,\sv)=3$ with the same cocycles $c_1,c_2,c_3$, since
these do not allow 'parity-changing' terms such as $L\times Y\to M$ or
$Y\times Y\to Y$ for instance (the $(L,M)$-generators being considered as
'even' and the $Y$-generators as 'odd'). 

But  the computation of $H^2(\sv_{\lambda},\R)$ will
yield very different results compared to Theorem 5.7, since 'parity'
is not conserved for all the cocycles we found, so we shall start all over
again.  Let us use
 the adjoint action of $L_0$ to simplify computations : all cohomologies are generated by cocycles $c$ such
that $ad L_0  \ .\  c = 0$, i.e.  such that $c (A_k, B_l) = 0$ for $ k
+ l \neq 0$, $A$ and $B$ being $L, Y$ or $M$. So, for non-trivial cocycles,
one must have $c (Y_n, L_p) = 0, c (Y_n, M_p) = 0$ for all $Y_n, L_p, M_p$
; in $H^2 (\sv_{\lambda}, \R)$, terms of the  type $H^1 ({\cal G}, H^1
(\h_\lambda))$ will automatically vanish. The Virasoro class in
$H^2 ({\cal G}, \R)$ will always survive, and one has to check what
happens with the terms of type $Inv_{\cal G} H^2 (\h_\lambda)$. As in the
proof of Lemma 5.3,
the only possibilities come from the short exact sequence:
$$0 \longrightarrow \underline{\n}^\ast_\lambda \longrightarrow \Lambda^2
y^\ast_\lambda \longrightarrow Coker \partial \longrightarrow 0$$
which induces: $0 \longrightarrow Inv_{\cal G} Coker \partial
\longrightarrow H^1 ({\cal G}, \underline{\n}^\ast_\lambda )
\longrightarrow H^1 ({\cal G}, \Lambda^2 y^\ast_\lambda )$ and one obtains
the same equation (5.6) as above:
$$(ad_{L_{n}} c) (Y_p, Y_q) + (q-p) \gamma (L_n) (M_{p+q}) = 0$$
If $c (Y_p, Y_q) = a_p \delta^0_{p+q}$, the equation gives:
$$-a_p (p + \frac{\lambda + 3}{2} n) + a_{p+n} (p- \frac{\lambda + 1}{2}
n) - ( p - q) \gamma (L_n) (M_{p+q}) = 0$$
One finds two exceptional cases with non-trivial solutions:
\begin{itemize}
\item for $\lambda = 1, \ \ a_p = p^3$ and $c (L_n, M_m) = n^3
\delta^0_{n+m}$ gives a two-cocycle, very much analogous to the $\Vect
(S^1) \otimes \R [\varepsilon] /_{ (\varepsilon^3=0)}$ case, except that one
has no term in $c (L_n, Y_p)$.
\item for $\lambda = - 3$, if $\gamma \equiv 0$, the above equation gives $p a_p
= (p+n) a_{p+n}$ for every $p$ and $n$.
So $a_p = \frac{1}{p}$ is a solution, and one sees why this solution was
not available in the twisted case. 

Let us summarize:
\end{itemize}

\vspace{0.5cm}
\noindent
{\bf Proposition 5.9.} {\em The space $H^2 (\sv_\lambda, \R)$ is one-dimensional, 
generated by the Virasoro cocycle, save for two exceptional
values of $\lambda$, for which one has one more independent cocycle,
denoted by $c_1$, with the following non-vanishing components:
\begin{itemize}
\item for $\lambda = 1$: \ \ $c_1 (Y_p, Y_q) = p^3 \delta^0_{p+q}$ \ and  \
 $c_1
(L_p, M_q) = p^3 \delta^0_{p+q}$,
\item for $\lambda = - 3$ :  \ \ $ c_1 (Y_p, Y_q) = \frac{\delta^{0}_{p+q}}{p}$.
\end{itemize} }

\vspace{0.5cm}
\noindent
{\bf Remark:} The latter case is the most surprising one, since it
contradicts the well-established dogma asserting that only  local classes
are interesting. This principle of locality has its roots in quantum field
theory (see e.g \cite{Kak}  for basic principles of axiomatic field theory); its
mathematical status has its foundations in  the famous theorem of J. Peetre,
asserting that local mappings are given by differential operators, so -- in
terms of modes -- the coefficients are polynomial in $n$. Moreover, there
is a general theorem in the
 theory  of cohomology of Lie algebras of vector fields (see \cite{Fuk})
which states that  continous cohomology is in general multiplicatively generated
by local cochains, called diagonal in \cite{Fuk}. Here our cocycle contains an
anti-derivative, so there could be applications in integrable systems,
considered as Hamiltonian systems, the symplectic manifold given by the
dual of (usually centrally extended) infinite dimensional Lie algebras (see
for example \cite{GuiRog05}, Chapters VI and X).

\section{Verma modules of $\sv$ and Kac determinants}

\subsection{Introduction}

There are a priori infinitely many   ways to define Verma modules on $\sv$,
corresponding
to the two natural graduations: one of them (called {\it degree} and denote
by $deg$ in the following) corresponds to the adjoint action of $L_0$, so
that $\deg(X_n)=-n$ for $X=L,Y$ or $M$; it is given by $-\del_1$ in the
notation of Definition 1.6. The other one, corresponding to the
graduation of the Cartan prolongation (see Section 4.1), is given by the outer
derivation $\del_2$. The action of both graduations is diagonal on the
generators
$(X_n)$; the subalgebra of weight $0$ is two-dimensional abelian, generated
by $L_0$ and $M_0$, in the former case, and three-dimensional solvable,
generated by $L_0,Y_{\half},M_1$ in the latter case.

Since Verma modules are usually defined by inducing a character of an
abelian subalgebra to the whole Lie algebra (although this is by no means
necessary), we shall forget altogether the graduation given by $\del_2$ in this section and consider
representations of $\sv$ that are induced from $\langle L_0,M_0\rangle$.

Let $\sv_{(n)}=\{Z\in\sv\ |\ ad L_0.Z=nZ\}$ ($n\in\half\Z$),
$\sv_{>0}=\oplus_{n>0}
\sv_{(n)}$, $\sv_{<0}=\oplus_{n<0} \sv_{(n)}$, $\sv_{\le
0}=\sv_{<0}\oplus\sv_{(0)}$.
Define $\C_{h,\mu}=\C \psi$ $(h,\mu\in\C)$ to be the character of
$\sv_{(0)}=\langle L_0,M_0\rangle$ such that $L_0 \psi=h\psi$,
$M_0\psi=\mu\psi$. Following
the usual definition of Verma modules (see \cite{KacRai} or 
\cite{MooPia}), we
extend $\C_{h,\mu}$
trivially to $\sv_{\le 0}$ by putting $\sv_{<0}.\psi=0$ and call
${\cal V}_{h,\mu}$ the induction of the representation $\C_{h,\mu}$ to
$\sv$:
\BEQ
{\cal V}_{h,\mu}={\cal U}(\sv)\otimes_{{\cal U}(\sv_{\le 0})}
\C_{h,\mu}.
\EEQ
Take notice that with this choice of signs, {\it negative} degree elements
$X_n,Y_n,M_n$ ($n>0$) applied to $\psi$ yield zero.

The Verma module ${\cal V}_{h,\mu}$ is positively graded through the
natural extension
of $deg$ from $\g$ to ${\cal U}(\g)$, namely, we put $({\cal
V}_{h,\mu})_{(n)}={\cal V}_{(n)}=
{\cal U}(\sv_{>0})_{(n)}\otimes\C_{h,\mu}$.

There exists exactly one bilinear form $\langle \ | \ \rangle$ (called the
{\it contravariant Hermitian form} or, as
we shall also say, 'scalar product', although it is neither necessarily
positive nor even necessarily non-degenerate)
 on
${\cal V}_{h,\mu}$ such that $\langle \psi\ |\ \psi\rangle=1$ and
$X_n^*=X_{-n},\ n\in\half\Z$ ($X=L,Y$ or $M$), where
the star means taking the adjoint with respect to the bilinear form (see
\cite{KacRai}). For the contravariant Hermitian
form, $\langle {\cal V}_{(j)}\ |\ {\cal V}_{(k)}\rangle=0$ if $j\not=k$.
It is well-known that the module
${\cal V}_{h,\mu}$ is indecomposable and possesses a unique maximal proper
sub-representation ${\cal K}_{h,\mu}$, which
is actually the kernel of Hermitian form, and such that the quotient
module ${\cal V}_{h,\mu}/{\cal K}_{h,\mu}$ is
irreducible.

Hence, in order to determine if ${\cal V}_{h,\mu}$ is irreducible, and
find the irreducible quotient representation
if it is not, one is naturally led to the computation of the {\it Kac
determinants}, by which we mean the determinants of the Hermitian form
restricted to ${\cal V}_{(n)}\times{\cal V}_{(n)}$ for each $n$.

\vskip 10 pt

Let us introduce some useful notations for {\it partitions}.

{\bf Definition 6.1.}
{\it A  partition $A=(a^1,a^2,\ldots)$ of {\it degree}
$n=\deg(A)\in\N^*$ is an ordered set $a^1,a^2,\ldots$ of non-negative
integers
such that $\sum_{i\ge 1} ia^i=n$. }

A partition can be represented as a Young tableau: one associates to $A$
a set of vertical stacks of boxes put side by side, with (from left to
right) $a^1$ stacks of
height 1, $a^2$ stacks of height 2, and so on.

The {\it width} $\wid(A)$ of the tableau is equal to $\sum_{i\ge 1} a^i$.

By convention, we shall say that there is exactly one partition of degree
$0$,
denoted by $\emptyset$, and such that $\wid(\emptyset)=0$.

Now any partition $A$ defines elements of ${\cal U}(\sv)$, namely,  let us
put $X^{-A}=X_{-1}^{A_1}X_{-2}^{A_2}\ldots$ ($X$ stands here for $L$ or
$M$)
and $Y^{-A}=Y_{-\half}^{A_1}Y_{-{3\over 2}}^{A_2}\ldots$, so that
$\deg(X^{-A})=\deg(A)$ and
$$\deg(Y^{-A})=\sum_{i\ge 1} (i-\half)A^i=\deg(A)-\half \wid(A)$$ (we shall
also call this expression the {\it shifted degree} of $A$, and
 write it $\widetilde{\deg}(A)$).

{\bf Definition 6.2.}
{\it We  denote by ${\cal P}(n)$ (resp. $\tilde{\cal P}(n)$)   the set of
partitions
 of degree (resp. shifted degree) $n$.}

By Poincar\'e-Birkhoff-Witt's theorem (PBW for short), ${\cal V}_{(n)}$ is
generated by the vectors \\
$Z=X^{-A}Y^{-B}M^{-C}\psi$ where $A,B,C$  range
among all partitions such that $\deg(A)+\widetilde{\deg}(B)+\deg(C)=n$. On
this basis of ${\cal V}_{(n)}$, that we shall call in the sequel the {\it
PBW basis at degree $n$}, it is possible to define three partial
graduations,
namely, $\deg_L(Z)=\deg(A),$ $\widetilde{\deg}_Y(Z)=\widetilde{\deg}(B)$
and $\deg_M(Z)
=\deg(C).$

It is then of course easy to express the dimension of ${\cal V}_{(n)}$ as
a (finite) sum of products of the  partition function $p$ of number
theory, but we do not
know how to simplify this (rather complicated) expression, so it is of
practically no use. Let us rather write the set of above generators for
degree $n=0,\half,1,{3\over 2}$ and $2$:

$${\cal V}_{(0)}=\langle\psi\rangle$$
$${\cal V}_{(\half)}=\langle Y_{-\half}\psi\rangle$$
$${\cal V}_{(1)}=\langle (M_{-1}\psi), (Y_{-\half}^2\psi,
L_{-1}\psi)\rangle$$
$${\cal V}_{({3\over 2})}=\langle (M_{-1}Y_{-\half}\psi),(Y_{-\half}^3\psi,
Y_{-{3\over 2}}\psi, L_{-1}Y_{-\half}\psi)\rangle$$
\BEQ
{\cal V}_{(2)}=\langle (M_{-1}^2\psi,M_{-2}\psi), \
(M_{-1}Y_{-\half}^2\psi,
M_{-1}L_{-1}\psi), \ (Y_{-\half}^4 \psi,Y_{-\half}Y_{-{3\over 2}}\psi,
X_{-1}Y_{-\half}^2 \psi, X_{-2}\psi,X_{-1}^2\psi )\rangle.
\EEQ

The elements of these Poincar\'e-Birkhoff-Witt bases have been written in
the {\it $M$-order} and separated
into {\it blocks} (see below paragraph 6.3 for a definition of these terms).

The Kac determinants are quite easy to compute in the above bases at degree
$0,\half,1$. If $\{x_1,\ldots,x_{\dim({\cal V}_n)}\}$ is the PBW  basis at
degree $n$,  put
$$\Del_n^{\sv}=\det (\langle x_i|x_j\rangle)_{i,j=1,\ldots,dim({\cal
V}_n)}.$$ Note that
$\Del_n^{\sv}$ does not depend
on the ordering of the elements of the basis $\{x_1,\ldots,x_{\dim({\cal
V}_n)}\}$.

Then
$$\Del_0^{\sv}=1$$
$$\Del_{\half}^{\sv}=\langle Y_{-\half}\psi,Y_{-\half}\psi\rangle=\mu$$
$$\Del_1^{\sv}=\det \left(\begin{array}{ccc} 0& 0&\mu \\ 0 & 2\mu^2 & \mu\\
\mu&\mu&2h\end{array}\right)=-2\mu^4.$$

For higher degrees, straigthforward computations become quickly  dull:
even at degree 2, one gets a $9\times 9$ determinant, to be compared with
the simple $2\times 2$-determinant
that one gets when computing the Kac determinant of the Virasoro algebra
at level 2.

The essential idea for calculating this determinant at degree $n$ is to
find
two permutations  $\sigma,\tau$ of the set of elements
of the basis, ${x_1,\ldots,x_{\dim({\cal V}_n)}}$, in such a way that the
matrix
$(\langle x_{\sigma_i}|x_{\tau_j}\rangle)_{i,j=1,\ldots,p}$ be
upper-triangular. Then the Kac determinant $\Del_n^{\sv}$, as computed in
the basis
$\{x_1,\ldots,x_{\dim({\cal V}_n)}\}$, is equal (up to a sign) to the
product of diagonal elements
of that matrix, which leads finally to the following theorem.

{\bf Theorem 6.2.}

{\it The Kac determinant $\Del_n^{\sv}$ is given (up to a non-zero
constant) by
\BEQ
\Del_n^{\sv}=\mu^{\sum_{0\le j\le n} \sum_{B\in\tilde{\cal P}(j)}
\left( wid(B)+2\sum_{0\le i\le n-j} p(n-j-i)(
 \sum_{A\in{\cal P}(i)} wid(A))\right) }
\EEQ
where $p(k):=\# {\cal P}(k)$ is the usual partition function.

The same formula holds for the central extension of ${\sv}$.}

The proof is technical but conceptually easy, depending essentially
 on the fact that $\h$ contains a central subalgebra, namely $\langle
M_n\rangle_{n\in\Z}$, that is a module
of tensor densities for the action of the Virasoro algebra. This fact
implies that, in the course of the computations,
the $(L,M)$-generators decouple from the other generators (see Lemma 6.6).

As a matter of fact, we shall need on our way to compute the Kac
determinants
for the subalgebra $\langle L_n,M_n\rangle_{n\in\Z}\simeq
\Vect(S^1)\ltimes{\cal F}_0$ or $\Vir\ltimes{\cal F}_0$.
 The result is
very similar and encaptures, so we think, the main characteristics of the
Kac determinants of Lie algebras $\Vect(S^1)\ltimes \kk$ or
$\Vir\ltimes\kk$ such that $\kk$ contains in its center a module of tensor
densities. A contrario, the very
first computations for the deformations and central extensions of
$\Vect(S^1)\ltimes{\cal F}_0$  obtained through
the cohomology spaces $H^2(\Vect(S^1),{\cal F}_0)$ and
$H^2(\Vect(S^1)\widetilde{\ltimes}{\cal F}_0,\C)$ (denoting by
$\widetilde{\ltimes}$ any deformed product)  show that
the Kac determinants look completely different as soon as the image of
$\langle M_n\rangle_{n\in\Z}$ is not central
any more in $\kk$.

 We state the result for $\Vir\ltimes{\cal F}_0$ as follows.

{\bf Theorem 6.1}
{\it Let $\Vir_c$ be the Virasoro algebra with central charge $C\in\R$ and
 ${\cal V}'={\cal V}'_{h,\mu}={\cal U}((\Vir_c\ltimes{\cal
F}_0)_{0})\otimes_{{\cal U}((\Vir_c\ltimes{\cal F}_0)_{\le
0})} \C_{h,\mu}\subset{\cal V}_{h,\mu}$ be the Verma module representation
of $\Vir_c\ltimes{\cal F}_0$ induced
from the character $\C_{h,\mu}$, with the graduation naturally inherited
from that of ${\cal V}_{h,\mu}$.

 Then the Kac determinant (computed in the PBW bases)
$\Del_n^{\Vir\ltimes{\cal F}_0}$
 of ${\cal V}'_{h,\mu}$ at degree $n$ is equal (up to
a positive constant) to
\BEQ
\Del_n^{\Vir_c\ltimes{\cal F}_0}=(-1)^{\dim({\cal V}'_{\mu})}
\mu^{2\sum_{0\le i\le n}p(n-i)(\sum_{A\in{\cal P}(i)} wid(A))}.
\EEQ

It does not depend on $C$. }

\subsection{ Kac determinant formula for $\Vir\ltimes{\cal F}_0$}

We first need to introduce a few notations and define two different
 orderings for the PBW bases of $\Vir\otimes {\cal F}_0$.

{\bf Definition 6.3.}
{\it  Let $A,B$ be two partitions of $n\in\N^*$. One says that
$A$ is {\it finer} than $B$, and write $A\preceq B$, if $A$ can be gotten
from $B$ by a finite number of transformations $B\to \cdots \to D\to
D'\to\cdots A$ where
$$D'^i=D^i+C^i\quad (i\not=p),\quad D'^p=D^i-1,$$
$C=(C^i)$ being a non-trivial partition of degree $p$.}

Graphically, this means that the Young tableau of $A$ is obtained from the
Young tableau of $B$ by splitting some of the stacks of boxes into several
stacks.

The relation $\preceq$ gives a {\it partial} order on the set of partitions
of fixed degree $n$, with smallest element $(1,1,\ldots,1)$ and largest
element the trivial partition $(n)$. One {\it chooses} arbitrarily, for
every degree $n$, a
total ordering $\le$ of ${\cal P}(n)$ compatible with $\preceq$, i.e. such that
$(A\preceq B)\Rightarrow(A\le B)$.

{\bf Definition 6.4.}
{\it  If $A$ is a partition, then $X^A:=(X^{-A})^*$ is given by
$X^A=\ldots X_2^{A_2}X_1^{A_1}$ (where $X$ stands for $L$ or $M$).}

Let $n\in\N$. By Poincar\'e-Birkhoff-Witt's theorem, the $L^{-A}M^{-C}$
($A$ partition of degree $p$, $C$ partition
of degree $q$, $p,q\ge 0$, $p+q=n$) form a basis ${\cal B}'$ of ${\cal
V}'_n$, the subspace of ${\cal V}'$ made
up of the vectors of degree $n$.

We now give two different orderings of the set ${\cal B}'$, that we call
{\it horizontal ordering} (or {\it M-ordering})
 and {\it vertical ordering} (or {\it L-ordering}). For the {\it
horizontal ordering}, we proceed as follows:

-we split ${\cal B}'$ into $(n+1)$ {\it blocks} ${\cal B}'_0,\ldots,{\cal
B}'_n$ such that
\BEQ
{\cal B}'_j=\{L^{-A}M^{-C}\psi\ |\ \deg(A)=j,\deg(C)=n-j\};
\EEQ

-we split each block ${\cal B}'_j$ into {\it sub-blocks} ${\cal B}'_{j,C}$
(also called $j$-sub-blocks if one
wants to be more explicit) such that $C$ runs among the set of partitions
of $n-j$ in the {\it increasing} order
chosen above, and $${\cal B}'_{j,C}=\{L^{-A}M^{-C}\psi\ |\ \deg(A)=j\}.$$

-finally, inside each sub-block ${\cal B}'_{j,C}$, we take the elements
$L^{-A}M^{-C}$, $A\in {\cal P}_j$, in
the {\it decreasing order}.

For the {\it vertical ordering}, we split ${\cal B}'$ into blocks
$$\widetilde{\cal B}'_j=\{L^{-A}M^{-C}\psi\ |\ \deg(A)=n-j,\deg(C)=j\},$$
each of these blocks into sub-blocks $\widetilde{\cal B}'_{j,A}$ where $A$
runs among all partitions of $n-j$ in
the increasing order, and take the elements $L^{-A}M^{-C}\in
\widetilde{\cal B}'_{j,A}$ according to the decreasing
order of the partitions $C$ of degree $j$.

As one easily checks, $L^{-C}M^{-A}\psi$ is at the same place in the
vertical ordering as $L^{-A}M^{-C}$ in the
horizontal ordering. Note that the vertical ordering can also be obtained
by reversing  the horizontal ordering. Yet we
maintain the separate definitions both for clarity and because these
definitions will be extended in the next paragraph
to the case of $\sv$, where there is no simple relation between the two
orderings.

Roughly speaking, one may say that, for the horizontal ordering, the
degree in $M$ decreases from one block to the
next one, the $(M^C)_{C\in{\cal P}(n-j)}$
 are chosen in the increasing order inside each block, and then the $L^A$
are chosen in the
decreasing order inside each sub-block;  the vertical ordering is defined
in exactly the same way, except
that $L$ and $M$ are exchanged.

We shall compute the Kac determinant $\Del'_n:=\Del_n^{\Vir\ltimes{\cal
F}_0}$ relative to ${\cal B}'$
by representing it as
\BEQ
\Del'_n=\pm \det {\cal A}'_n,\quad
{\cal A}'_n=(\langle H_j|V_i\rangle)_{i,j}
\EEQ
where the $(H_j)_j\in{\cal B}'$ are chosen in the horizontal order and the
$(V_i)_i\in{\cal B}'$ in the vertical
order.

The following facts are clear from the above definitions:

-- the horizontal and vertical blocks ${\cal B}'_j, \widetilde{\cal B}'_j$
($j$ fixed)
 and sub-blocks ${\cal B}'_{j,C},\widetilde{\cal B}'_{j,C}$ ($j,C$ fixed)
have same size, so one has matrix diagonal blocks and sub-blocks;

-- diagonal elements are of the form $\langle L^{-A}M^{-C}\psi\ |\
L^{-C}M^{-A}\psi\rangle$;

-- define {\it sub-diagonal} elements to be the $\langle H|V\rangle$,
$H\in{\cal B}'_j, V\in{\cal B}'_i$ such
that $i>j$; then $\deg_L (V)<\deg_M (H)$;

-- define $j$-{\it sub-sub-diagonal} elements (or simply sub-sub-diagonal
elements if one doesn't need to be very
definite) to be the $\langle H|V\rangle$, $H\in{\cal B}'_{j,C_1},
V\in{\cal B}'_{j,C_2}$ with $C_2>C_1$. Then
$\deg_L(V)=\deg_M(H)=n-j$ and $H=L^{-A_1}M^{-C_1}, V=L^{-C_2}M^{-A_2}$ for
certain partitions $A_1,A_2$ of degree $j$;

-- define $(j,C)$-{\it sub$^3$}-diagonal elements to be the $\langle
H|V\rangle$, $H,V\in {\cal B}'_{j,C}$, such
that $H=L^{-A_1}M^{-C}, V=L^{-C}M^{-A_2}$ with $A_1>A_2$.

Then the set of sub-diagonal elements is the union of the matrix blocks
situated under the diagonal; the set
of $j$-sub-sub-diagonal elements is the union of the matrix sub-blocks
situated under the diagonal
of the $j$-th matrix diagonal blocks; the
set of $(j,C)$-sub$^3$-diagonal elements is the union of the elements
situated under the diagonal of the
$(j,C)$-diagonal sub-block. All these elements together form the set of
lower-diagonal elements of the matrix
 ${\cal A}'_n$.

Elementary computations show that
$${\cal A}'_1=\left(\begin{array}{cc} \mu& 2h\\ 0&\mu \end{array}\right)$$
with horizontal ordering $(M^{-1}\psi,L^{-1}\psi)$;
 $${\cal A}'_2=
\left(\begin{array}{ccccc} 2\mu^2& 2\mu & 2\mu(1+2h)& 6h& 4h(2h+1)
\\
  0& 2\mu & 3\mu & 4h+c/2 & 6h \\
0 &0& \mu^2&3\mu&2\mu(1+2h)\\
0&0&0&2\mu&2\mu\\
0&0&0&0&2\mu^2 \end{array}\right)$$
with horizontal ordering $((M_{-1}^2\psi, M_{-2}\psi),(L_{-1}M_{-1}\psi),
(L_{-2}\psi,L_{-1}^2\psi))$
(the blocks being separated by parentheses). Note that the Kac determinant
of $\Vir_c$ at level 2 appears as the
top rightmost $2\times 2$ upper-diagonal block of ${\cal A}'_2$, and hence
does not play any role in the computation
of $\Del'_n$.

So ${\cal A}'_1$, ${\cal A}'_2$ are upper-diagonal matrices, with diagonal
elements that are (up to a coefficient) simply
powers of $\mu$. More specifically, $\Del'_1=-\det {\cal A}'_1=-\mu^2$ and
$\Del'_2=\det {\cal A}'_2=16\mu^8.$

The essential technical lemmas for the proof of Theorem 6.1 and Theorem
6.2  are
Lemma 6.1 and Lemma 6.2, which
show, roughly speaking, how to move the $M$'s through the $L$'s.

{\bf Lemma 6.1}

{\it Let $A,C$ be two partitions of degree $n$. Then:

\begin{enumerate}
\item[(i)] If $A\npreceq C$, then $\langle L^{-A}\psi\ |\
M^{-C}\psi\rangle=0$.
\item[(ii)] If $A\preceq C$, then
\BEQ
\langle L^{-A}\psi\ |\ M^{-C}\psi\rangle=a. \ \mu^{\wid(C)}
\EEQ
for a certain positive constant $a$ depending only on $A$ and $C$.
\end{enumerate} }

{\bf Proof of Lemma 6.1}

We use induction on $n$. Take $A=(a_i),C=(c_i)$ of degree $n$, then
$$\langle L^{-A}\psi|M^{-C}\psi\rangle=\langle \psi\ |\ \left(
\prod_{i=\infty}^1 L_i^{a_i}\right)
\left( \prod_{j=1}^{\infty} M_{-i}^{c_i}\right)\psi\rangle.$$
We shall compute $\langle L^{-A}\psi|M^{-C}\psi\rangle$ by moving
successively to the left the $M_{-1}$'s,
then the $M_{-2}$'s and so on, and taking care of the commutators that
show up in the process.

\begin{itemize}
\item
Suppose $c_1>0$. By commuting $M_{-1}$ with the $L$'s, there appear terms
of two types (modulo positive constants) :

- either of type $$\langle \psi\ | \left( \prod_{i=\infty}^{k+1}
L_i^{a_i}\right) (L_k^{a'_k}M_{k-1}L_k^{a''_k})
(\prod_{i=k-1}^1 L_i^{a_i}) M_{-1}^{c_1-1} (\prod_{i=2}^{\infty}
M_{-i}^{c_i})\psi\rangle,$$ with $a'_k+a''_k=a_k-1$
(by commuting with $L_k$, $k\ge 2$).  But this is zero since transfering
$M_{k-1}$ (of negative degree $1-k$)
to the right through the $L$'s can only lower its degree.

- or of type   $$\langle \psi\ | \left( \prod_{i=\infty}^{2}
L_i^{a_i}\right) (L_1^{a'_1}M_{0}L_1^{a''_1})
 M_{-1}^{c_1-1} (\prod_{i=2}^{\infty} M_{-i}^{c_i})\psi\rangle,$$ with
$a'_1+a''_1=a_1-1$ (by commuting with
$L_1$). Since the central element $M_0\equiv \mu$ can be taken out of the
brackets, we may compute this as $\mu$
times
a scalar product between two elements of degree $n-1$. Removing one
$M_{-1}$ and one $L_{-1}$ means removing
the leftmost column of the Young tableaux  $C$ and $A$. Call $C',A'$ the
new tableaux: it is clear that $A\npreceq
C\Rightarrow A'\npreceq C'.$ So we may conclude by induction.

\item Suppose that all $c_1=\ldots=c_{j-1}=0$ and $c_j>0$. Then, by
similar arguments, one sees that all potentially
non-zero terms appearing on the way (while moving $M_{-j}$ to the left)
are of the form
$$\alpha\mu\ .\  \langle \psi\ |\ (\prod_{i=\infty}^1 L_i^{a'_i})
M_{-j}^{c_j-1} (\prod_{ij} M_{-i}^{c_i}) \psi\rangle,$$
$\sum_{i=\infty}^1 i(a_i-a'_i)=j$, with $\alpha$ defined by
$$\left(\prod_{i=\infty}^1 (ad L_i)^{a_i-a'_i}\right) \ .\ M_{-j}=\alpha
M_0.$$
Without computing $\alpha$ explicitly, it is clear that $\alpha>0$. On the
Young tableaux, this corresponds to
removing one stack of height $j$ from $C$ and $(a_i-a'_i)$ stacks of
height $i$ ($i=\infty,\ldots,1$) from $A$. Once
again, $A\npreceq C\Rightarrow A'\npreceq C'$. So one may conclude by
induction.

\end{itemize} 
\eop

{\bf Lemma 6.2.}

{\it Let $A_1,A_2,C_1,C_2$ be partitions such that $\deg A_1+\deg C_1=\deg
A_2+\deg C_2$. Then :

\begin{itemize}
\item[(i)] $$\langle L^{-C_2}M^{-A_2}\psi\ |\
L^{-A_1}M^{-C_1}\psi\rangle=0$$ if $\deg A_1<\deg A_2$.
\item[(ii)] If $\deg A_1=\deg A_2$, then
\BEQ
\langle L^{-C_2}M^{-A_2}\psi \ |\ L^{-A_1}M^{-C_1}\psi\rangle=\langle
L^{-C_2}\psi|M^{-C_1}\psi\rangle
\langle M^{-A_2}\psi|L^{-A_1}\psi\rangle.
\EEQ
\end{itemize} }

{\bf Remark.} The central argument in this Lemma can be trivially
generalized (just by using the fact that the $M$'s
are central in $\h$) in a form that will be used again and again
in the next section : namely,
$$\langle \psi\ |\ UL^{C_2}VL^{-A}WM^{-C_1}\psi\rangle=0\quad
(U,V,W\in{\cal U}(\h))$$
if $\deg(C_1)> \deg(C_2).$

{\bf Proof.}

Putting all generators on one side, one gets
$$\langle L^{-C_2}M^{-A_2}\psi\ |\
L^{-A_1}M^{-C_1}\psi\rangle=\langle\psi\ |\
M^{A_2}L^{C_2}L^{-A_1}M^{-C_1}\psi\rangle.$$
Assume that $\deg(A_1)\le \deg(A_2)$ (so that $\deg(C_2)\le \deg(C_1)$).
Let us move $M^{-C_1}$ to the left and consider the
successive commutators with the $L$'s: it is easy to see that one gets
(apart from the trivially commuted
term $M^{-C_1}L^{-A_1}$) a sum of terms of the type $M^{-C'_1}
L^{-A'_1}$ with $$\deg(C'_1)+\deg(A'_1)=\deg(C_1)+\deg(A_1),\quad
\deg(C'_1)>\deg(C_1), \ \deg(A'_1)<\deg(A_1).$$

Now comes the central argument : let us commute  $M^{-C'_1}$ through
$L^{C_2}$. It yields terms of the type
$M_0^{\al}M^{-C''_1}M^{C'''_1}L^{C'_2}$ with
$\deg(C_2)-\deg(C'_1)=\deg(C'_2)+\deg(C'''_1)-\deg(C''_1)$.

If $C''_1\neq\emptyset$, then $M^{-C''_1}$ commutes with $M^{A_2}$ and
gives $0$ when set against $\langle\psi$; so one may assume
$C''_1=\emptyset$. But this is impossible, since it would imply
\BEA
\deg(C'''_1)&=&\deg(C_2)-\deg(C'_1)-\deg(C'_2)\\
&\le &\deg(C_1)-\deg(C'_1) \quad ({\mathrm{even}}\ < \ {\mathrm{if}} \
\deg(C_2)<\deg(C_1)) \\
&<&0.
\EEA

So
$$\langle L^{-C_2}M^{-A_2}\psi\ |\
L^{-A_1}M^{-C_1}\psi\rangle=\langle\psi\ |\
M^{A_2}L^{C_2}M^{-C_1}L^{-A_1}\psi\rangle.$$
What's more,  the above argument applied to $M^{-C_1}$ instead of
$M^{-C'_1}$, shows that
$$\langle\psi\ |\ M^{A_2}(L^{C_2}M^{-C_1})L^{-A_1}\psi\rangle=\langle\psi\
|\ M^{A_2}(M^{-C_1}L^{C_2})L^{-A_1}\psi\rangle=0$$
if $\deg(C_2)<\deg(C_1)$. So (i) holds and one may assume that
$\deg(A_1)=\deg(A_2),\ \deg(C_1)=\deg(C_2)$ in the sequel.

Now move $M^{A_2}$ to the right: the same argument proves that
$$\langle\psi\ |\ M^{A_2}L^{C_2}M^{-C_1}L^{-A_1}\psi\rangle=\langle \psi\
|\ (L^{C_2}M^{-C_1})(M^{A_2}X^{-A_1})\psi\rangle.$$
But $M^{A_2}L^{-A_1}$ has degree $0$, so
\BEA
\langle\psi \ |\
(L^{C_2}M^{-C_1})(M^{A_2}L^{-A_1})\psi\rangle&=&\langle\psi\ |\
L^{C_2}M^{-C_1}\psi\rangle\langle\psi\ |\
M^{A_2}L^{-A_1}\psi\rangle\\
&=& \langle L^{-C_2}\psi\ |\ M^{-C_1}\psi\rangle\langle M^{-A_2}\psi\ |\
L^{-A_1}\psi\rangle.
\EEA
 \eop

{\bf Corollary 6.3.}

{\it The matrix ${\cal A}'_n$ is upper-diagonal.}

{\bf Proof.}

Lower-diagonal elements come in three classes: let us give an argument for
each class.

By Lemma 6.2, (i), sub-diagonal elements are zero.

Consider a $j$-sub-sub-diagonal element $\langle L^{-C_2}M^{-A_2}\psi\ |\
L^{-A_1}M^{-C_1}\psi\rangle$
with $\deg(C_1)=\deg(C_2)=n-j$, $\deg(A_1)=\deg(A_2)$, $C_2>C_1$. Then, by
Lemma 6.2., (ii), and Lemma 6.1, (i),
$$\langle L^{-C_2}M^{-A_2}\psi\ |\ L^{-A_1}M^{-C_1}\psi\rangle=\langle
L^{-C_2}\psi\ |\ M^{-C_1}\psi\rangle
\langle M^{-A_2}\psi\ |\ L^{-A_1}\psi\rangle=0.$$

Finally, consider a $(C,j)$-sub$^3$-diagonal element $\langle
L^{-C}M^{-A_2}\psi\ |\ L^{-A_1}M^{-C}\psi\rangle$ with
$A_1>A_2$. By the same arguments, this is zero. \eop

{\bf Proof of Theorem 6.1}

By Corollary 6.3, one has $$\Del'_n=(-1)^{\dim({\cal
V}'_n)}\prod_{i=1}^{\dim({\cal V}'_n)} ({\cal A}'_n)_{ii}.$$
By Lemma 6.2 (ii) and Lemma 6.1 (ii), the diagonal elements of ${\cal
A}'_n$ are equal (up to a positive
constant) to $\mu$ to a certain power. Now the total power of $\mu$ is
equal to
\BEQ
\sum_{0\le j\le n} \sum_{A\in{\cal P}(j)} \sum_{C\in{\cal P}(n-j)}
\left(\wid(A)+\wid(C)\right)
=2\sum_{0\le j\le n} p(n-j) \left( \sum_{A\in{\cal P}(j)} \wid(A)\right).
\EEQ
\eop

\subsection{Kac determinant formula for $\sv$}

Let $n\in\half\N$. We shall define in this case also a {\it horizontal
ordering} (also called {\it $M$-ordering})
 and a {\it vertical
ordering} (also called {\it $L$-ordering}) of the Poincar\'e-Birkhoff-Witt
basis ${\cal B}=\{L^{-A}Y^{-B}M^{-C}\psi \ |
\ \deg(A)+\widetilde{\deg}(B)+\deg(C)=n\}$ of ${\cal V}_{(n)}$.

The {\it $M$-ordering} is defined as follows (note that blocks and
sub-blocks are defined more or less as in the preceding sub-section):

-- split ${\cal B}$ into $(n+1)$-blocks ${\cal B}_0,\ldots,{\cal B}_n$ such
that ${\cal B}_j=\{L^{-A}Y^{-B}M^{-C}\psi\in{\cal B}\
|\ \deg(C)=n-j\}$;

-- split each block ${\cal B}_j$ into sub-blocks ${\cal B}_{jC}$ such that
${\cal B}_{jC}=\{L^{-A}Y^{-B}M^{-C}\psi\in{\cal B}\}$,
with $C$ running among the set of partitions of $n-j$ in the {\it
increasing} order;

-- split each sub-block ${\cal B}_{jC}$ into sub$^2$-blocks ${\cal
B}_{jC,\kappa}$ (for {\it decreasing} $\kappa$)
 with $L^{-A}Y^{-B}M^{-C}\in{\cal B}_{jC,\kappa}
\Leftrightarrow L^{-A}Y^{-B}M^{-C}\in{\cal B}_{jC}$ and
$\widetilde{\deg}(B)=\kappa$;

-- split each sub$^2$-block ${\cal B}_{jC,\kappa}$ into sub$^3$-blocks
${\cal B}_{jC,\kappa B}$ where $B$ runs among all
partitions of shifted degree $\kappa$ (in any randomly chosen order);

-- finally, order the elements $L^{-A}Y^{-B}M^{-C}\psi$ of ${\cal
B}_{jC,\kappa B}$  $(A\in{\cal P}(n-\deg(C)-\widetilde{\deg}(B))={\cal
P}(j-\kappa))$
so that the $A$'s appear in the {\it decreasing} order.

Then the {\it $L$-ordering} is chosen in such a way that
$L^{-C}Y^{-B}M^{-A}$ appears vertically in the same place as
$L^{-A}Y^{-B}M^{-C}$ in the $M$-ordering. From formula (6.2) giving the
$M$-ordering of ${\cal V}_{(2)}$, it
is clear that the $L$-ordering is {\it not} the opposite of the
$M$-ordering.

These two orderings define as in paragraph 6.2  a block matrix ${\cal A}_n$
whose determinant is equal to $\pm \Del_n^{\sv}$.

We shall need three preliminary lemmas.

{\bf Lemma 6.4.}

{\it Let $B=(b_j),B'=(b'_j)$ be two partitions with same shifted degree:
then
\BEQ
\langle Y^{-B}\psi\ |\ Y^{-B'}\psi\rangle=\delta_{B,B'} \left[ \prod_{j\ge
1} (b_j)!(2j-1)^{b_j} \right]
\ .\ \mu^{\wid(B)}.
\EEQ }

{\bf Proof.}

Consider the sub-module ${\cal W}\subset{\cal V}$ generated by the
$Y^{-P}M^{-Q}\psi$ ($P,Q$ partitions).  Then, inside this
module, and as long as vacuum expectation values $\langle \psi \ |\
Y_{\pm(j_1-\half)}\ldots
Y_{\pm(j_k-\half)}\psi\rangle$ are  concerned, the
$(Y_{-(j-\half)},Y_{j-\half})$ can be considered as independent couples of
creation/annihilation operators with
$[Y_{(j-\half)},Y_{-(j-\half)}]=(2j-1)M_0.$ Namely, other commutators
$[Y_k,Y_l]$ with $k+l\not=0$ yield $(k-l)M_{k+l}$ which commutes with all
other generators $Y$'s and $M$'s and gives $0$ when applied to
$\psi\rangle$ or $\psi\langle$ according to the sign
of $k+l$. The result now follows for instance by an easy application of
Wick's theorem, or by induction.
\eop

{\bf Lemma 6.5.}

{\it Let $B_1,B_2,C_1,C_2$ be partitions such that $C_1\neq\emptyset$ or
$C_2\neq\emptyset.$ Then
$$\langle Y^{-B_1}M^{-C_1}\psi\  |\
Y^{-B_2}M^{-C_2}\psi\rangle=0.$$}

{\bf Proof.} Obvious (the $M^{\pm C}$'s are central in the subalgebra
$\h\subset\sv$
 and can thus be commuted freely with the $Y$'s and the $M$'s; set against
$\langle\psi$ or $\psi\rangle$
 according to the sign of their degree,
they give zero). \eop

{\bf Lemma 6.6.}

{\it Let $A_1,A_2,B_1,B_2,C_1,C_2$ be partitions such that
$\deg(A_1)+\deg(B_1)+\deg(C_1)=\deg(A_2)+\deg(B_2)+\deg(C_2).$ Then

\begin{enumerate}
\item[(i)] If $\deg(A_1)<\deg(A_2)$ or $\deg(C_2)<\deg(C_1)$, then
$$\langle L^{-C_2}Y^{-B_2}M^{-A_2}\psi \ |\
L^{-A_1}Y^{-B_1}M^{-C_1}\psi\rangle=0.$$
\item[(ii)] If $\deg(A_1)=\deg(A_2)$ and $\deg(C_1)=\deg(C_2)$, then
$$\langle L^{-C_2}Y^{-B_2}M^{-A_2}\psi \ |\
L^{-A_1}Y^{-B_1}M^{-C_1}\psi\rangle=\langle Y^{-B_2}\psi\ |\
Y^{-B_1}\psi\rangle
\langle M^{-A_2}\psi \ |\ L^{-A_1}\psi\rangle \langle L^{-C_2}\psi \ |\
M^{-C_1}\psi\rangle.$$
\end{enumerate} }

{\bf Proof of Lemma 6.6.}

 (i) is a direct consequence of the remark following Lemma 6.2. So we may
assume that
$\deg(A_1)=\deg(A_2)$ and $\deg(C_1)=\deg(C_2)$. Using the hypothesis
$\deg(A_1)=\deg(A_2)$, we may choose this
time to move $L^{-A_1}$ to the left in the expression $\langle \psi\ |\
M^{A_2}Y^{B_2}L^{C_2}L^{-A_1}Y^{-B_1}M^{-C_1}
\psi\rangle.$ Commuting $L^{-A_1}$ through $L^{C_2}$ and $Y^{B_2}$, one
obtains terms of the type
$$\langle \psi\ |\ M^{A_2} (L^{-A'_1} Y^{B'_2}
L^{C'_2})Y^{-B_1}M^{-C_1}\psi\rangle,$$
with $\deg(A'_1)\le \deg(A_1)$, and $(\deg(A'_1)=\deg(A_1))\Rightarrow
(A'_1=A_1,B'_2=B_2,C'_2=C_2)$. So, by
the Remark following Lemma 6.2,

\BEA
\langle L^{-C_2}Y^{-B_2}M^{-A_2}\psi \ |\
L^{-A_1}Y^{-B_1}M^{-C_1}\psi\rangle &=& \langle\psi \ |\
M^{A_2}Y^{B_2}L^{C_2}L^{-A_1}
Y^{-B_1}M^{-C_1}\psi\rangle\\
&=& \langle \psi\ |\
Y^{B_2}L^{C_2}M^{A_2}L^{-A_1}Y^{-B_1}M^{-C_1}\psi\rangle\\
 &=& \langle \psi\ | M^{A_2}L^{-A_1}\psi\rangle \langle \psi\ |\
Y^{B_2}L^{C_2}Y^{-B_1}M^{-C_1}\psi\rangle.
\EEA

Moving $L^{C_2}$ to the right in the same way leads to (ii),  thanks to
the hypothesis $\deg(C_1)=\deg(C_2)$ this time.
\eop

{\bf Corollary 6.7.}

{\it The matrix ${\cal A}_n$ is upper-diagonal.}

{\bf Proof.}

Lower-diagonal elements $f=\langle L^{-C_2}Y^{-B_2}M^{-A_2}\psi \ |\
L^{-A_1}Y^{-B_1}M^{-C_1}\psi\rangle$ come  this
time in five classes. Let us treat each class separately.

By Lemma 6.6 (i), sub-diagonal elements (characterized by
$\deg(C_2)<\deg(C_1)$) are zero.

Next, $j$-sub$^2$-diagonal elements (characterized by
$\deg(C_2)=\deg(C_1)=n-j$, $C_2>C_1$) are also zero: namely,
the hypothesis $\deg(C_1)=\deg(C_2)$ gives as in the proof of Lemma 6.6.
\BEA
f&=& \langle \psi\ |\ M^{A_2}Y^{B_2}L^{C_2}L^{-A_1}Y^{-B_1}M^{-C_1}\psi
\rangle\\
&=& \langle \psi\ |\ L^{C_2} M^{-C_1}\psi\rangle\langle\psi\ |\
M^{A_2}Y^{B_2}L^{-A_1}Y^{-B_1}\psi\rangle\\
&=& 0 \ ({\mathrm{by\ Lemma\ 6.1}}).
\EEA

Now $(j,C)$-sub$^3$-diagonal elements (characterized by $C_1=C_2:=C$,
$\widetilde{\deg}(B_2)<\widetilde{\deg}(B_1)$) are again
zero because, as we have just proved,
$$f=\langle \psi\ |\ L^{C_2}M^{-C_1}\psi\rangle\langle\psi\ |\
M^{A_2}Y^{B_2}L^{-A_1}Y^{-B_1}\psi\rangle$$
and this time $\langle\psi\ |\
M^{A_2}Y^{B_2}L^{-A_1}Y^{-B_1}\psi\rangle=0$ since
$$\deg(A_1)=j-\widetilde{\deg}(B_1)<j-\widetilde{\deg}(B_2)=\deg(A_2).$$

Finally, $(jC,\kappa)$-sub$^4$-diagonal elements (with $C_1=C_2=C$, $\deg
B_1=\deg B_2$, $\deg A_1=\deg A_2$, $B_1\not=B_2$)
are $0$ by Lemma 6.6 (ii) and Lemma 6.4 since $\langle Y^{-B_2}\psi\ |\
Y^{-B_1}\psi\rangle=0$, and
$(jC,\kappa B)$-sub$^5$-diagonal elements (such that $B_1=B_2$ but
$A_2<A_1$)
are $0$ by Lemma 6.6 (ii) and Lemma 6.1. (i) since $\langle M^{-A_2}\psi\
|\
L^{-A_1}\psi\rangle=0.$
\eop

{\bf Proof of Theorem 6.2}

By Corollary 6.7., $\Del_n^{\sv}=\pm \prod_{i=1}^{\dim({\cal V}_{(n)})}
({\cal A}_n)_{ii}.$ By Lemma 6.6. (ii),
Lemma 6.1 (ii) and Lemma 6.4., the diagonal element $\langle
L^{-C}Y^{-B}M^{-A}\psi\ |\ L^{-A}Y^{-B}M^{-C}\psi\rangle$
is equal (up to a positive constant) to $\mu^{\wid(A)+\wid(B)+\wid(C)}$,
so (see proof of Theorem 6.1) $\Del_n^{\sv}
=c_n \mu^{a_n}$ ($c_n$ non-zero constant, $a_n\in\N$) with
$$a_n=\sum_{0\le j\le n} \sum_{B\in\tilde{\cal P}(j)} (
\wid(B)+a'_{n-j}),$$
where $a'_{n-j}$ is the power of $\mu$ appearing in
$\Del_{n-j}^{\Vir\ltimes{\cal F}_0}.$ Hence the final result.
\eop

\end{document}